%% file: main_revised.tex
\documentclass[conference]{IEEEtran}

\usepackage{comment}
\usepackage{subfig}
\usepackage{subcaption} 
\usepackage{listings}
\usepackage{multirow}
\usepackage[utf8]{inputenc}
\usepackage{textcomp}
\usepackage{fancybox}
\usepackage{hyperref}
\usepackage{amsmath,amssymb,amsfonts}
\usepackage{algorithmic}
\usepackage{graphicx}
\usepackage{tcolorbox}
\usepackage{textcomp}
\usepackage{booktabs}
\usepackage{xcolor}
\usepackage{float}
\usepackage{mdframed}
\usepackage{soul} 
\usepackage{booktabs}
\usepackage{array}
\usepackage{tabularx}
\usepackage{makecell}
\usepackage{amssymb}

\newif\ifcomment
\commenttrue
\commentfalse
\ifcomment
\newcommand{\shirin}[1]{{\bf \textcolor{blue}{Shirin: #1}}}
\newcommand{\nowshin}[1]{{\bf \textcolor{violet}{Nowshin: #1}}}
\newcommand{\mohit}[1]{{\bf \textcolor{red}{Mohit: #1}}}
\else
\newcommand{\shirin}[1]{}
\newcommand{\nowshin}[1]{}
\newcommand{\mohit}[1]{}
\fi
\ifCLASSINFOpdf
\else
\fi
\begin{document}
%
\title{When Technically Plausible Advice Is Unsafe: A Cross-Ecosystem Measurement of Online Support for Technology-Facilitated Abuse}


\author{
\IEEEauthorblockN{Nowshin Tabassum\IEEEauthorrefmark{2},
Solomon G. Dandekar\IEEEauthorrefmark{2},
Morgan PettyJohn\IEEEauthorrefmark{2},
Tim Ryan\IEEEauthorrefmark{2}, \\
Minjaal Raval\IEEEauthorrefmark{2},
Rachel Voth Schrag\IEEEauthorrefmark{2}, 
Mohit Singhal\IEEEauthorrefmark{3},
Shirin Nilizadeh\IEEEauthorrefmark{2}}
\IEEEauthorblockA{\IEEEauthorrefmark{2} The University of Texas at Arlington 
\IEEEauthorblockA{\IEEEauthorrefmark{3} Northeastern University \\
(nxt6577, sxd6555, tim.ryan, mmr8972)@mavs.uta.edu \\ (morgan.pettyjohn, rachel.vothschrag, shirin.nilizadeh)@uta.edu} m.singhal@northeastern.edu}
}


%


\IEEEoverridecommandlockouts
\makeatletter\def\@IEEEpubidpullup{6.5\baselineskip}\makeatother
\IEEEpubid{\parbox{\columnwidth}{
		Network and Distributed System Security (NDSS) Symposium 2027\\
		22--26 March 2027, Seoul, Republic of Korea\\
		ISBN 978-1-970672-09-1\\  
		https://dx.doi.org/10.14722/ndss.2027.242692,\\
		www.ndss-symposium.org
}
\hspace{\columnsep}\makebox[\columnwidth]{}}

\maketitle

\begin{abstract}

Technology-facilitated abuse (TFA) creates an adversarial interpersonal setting in which conventionally sound cybersecurity advice can be unsafe: changing credentials, removing access, or resetting devices may alert an abuser, destroy evidence, or increase escalation risk. Yet victims seek guidance from search engines, peer support forums, and conversational AI systems, whose evaluations typically emphasize relevance and technical correctness over contextual safety. We conduct a measurement across these ecosystems to determine whether they safely and effectively meet TFA victims' help-seeking needs. Drawing on a decade of narratives from \emph{r/Stalking}, we construct 2,797 victim-derived queries spanning 11 named categories of technology misuse. We use these queries to analyze 27,162 Google-retrieved webpages, 2,476 Reddit query--thread responses, and 250 responses from three general-purpose LLMs and two survivor-support chatbots. Our expert-informed framework applies common measures of technical quality and damaging guidance across the three ecosystems, then examines additional risks specific to each one, including secondary-link integrity for webpages, toxicity for Reddit, and trauma-informed support for conversational systems. 
Across all three ecosystems, we identify conventional quality failures and safety risks that relevance, accuracy, and actionability alone do not capture. Google Search and most conversational systems frequently return relevant information, whereas Reddit responses are less consistently relevant and actionable. In our manually evaluated accuracy sample, 17.3\% of webpages, 13.3\% of Reddit threads, and 19.6\% of conversational AI responses contained damaging guidance. Moreover, 65.5\% of victim queries led to at least one webpage containing a secondary URL flagged by multiple VirusTotal engines, over 20\% received at least one toxic Reddit comment, and every conversational system produced some guidance that failed to account for escalation risk. Domain specialization did not guarantee better support: HopeChat underperformed general-purpose LLMs across several dimensions, while Ruth remained limited in trauma-informed support. These findings expose a gap between technical quality and contextual safety across the online support ecosystem. Safe TFA assistance requires risk-aware recommendations, trustworthy sources, uncertainty communication, and appropriate human support, not merely relevant or technically plausible answers.
\end{abstract}


%
\IEEEpeerreviewmaketitle

\input{introduction_revised}
\input{reddit_posts}

\input{modelling_victim_queries}

\input{evaluation}

\input{technical_evaluation}
\input{social_evaluation}
\input{results_tech}
\input{results_social}
\input{discussions}

\input{background}

\input{conclusion}
\input{ethics}

\bibliographystyle{IEEEtran}

\bibliography{refs}

\input{appendix}

\end{document}

%% file: introduction_revised.tex
\section{Introduction}

Technology-facilitated abuse (TFA) is the misuse of digital technologies to stalk, monitor, harass, threaten, or control another person~\cite{stalkerparadise,techstalking,TFAdefinition}. It exploits everyday systems that victims must continue to navigate, including smartphones, online accounts, location-sharing services, financial applications, and smart-home devices, and embeds security decisions within an adversarial interpersonal relationship. In this setting, an action that would be routine after an ordinary account or device compromise can carry different consequences. Changing a password, removing an account, resetting a device, disabling tracking, or blocking another person may alert an abuser, destroy evidence, or provoke escalation. Thus, technically correct advice is not necessarily safe advice; the safety consequences of acting on it are part of the security problem~\cite{clinicalTFA,frydmanTIA,prakash2026assessing}.

Obtaining such context-sensitive assistance remains difficult. Advocates and social workers may lack specialized digital-safety expertise, while risk-assessment and safety-planning practices struggle to keep pace with changing technologies~\cite{lacoftechadvocates,lackofexpertise,barriershelp-seeking}. Victims may also avoid formal assistance because of privacy concerns, retaliation, stigma, or fear of not being believed~\cite{helpseekingyouth,barriershelp-seeking,techabusepersonas,disclosureofabuse}. Consequently, many seek information through web search, forums, and conversational AI systems to interpret what is happening, assess risk, and choose protective actions~\cite{webofabuse,chatbotdesigning,chatbotevaluation,prakash2026assessing}. These systems collectively form a digital support ecosystem, but they were not necessarily designed to reason about the interaction between technical remediation and interpersonal safety.

Existing studies offer important but partial views of this ecosystem. Prior work has examined TFA-related resources returned by search engines~\cite{webofabuse}, social support in Reddit communities~\cite{whiting2023online,hui2023examining}, and the quality of responses from conversational systems~\cite{sanz2025empathy,prakash2026assessing,kim2026ai}. Most evaluate one class of system or emphasize either technical quality or social support. Closest to our work, Prakash et al.~\cite{prakash2026assessing} evaluate general-purpose and survivor-oriented LLMs for TFA questions. We instead study the broader help-seeking ecosystem: whether relevant support is available, whether retrieved sources and their secondary links are trustworthy, how peer discussions respond, and whether general-purpose and domain-specific conversational systems provide guidance that is both technically useful and contextually safe. A cross-ecosystem view is necessary because victims may move among these systems, and each introduces a distinct failure mode.

We conduct a victim-centered measurement study grounded in a decade of narratives from \emph{r/Stalking}. Through qualitative coding, human annotation, and validated automated methods, we construct 2,797 victim-derived TFA queries and use them to analyze Google-retrieved webpages, Reddit comment threads, three general-purpose LLMs, and two survivor-support chatbots. Although the source community centers on stalking, the resulting questions span all 11 predefined categories of technology misuse, plus an \emph{Other} category. We use the category distribution to characterize the breadth of help-seeking needs represented in our evaluation corpus. 
Our layered evaluation first applies common measures of technical quality, including relevance, accuracy, actionability, persuasiveness, and understandability, together with a risk-sensitive \emph{Damaging Guidance} label to webpages, Reddit threads, and conversational-system responses. It then examines additional risks specific to each ecosystem, including secondary-link integrity for webpages, toxic discourse for Reddit, and trauma-informed support for conversational systems. These supplementary measures characterize complementary failure modes and are not used to produce an overall safety ranking across ecosystems. This layered design reveals the paper's central result: systems can appear useful under conventional quality measures while still producing or exposing users to guidance that is unsafe in an adversarial interpersonal context.

We investigate the following research questions:

\noindent\textbf{RQ1: Technical support:} How consistently do web search, peer-support forums, and conversational AI provide relevant, accurate, actionable, persuasive, and understandable guidance for TFA-related help-seeking needs?

\noindent\textbf{RQ2: Contextual safety:} What safety failures remain when guidance appears useful, including damaging recommendations and additional risks specific to search, Reddit, and conversational AI?

\noindent\textbf{RQ3: Cross-ecosystem differences:} How do web search, peer-support forums, general-purpose LLMs, and survivor-support chatbots differ in information availability, technical quality, and contextual-safety failures?

Our work makes three contributions:

\noindent (1)~\textbf{Victim-centered, cross-ecosystem measurement.} We derive 2,797 help-seeking queries from authentic victim narratives spanning all 11 predefined technology-misuse categories, plus an \emph{Other} category, and use them to measure the support encountered across web search, peer discussion, and conversational AI. Although the source community centers on stalking, the breadth of represented concerns supports a diverse evaluation of online assistance. We treat the observed category distribution as corpus characterization rather than an estimate of TFA prevalence. This design connects victims' expressed information needs to the responses and risks encountered across the broader digital support ecosystem.

\noindent (2)~\textbf{A layered safety evaluation framework.} Developed through interdisciplinary collaboration between Computer Science and Social Work researchers with victim-advocacy experience, our framework combines technical measures and a risk-sensitive \emph{Damaging Guidance} audit applied across ecosystems with additional analyses tailored to the structural risks of search, community, and conversational systems. These ecosystem-specific analyses characterize complementary failure modes rather than provide interchangeable safety scores.

\noindent (3)~\textbf{Evidence of a gap between technical quality and contextual safety.} We identify distinct but connected failures across the ecosystem: search results are often relevant but can lead to flagged secondary links; Reddit discussions provide inconsistent technical help and sometimes toxic responses; and conversational systems can generate polished, actionable advice while omitting escalation risks or appropriate support resources. We further show that domain specialization alone does not ensure safer support: HopeChat underperforms general-purpose LLMs on several dimensions, whereas Ruth is competitive on some technical measures but remains limited on trauma-informed support.

Together, these findings show why relevance, correctness, and actionability are insufficient measures of support quality in TFA and other safety-sensitive security contexts. Safe assistance must also account for escalation, evidence preservation, uncertainty, source integrity, and appropriate human support. Our codebook, prompts, and code are available at \url{https://shorturl.at/UVsZU} to support reproducibility.

%% file: reddit_posts.tex
\section{Constructing Victim-Derived TFA Queries}
\label{sec3}

\begin{figure*}[t]
\centering
    \includegraphics[width=0.9\linewidth]{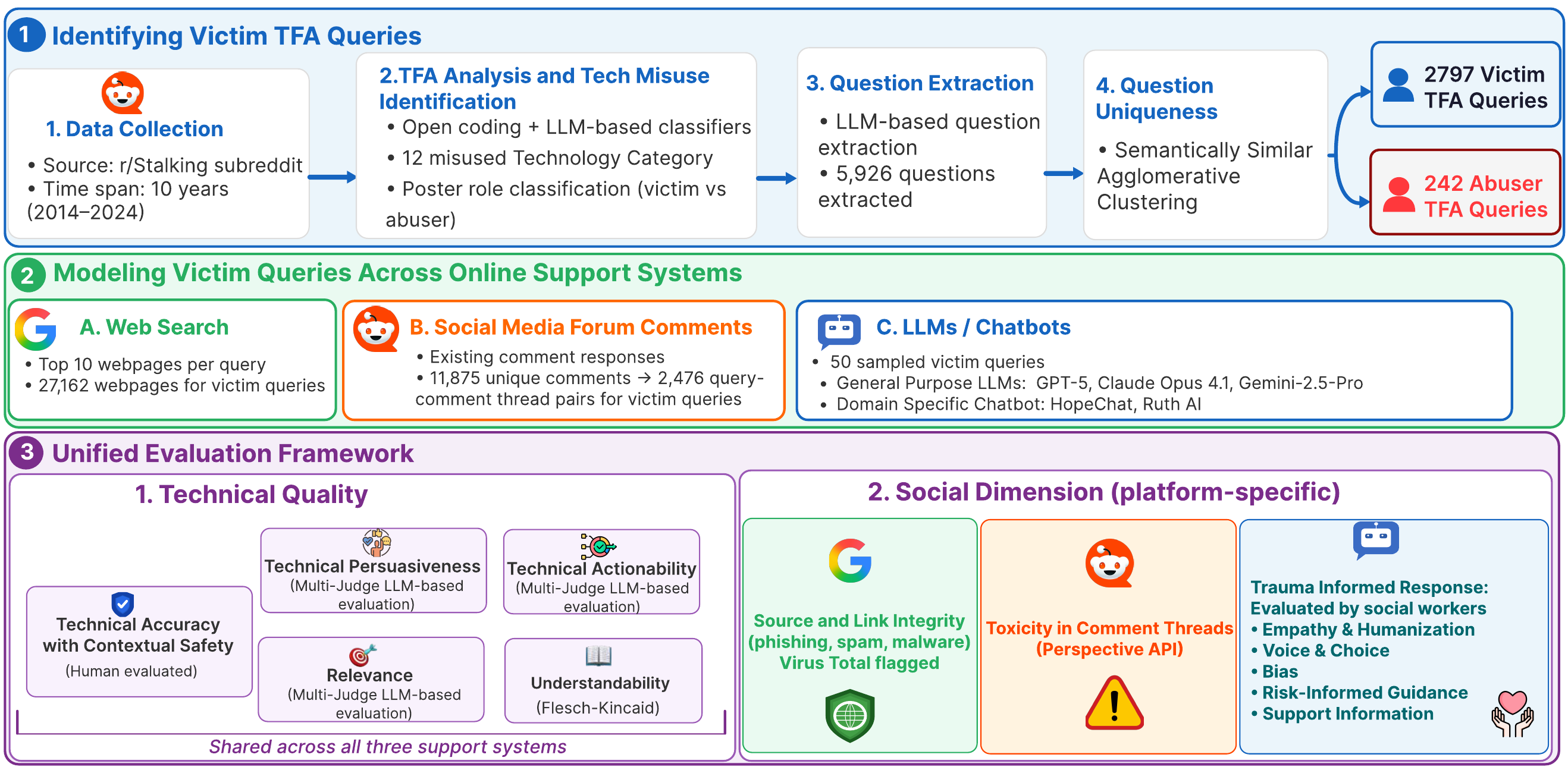}
    \caption{Overview of the study methodology.}
    \label{fig:methodology}
\end{figure*}

Figure~\ref{fig:methodology} summarizes our study. We first construct an evaluation corpus from narratives posted to \emph{r/Stalking}. We identify the misused technologies described in the source posts, extract explicit and implicit help-seeking questions, and retain victim-oriented questions involving TFA. We then use these victim-derived queries to evaluate web search, peer-support forums, and conversational AI. The category labels characterize the breadth of concerns represented in the corpus and support stratified analyses. They are not used to estimate the prevalence of different forms of TFA.

\textbf{Data Source and Collection.}
We selected the public \emph{r/Stalking} subreddit~\cite{stalking-subreddit} because its posts combine interpersonal abuse narratives with questions about digital technologies, safety, and help-seeking. Although the community is organized around stalking, its posts describe a broader range of TFA concerns. We do not assume that this community represents all people experiencing TFA. Instead, we use it as a source of diverse, context-rich questions for evaluating online support systems. 
We collected public posts using the Reddit API through AsyncPRAW~\cite{asyncpraw} and historical Pushshift archives~\cite{pushshift}. The resulting dataset contains 3,267 posts from 2014 through 2024 and 21,323 associated comments and replies.

\subsection{Characterizing Technology Misuse}
We developed a taxonomy of \textit{technology misuse} through iterative open coding~\cite{glaser2017discovery}. Coders reviewed Reddit posts to identify the technologies, platforms, and digital services being misused for TFA. Because one narrative could involve several technologies, the labels were not mutually exclusive.

Two interdisciplinary teams contributed to taxonomy development. Two Computer Science researchers independently annotated 210 posts selected using uniform sampling for 2015 through 2020 and proportional sampling for 2021 through 2024. Their average Cohen's $\kappa$ was 0.73 for technology-misuse labels, indicating substantial agreement~\cite{schuster2004note}. Two Social Work researchers independently annotated a separate sample of 398 posts, contributing expertise on abuse dynamics. Within this team, agreement for individual labels ranged from $\kappa=0.66$ to $\kappa=1.0$. The teams annotated different samples, so agreement was calculated within each team. We used separate samples to broaden coverage during taxonomy development. Disagreements were resolved through discussion. We also used targeted keyword sampling to obtain development examples for categories that were sparsely represented in the initial samples, such as gaming platforms, financial applications, and image or video manipulation. These targeted examples supported taxonomy and classifier development and were not used to infer category prevalence.

This process produced 11 named technology-misuse categories, and an \emph{Other} category. We use the technology labels to characterize corpus coverage, construct the evaluation sample, and examine variation in support quality.
 \subsubsection{Technology Misuse Categories}
The 11 named technology-misuse categories
are: (1) \emph{Phone Communications}; (2) \emph{Surveillance and Tracking Technologies}; (3) \emph{Unauthorized Access and Device Compromise}; (4) \emph{Identity Obfuscation and Spoofing Tools}; (5) \emph{Email-Based Abuse}; (6) \emph{Social Media and Messaging Platforms}; (7) \emph{Other Online Service Accounts}; (8) \emph{Data Broker and People-Search Websites}; (9) \emph{Financial and Payment Platforms}; (10) \emph{Online Gaming Platforms}; and (11) \emph{Image and Video Manipulation}. The \emph{Other} category captures TFA questions that do not identify a particular technology, such as questions about online harassment or online stalking.

\subsubsection{Automated Classification of Technology Misuse}

Using the manually labeled posts, we developed prompt-based classifiers to assign technology-misuse labels to the remaining corpus. Since narratives often involved multiple technologies, we treated this as a multi-label task. Preliminary experiments found that a single joint classifier produced more errors for overlapping concepts. We therefore used 12 independent binary labels, one for each of the 11 named categories and one for \emph{Other} (described in details below).
We used few-shot prompting~\cite{brown2020language} with representative labeled examples and implemented the classifiers using \textit{GPT-OSS 20B}. We refined the prompts using development examples and evaluated them on held-out manual annotations of 395 samples that were not used during prompt development. 
Appendix~\ref{prompt2} provides an example prompt and Appendix Table~\ref{tab:tech_misused} reports  detailed performance for each label. F1-scores ranged from 0.81 to 0.96, with with \textit{Social Media and Messaging Platforms} and \textit{Financial and Payment Platforms} jointly achieving the highest F1-score of 0.96 achieving the highest F1-score of 0.96 and \textit{Data-Broker and People Search Sites} the lowest at 0.81.

\subsection{Filtering TFA Queries and Poster Role}
Manual review showed that not every extracted question involved technology and that some source posts appeared to seek assistance for abusive behavior. 
For this purpose, two Computer Science researchers independently annotated a sample of 132 source posts selected through proportional and uniform sampling along two dimensions: (1)~\emph{TFA relevance}, indicating whether the post concerned technology-facilitated abuse, and (2)~\emph{poster role}, indicating whether it was victim-authored or reflected abusive intent. Agreement was near perfect for Poster-Role ($\kappa=0.90$) and substantial for TFA relevance ($\kappa=0.62$), and a third researcher adjudicated disagreements. Using these labels, we developed Llama3.3 prompt-based classifier with definitions, decision rules, edge cases, and representative examples. The prompt appears in Appendix Table~\ref{tab:prompt_poster}. On held-out manual annotations(100 samples), the classifiers achieved an F1-score of 0.86 \& 0.96 for TFA-Relevance (`Other') and Poster-Role, as shown in Table~\ref{tab:tech_misused}. 
We finally used the TFA-relevance classifier to identify posts that fall on the \emph{`Other'} technology-misuse category. Specifically, if a question was classified as TFA-related but none of the 11 named technology-misuse categories was labeled as present, we assigned it to \emph{`Other'}.

\subsection{Question Extraction}
Questions in the Reddit posts were often implicit within longer descriptions of ongoing abuse. We therefore used prompt-based extraction to transform each narrative into one or more concise help-seeking questions. The prompt instructed the model to preserve details that could affect the interpretation of a question, including the technology involved and the poster's stated concern. When a post contained no explicit or implicit information need, the model returned \textit{No questions implied by the poster}. The complete prompt appears in Appendix~\ref{prompt1}. 
We used Llama3.3 for question extraction because it could be deployed locally and consistently followed the required extraction format during pilot testing. We assessed the suitability of its outputs through the human audit.

\textbf{Human Audit.} Two researchers independently reviewed 40 source posts randomly selected from a broader 108-post sample. The broader sample used uniform sampling for posts from 2014 through 2020 and proportional sampling for posts from 2021 through 2024. Each audit item paired a source post with the full set of questions extracted from that post. Across the 40 audited posts, Llama 3.3 extracted a total of 77 questions. The researchers applied five binary criteria for the audit. 
Since individual posts could yield multiple questions, the coders examined every extracted question associated with each post but assigned one final judgment to the complete question set for that post. \emph{Faithfulness} assessed whether every extracted question was supported by the post and introduced no unsupported details. \emph{Help-seeking} assessed whether each question represented an explicit or implicit information need expressed by the poster. \emph{Self-containedness} assessed whether each question could be understood without consulting the source post. \emph{Retention decision} assessed whether the question set was correctly retained as victim-oriented and TFA-related. \emph{Information-need coverage} assessed whether the complete question set captured all major help-seeking needs expressed in the post. Appendix~\ref{sec:question_audit} provides the complete audit criteria.

Table~\ref{tab:question_extraction_validation} reports Cohen's $\kappa$, Gwet's AC1, and the percentage of audited items satisfying each criterion after consensus. Because positive labels dominated several criteria, Cohen's $\kappa$ was low or negative despite strong agreement on the dominant class. We therefore also report Gwet's AC1, which is less sensitive to label imbalance~\cite{gwet2002kappa}. AC1 was at least 0.82 for every criterion. The researchers resolved all disagreements through discussion. After consensus, at least 87.5\% of the audited post and question-set pairs satisfied each criterion. Overall, 35 of the 40 audited post-level question sets (87.5\%) simultaneously satisfied the three core question-extraction quality criteria: Faithfulness, Help-seeking, and Self-containedness. 

\begin{table}[t]
\centering
\caption{Human audit of Llama3.3 question extraction.}
\label{tab:question_extraction_validation}
\resizebox{0.8\columnwidth}{!}{%
\begin{tabular}{lccc}
\toprule
\textbf{Metric} & 
\textbf{Cohen's $\boldsymbol{\kappa}$} & 
\textbf{Gwet's AC1} & 
\textbf{\% Satisfied} \\
\midrule
Faithfulness               & 0.172  & 0.82 & 87.5\% \\
Help-seeking                & $-0.043$ & 0.83 & 90.0\% \\
Self-containedness          & N/A     & 1.000   & 100.0\% \\
Retention decision          & 0.875  & 0.97 & 90.0\% \\
Information-need coverage   & 0.172  & 0.82 & 87.5\% \\
\bottomrule
\end{tabular}%
}
\end{table}

The audit provides evidence that Llama3.3 is suitable for extracting victim help-seeking questions from the narratives in our corpus. The extracted question sets were generally faithful, self-contained, and representative of the posters' explicit or implicit information needs. We therefore used Llama3.3 to construct the evaluation query corpus.Llama3.3 extracted 5,926 questions from the source posts. After applying the TFA-relevance and poster-role filters, the final corpus contained 2,797 victim-derived TFA queries. We also identified 242 TFA-related queries associated with abusive intent and excluded them from the victim-centered evaluation. The retained 2,797 queries originated from 1,830 unique posts.

\textbf{Question Uniqueness Analysis.} 
Because automated extraction and recurring abuse scenarios could produce many highly similar questions, we examined whether a small number of repeated information needs disproportionately shaped the evaluation corpus. Our semantic-clustering analysis found that 74.2\% of the queries formed singleton clusters. The remaining clusters were small, containing two to eight queries with a median of two. Although questions within the same cluster reflected similar information needs, they often retained different details about technologies, accounts, devices, or threat circumstances. Prior work has shown that different formulations of the same information need can produce different search and retrieval outcomes~\cite{alaofi2022queries,penha2022evaluating}. In the TFA context, these differences may also affect how a support system interprets the situation and whether the resulting guidance is relevant, appropriate, and safe. We therefore retained all 2,797 queries. Appendix~\ref{sec:q_quality} provides additional details.

\subsection{Characterizing the Evaluation Corpus}
\begin{figure}[]
\centering

\subfloat[Observed technology categories]{
    \includegraphics[width=0.46\linewidth]{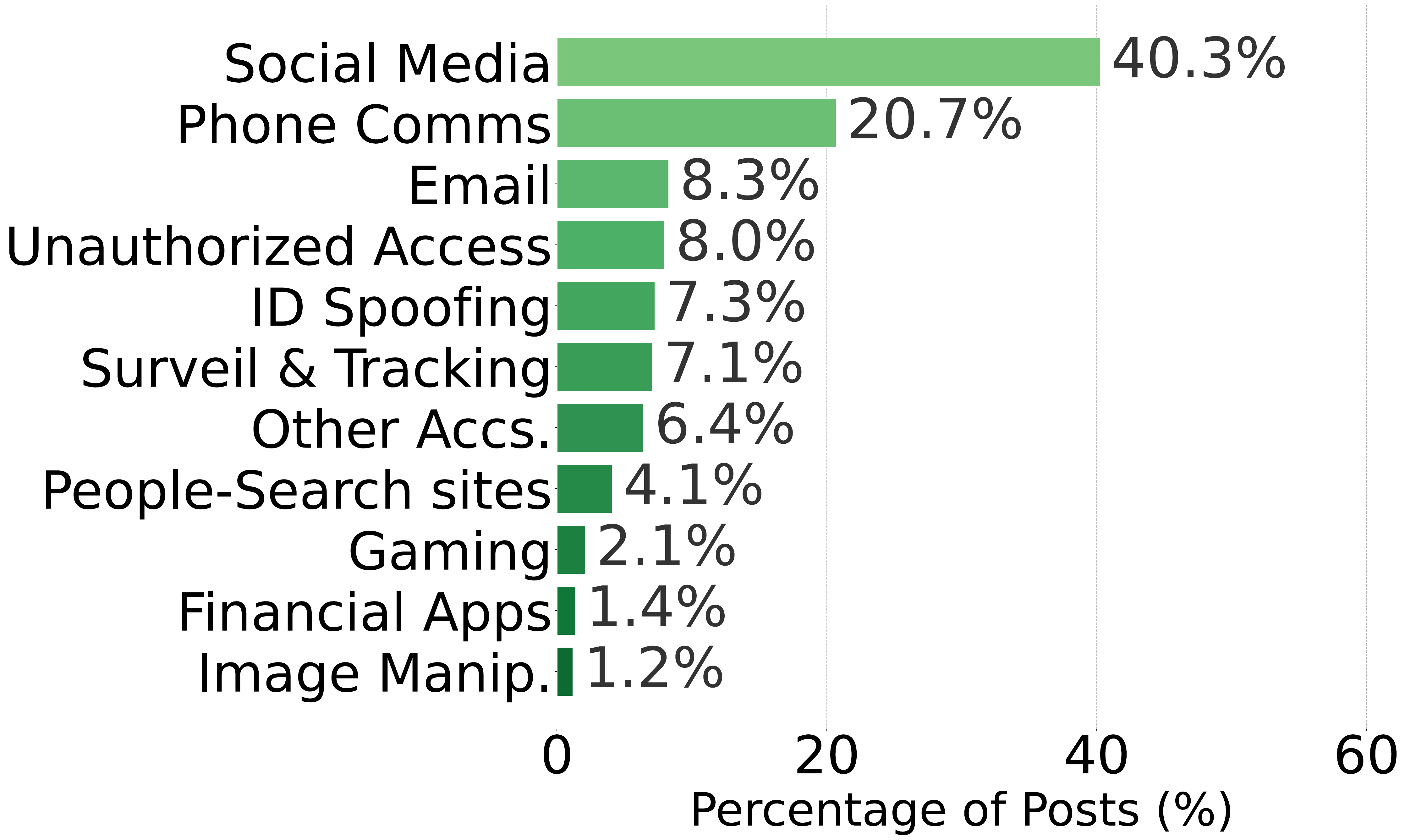}
    \label{fig:distrn1}
}
\hfill
\subfloat[TFA relevance of questions]{
    \includegraphics[width=0.46\linewidth]{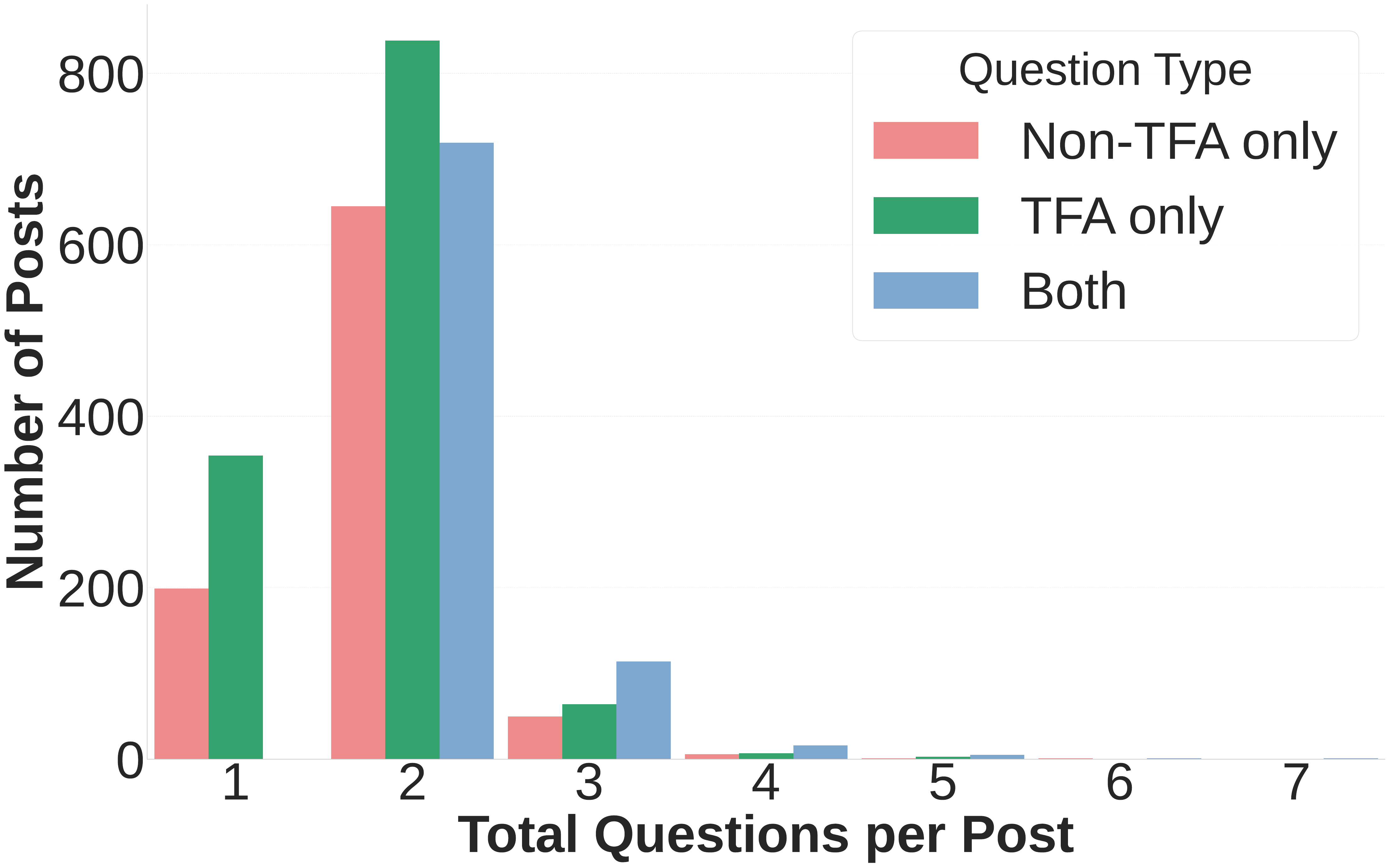}
    \label{fig:distrn3}
}
\caption{Coverage and characteristics of the evaluation corpus.}

\label{fig:postsvstech}
\end{figure}

We use the category analysis to assess the breadth of the evaluation corpus. As shown in Figure~\ref{fig:distrn1}, the collected posts included 11 technology-misuse categories and the \emph{Other} category. Within this subreddit corpus, \emph{Social Media and Messaging Platforms} appeared in 40.3\% of posts, while \emph{Image and Video Manipulation} appeared in 1.2\%. These percentages describe only the collected \emph{r/Stalking} posts and should not be interpreted as estimates of how commonly victims experience each form of abuse.

Figure~\ref{fig:distrn3} shows that posts often expressed one or two help-seeking questions and sometimes combined TFA concerns with broader needs involving legal processes, safety planning, or formal support. Thus, although the source community centers on stalking, it provides questions that cover a wide range of technologies, and support needs for the subsequent evaluation.


%% file: modelling_victim_queries.tex
\section{Collecting Responses Across Online Support Systems}
\label{sec4}
Figure~\ref{fig:methodology} shows the response-collection stage. Using the 2,797 victim-derived TFA queries, we collected guidance from web search, peer-support forums, and conversational AI systems. Each collection procedure reflects how the corresponding system presents information to a person seeking help.

\textbf{Google Search and Retrieved Webpages.}
We queried Google Search through the \textit{googlesearch} Python module~\cite{googlesearch-python} and requested the top 10 results for each query, approximating the first page of results. We disabled personalized search to reduce the influence of search history. We used Selenium~\cite{selenium} to retrieve each page and \textit{trafilatura}~\cite{trafilatura} to extract the main body of the webpage while removing advertisements, headers, navigation elements, and other boilerplate. We implemented preprocessing rules for frequently returned domains, such as Google Support, Apple Discussions, Reddit, and YouTube.

The searches returned 27,970 result URLs for the 2,797 queries. Webpages that were inaccessible
due to paywalls or login requirements, as well as non-English webpages, were excluded from analysis, removing 808 of 27,970 webpages (2.9\%). We also extracted hyperlinks embedded in each retained webpage. We refer to these as \emph{secondary URLs} and evaluate whether retrieved webpages may direct help seekers to unsafe destinations.

\textbf{Reddit Comment Threads.}
To study peer responses, we used the comments attached to each query's source post. We collected 21,323 unique comments associated with the
3,267 Reddit posts in our corpus. 
After linking these posts to the 2,797 victim TFA queries, 11,875 unique comments were retained. Among the 2,797 queries, 2,476 (88.52\%) were associated with at least one comment. For each query, we ordered the associated comments by timestamp and concatenated them without truncation to form a thread-level response. This procedure produced 2,476 query-thread pairs for evaluating peer support. These pairs originated from 1,616 unique underlying Reddit discussion threads, while 214 source posts had no comments. Since a single Reddit post could yield multiple victim-TFA questions, the same underlying comment thread could therefore be associated with more than one query.

\textbf{General-Purpose LLMs and Specialized Chatbots.}
We evaluated three general-purpose LLMs, GPT-5, Claude Opus 4.1, and Gemini 2.5 Pro, and two specialized chatbots developed for domestic violence support, HopeChat~\cite{hopechat} and Ruth AI~\cite{ruth}. These systems allow us to compare general conversational tools with systems explicitly presented as survivor support. 
We selected 50 victim queries through proportional sampling over the observed technology-misuse labels. This sampling preserves the category diversity represented in our corpus. Since a query could receive more than one technology label, the category counts were not mutually exclusive. 
Responses were collected in September 2025 through the publicly available free, user-facing web interfaces of all five systems and did not modify any configurable generation parameters, using the systems' default settings. We submitted every query independently to each system using a new session for every interaction, yielding 250 query-response pairs.

%% file: evaluation.tex
\section{Evaluation Framework: Technical Quality and Contextual Safety}
\label{sec5}

TFA support can fail in two distinct ways. A response may fail to provide relevant, correct, or usable technical information. It may also recommend a technically plausible action without considering surveillance, evidence preservation, retaliation, or escalation risk. We therefore use a layered evaluation, as shown in Figure~\ref{fig:methodology}, Stage~3. First, we apply common measures of technical quality and damaging guidance to webpages, Reddit threads, and conversational-system responses. Second, we examine additional safety and support risks specific to each ecosystem. This design supports comparisons using the common measures while accounting for risks that arise from the different structures of search, community, and conversational systems.

\subsection{Technical Quality}
We apply five technical metrics across webpages, Reddit comment threads, and conversational-system responses. The metrics draw on prior work on LLM evaluation, retrieval-augmented generation, and cybersecurity advice~\cite{ragas,memerag,redmiles2020comprehensive,cybersecurityqa}, and were adapted to TFA through consultation with social work professionals experienced in victim advocacy.

\emph{Relevance} assesses whether a response addresses the specific information need expressed in the query~\cite{ragas,memerag,cybersecurityqa}. For example, a webpage describing Snapchat community guidelines is topically related to a query about unauthorized access to Snapchat, but it does not address how the access could occur or how the victim could investigate it. \emph{Technical Accuracy} assesses whether the technical information is factually correct, sufficiently complete, and procedurally sound. \emph{Technical Actionability} assesses whether a response translates advice into clear and feasible steps that a help seeker could follow~\cite{clinicalactionability,lawactionability,cybersecurityqa}. \emph{Technical Persuasiveness} assesses whether the response explains why a recommended action is useful in the stated situation~\cite{persuasiveness,cybersecurityqa}. \emph{Understandability} assesses whether the response is accessible to a nontechnical audience. We operationalize understandability using the Flesch--Kincaid Grade Level~\cite{flesch-kincaid}. 
These metrics capture different failure modes and should not be collapsed into a single quality score. In particular, relevance is a prerequisite for evaluating the content returned by retrieval and community systems. Within the manual accuracy audit, we also apply a risk-sensitive \emph{Damaging Guidance} label to all three ecosystems. This label identifies recommendations that may increase danger, destroy evidence, alert an abuser, or interfere with safety planning. The complete coding procedure appears in Section~\ref{sec:tech_accuracy}.

\subsection{Contextual Safety and Support}
Beyond the common damaging-guidance audit, we examine risks that are not represented by ordinary answer-quality metrics and do not arise in the same form across the three ecosystems. These additional measures are complementary. We do not treat them as equivalent quantities, combine them into a single safety score, or use them to rank the ecosystems.

For \textbf{Google Search}, we assess \emph{Source and Link Integrity} by examining whether retrieved webpages direct help seekers to secondary URLs flagged by multiple security engines. For \textbf{Reddit}, we assess \emph{Toxicity and Harmful Discourse}, including hostile, insulting, threatening, or identity-targeted comments that may discourage further help-seeking~\cite{aleksandric2024users,almerekhi2020these,aleksandric2022twitter,salehabadi2022user}.

For \textbf{general-purpose LLMs and specialized chatbots}, we manually assess five trauma-informed dimensions informed by prior research~\cite{frydmanTIA,chenTIA,saglam2024empathy,sanz2025empathy} and consultation with social workers experience in victim advocacy. \emph{Empathy and Humanization} captures compassionate acknowledgment of the victim's experience. \emph{Voice and Choice} captures whether the response preserves autonomy by presenting options. \emph{Bias} identifies stereotypes, victim-blaming, and unsupported assumptions. \emph{Risk-Informed Guidance} assesses whether recommendations consider escalation, evidence preservation, and other safety consequences. \emph{Support Information} assesses whether the response connects the user with appropriate human or organizational support. These trauma-informed measures support conclusions about conversational systems only. 
The technical evaluation procedures appear in Section~\ref{sec:tech_eval}, and the contextual safety and support procedures appear in Section~\ref{sec:social_eval}. Appendix Tables~\ref{tab:techmetrics} and~\ref{tab:socialmetrics} provide the complete rubrics.

%% file: technical_evaluation.tex
\section{Technical Evaluation Procedures}
\label{sec:tech_eval}

Evaluating thousands of webpages, comment threads, and conversational responses entirely by hand was infeasible. We therefore combined manual annotation, reference-answer assessment, established computational measures, and validated LLM-assisted evaluation. Human coders developed the rubrics and consensus labels used for validation. The automated evaluators were applied only after comparison with those labels. All technical annotations were conducted by researchers with computer science backgrounds, including one researcher with training in TFA victim advocacy.

For Google Search and Reddit, we first evaluated relevance. We assessed accuracy, actionability, and persuasiveness only for relevant responses, so these results measure quality conditional on retrieving or receiving relevant information. For conversational systems, we evaluated every response because each system returned one direct answer per query. We report this denominator distinction whenever comparing systems. 

\subsection{Relevance}
\label{sec:tech_relevance}

We used a common definition of relevance across systems, namely whether a response fully or partially addressed the victim's question. The response unit differed by ecosystem, with webpage text for Google, a concatenated comment thread for Reddit, and a single answer for each conversational system.

\textbf{Google Search.} To construct a validation set, we sampled 20 victim queries and five of their top ten results, yielding 100 query-webpage pairs. After applying our exclusion criteria mentioned in Section~\ref{sec4}, 90 pairs remained. Three coders individually labeled 30 pairs each using a shared codebook developed through open discussion in the pilot annotation. Then, Llama 3.3 with a chain-of-thought prompting strategy along with the original query and cleaned webpage content is used to
assign a binary relevance label to each webpage. Validated against the human annotations, it achieved 87\% accuracy (Table~\ref{tab:judge_llm_performance}), after which we applied the classification pipeline to all retained webpages. 

\textbf{Reddit.} We evaluated Reddit comment threads using the same query-level relevance framework. During validation of the Google Search relevance pipeline, we observed that many search results linked to forum-style discussions, including Reddit and Quora; specifically, 25 of the 90 validation webpages (28\%) originated from these platforms.
Because these pages contained user-generated discussions similar in structure to Reddit comment threads, we transferred the same relevance classifier and prompt design to the 2,476 concatenated r/Stalking query-thread pairs.

\textbf{Conversational Systems.} We randomly sampled ten responses from each of the five systems, yielding 50 question and answer pairs. Two coders independently labeled every pair and resolved disagreements through discussion. Cohen's $\kappa=$ 0.62. We used LLM-as-a-Judge approach comprising of three judge models, Llama~3.3, GPT-OSS 20B, and Gemma3:12b, with a majority voting procedure.  The LLM-as-a-judge model was validated against human annotations and achieved an accuracy of 90\% (Table~\ref{tab:judge_llm_performance}), after which we applied the ensemble to the remaining 200 responses.

\begin{table}[t]
\centering
\caption{Performance of automated evaluators validated against human consensus labels (macro averages).}
\label{tab:judge_llm_performance}
\resizebox{0.95\columnwidth}{!}{%
\begin{tabular}{p{0.25\linewidth}p{0.18\linewidth}cccc}
\toprule
\textbf{Response set} & \textbf{Metric} & \textbf{Acc.} & \textbf{Prec.} & \textbf{Rec.} & \textbf{F1} \\
\midrule
\multirow{3}{=}{Conversational-system responses}
& Relevance      & 0.90 & 0.80 & 0.76 & 0.78 \\
& Persuasiveness & 0.88 & 0.88 & 0.79 & 0.83 \\
& Actionability  & 0.82 & 0.82 & 0.80 & 0.81 \\
\midrule
\multirow{3}{=}{Webpages and forum-style responses}
& Relevance      & 0.87 & 0.85 & 0.88 & 0.86 \\
& Persuasiveness & 0.89 & 0.81 & 0.93 & 0.84 \\
& Actionability  & 0.82 & 0.81 & 0.82 & 0.81 \\
\bottomrule
\end{tabular}%
}
\end{table}

\subsection{Technical Accuracy and Contextual Safety}
\label{sec:tech_accuracy}

We evaluated accuracy on the 50 victim queries submitted to the conversational systems. Two computer science researchers constructed reference answers, with each researcher drafting answers for 25 queries. The answers were grounded in official platform documentation, CETA Cornell Tech resources~\cite{cetaresources}, TechSafety.org~\cite{safetynet}, and other cybersecurity and digital-safety sources. 
The 50 queries yielded 236 relevant Google webpages. Thirty queries, each from a distinct Reddit post, had relevant comment threads, comprising 279 comments and ten of the 11 named technology-misuse categories. No relevant thread was identified for Financial and Payment Platform queries. We evaluated one concatenated thread per query. And, the five conversational systems yielded 250 responses.

We use a risk-sensitive accuracy rubric with five mutually exclusive labels: \emph{Fully Supported}, \emph{Partially Supported}, \emph{Contradictory}, \emph{Not Present}, and \emph{Damaging Guidance}. Fully supported responses cover all critical elements of the reference answer. Partially supported responses provide correct but incomplete guidance. Contradictory responses contain substantive technical claims that conflict with the reference. Not present indicates that the response provides no substantive technical guidance addressing the question.

\emph{Damaging Guidance} captures advice that may increase danger or destroy useful evidence, such as immediately blocking an abuser, deleting communications, immediately destroying a GPS tracker, or resetting a device without first considering surveillance, escalation, documentation, or individualized safety planning. We assign this label precedence whenever a response contains such a recommendation, even if other parts of the response contain technically supported information. This precedence rule ensures that consequential safety concerns are not obscured by otherwise useful content. Because each response receives one final label, the analysis identifies the prevalence of damaging guidance across the three ecosystems but does not quantify how often technically supported and damaging content co-occur within the same response.

Two coders independently evaluated the webpages, Reddit threads, and conversational responses. Cohen's $\kappa$ \& Gwet's AC1 is 0.48 \& 0.62 for webpages, 0.88 \& 0.89 for Reddit, and 0.60 \& 0.67 for conversational responses respectively. The coders resolved disagreements to establish final labels.

\subsection{Technical Actionability}
\label{sec:tech_actionability}

We assign one of three labels. \emph{Actionable} responses provide concrete instructions that a help seeker could follow. \emph{Informative} responses recommend a useful action but do not explain how to carry it out. \emph{Not Actionable} responses provide only vague advice or no concrete technical guidance. 
Across all three support systems, Actionability was assessed relative to the technical guidance for victim's specific question using a hierarchical decision rule. Responses lacking meaningful question-relevant technical guidance, or dominated by unrelated, anecdotal, legal, emotional, promotional, or unsafe content, were labeled \textit{Not Actionable}. Among responses containing relevant technical guidance, responses were labeled \textit{Actionable} when the majority of that guidance included specific, executable implementation details. Otherwise, when the majority of the technical guidance provided recommendations or explanations but lacked sufficient execution steps, the response was labeled \textit{Informative}. Thus, labels reflected the dominant technical guidance in the response rather than isolated actionable statements.

We used the same three-model judging ensemble and majority vote described for relevance. For validation, two coders labeled 50 conversational responses, comprising ten responses from each system. Three coders independently labeled all 100 webpages, with two webpages evaluated for each of randomly sampled 50 queries. Twenty-five of those webpages were Reddit or Quora discussions and served as the only forum-style examples in the current validation set. Inter-rater reliability was $\kappa=0.40$ for the two-coder set and Krippendorff's $\alpha=0.60$ for the three-coder set. After consensus, the ensemble achieved 0.81 F1 on both validation sets, as shown in Table~\ref{tab:judge_llm_performance}.

\subsection{Technical Persuasiveness}
\label{sec:tech_persuasiveness}

Persuasiveness captures whether a response explains why a recommended technical action is useful in the stated situation. A response is \emph{Persuasive} when it connects the action to a relevant benefit. For example, explaining that two-factor authentication can prevent account access even when an abuser knows the password provides a rationale. A response is \emph{Not Persuasive} when it lists an action, such as changing a password, without explaining why the action may help. Across all three support systems, Persuasiveness was assessed relative to the victim's specific question using a hierarchical decision rule. Among responses containing relevant technical recommendations specific to victim's question, responses were labeled \textit{Persuasive} when the majority of the recommended technical actions were supported by reasoning, justification, reassurance, or explanation of why the actions would be useful or beneficial in the victim's situation. Otherwise, when the majority of the technical recommendations were presented without explaining why they mattered or how they could help the victim, the response was labeled \textit{Not Persuasive}. For mixed responses, the label reflected the dominant substantive guidance rather than isolated persuasive statements.

We validated the same judging ensemble on the 100 webpage pairs and 50 conversational responses used for actionability. Inter-rater reliability was $\alpha=0.56$ for the three-coder set and $\kappa=0.50$ for the two-coder set. After consensus, the ensemble achieved F1 scores of 0.84 for webpages and forum-style responses and 0.83 for conversational responses. We then applied the validated evaluator to the remaining eligible responses.

\subsection{Understandability}
\label{sec:tech_understandability}

We computed Flesch--Kincaid Grade Level using the \textit{readability} Python module~\cite{readability-python}. Source-specific preprocessing removed Unicode artifacts and links from webpages, restored punctuation in YouTube transcripts using XLM-RoBERTa~\cite{1800badcode2023punctuators}, and removed URLs, hashtags, emojis, and punctuation-only text from Reddit comments. Appendix Section~\ref{sec:understandability} provides score interpretation and results. We use readability as a limited proxy for linguistic accessibility, not as a measure of whether a victim can safely follow the guidance.

%% file: social_evaluation.tex
\section{Ecosystem-Specific Safety and Support Analyses}
\label{sec:social_eval}

The technical metrics do not capture all risks encountered during TFA help-seeking. We therefore measure source integrity for web search, harmful discourse for Reddit, and trauma-informed support for conversational systems. These measures reflect platform-specific exposures and should not be interpreted as interchangeable safety scores.

\subsection{Google Search: Source and Link Integrity}

We examined secondary URLs extracted from the main text of the 27,162 retained webpages. The extraction produced approximately 185,200 unique secondary URLs. We removed approximately 215 allowlisted domains and URLs, including government and educational sites, official support pages from major platforms, and nonprofit resources compiled from the No More Directory~\cite{nomoredirectory}. We submitted the remaining 165,435 URLs to VirusTotal~\cite{virustotal}, using the VirusTotal API v3 URL scanning endpoint. 
VirusTotal returned verdicts from 97 security engines for every analyzed URL. Following prior work~\cite{roy2025darkgram,vafa2025learning,singhal2023cybersecurity}, we marked a URL as \emph{flagged} when at least two engines classified it as malicious or suspicious. We use \emph{flagged} rather than \emph{malicious} because VirusTotal aggregates vendor detections and does not establish ground truth. At the query level, exposure means that at least one webpage returned for the query contained at least one flagged secondary URL. 
We also examined the threat categories reported by VirusTotal engines for each flagged URL. Duplicate category labels across engines were counted only once per URL. If different engines assigned different categories to the same URL, such as phishing and malware, the URL was counted in both categories. 
This analysis measures the integrity of destinations linked from retrieved webpages. It does not show that Google search of a query itself returned a malicious page or that a victim clicked the link.

\subsection{Reddit: Toxicity and Harmful Discourse}

We applied the \textit{Perspective Comment Analyzer API v1alpha1}~\cite{perspectiveapi} to individual comments and measured six attributes, \emph{Toxicity}, \emph{Severe Toxicity}, \emph{Profanity}, \emph{Identity Attack}, \emph{Insult}, and \emph{Threat} during July 2025. We removed URLs, hashtags, emojis, punctuation-only comments, and non-English comments before scoring. After preprocessing and language identification using the Python \textit{langdetect} module~\cite{langdetect}, we retained 10,845 English comments from the 11,875 unique comments for analysis. Following prior work~\cite{habib2022exploring,gehman2020realtoxicityprompts,kumarswamy2025causal}, scores greater than 0.5 were treated as positive. We aggregate these comment-level labels to characterize whether a query's thread contained at least one comment flagged for each attribute and how many flagged comments appeared in the thread. 
Unlike the technical evaluation, where each comment
thread was evaluated as a whole for each query, we applied the Perspective API to individual unique comments. This allowed us to estimate, for each query, the prevalence of toxic or harmful comments received rather than assigning a single toxicity score to the entire
thread. 
Perspective scores are signals of potentially harmful language, not definitive judgments about author intent or the overall supportiveness of a thread. A thread may contain both supportive and harmful comments.

\subsection{Conversational Systems: Trauma-Informed Support}

Three coders manually evaluated all 250 conversational-system responses. Two coders had social work backgrounds and experience in victim advocacy. The third had a computer science background and training in TFA victim advocacy. The team refined the rubric through pilot coding and discussion. The five dimensions were \emph{Empathy and Humanization}, \emph{Voice and Choice}, \emph{Bias}, \emph{Risk-Informed Guidance}, and \emph{Support Information}. Appendix Table~\ref{tab:socialmetrics} defines the labels.

For \textit{Empathy and Humanization} and \textit{Voice and Choice}, coders assigned ratings on a 1--5 Likert scale; we report the distribution of ratings assigned by coders. For inter-coder reliability, ordinal Krippendorff's $\alpha$ is 0.63 for \textit{Empathy and Humanization} and 0.36 for \textit{Voice and Choice}. For Bias, Risk-Informed Guidance, and Support Information, we report Krippendorff's $\alpha$ and Gwet's AC1. Agreement was $\alpha=0.30$ and $\mathrm{AC1}=0.96$ for Bias, $\alpha=0.27$ and $\mathrm{AC1}=0.71$ for Risk-Informed Guidance, and $\alpha=0.70$ and $\mathrm{AC1}=0.75$ for Support Information. The coders resolved disagreements through discussion and reviewed labels for consistency.



We did not automate these assessments. Empathy, autonomy, victim-blaming, and escalation risk depend on subtle contextual interpretation, and the social work coders' expertise was central to the evaluation. The consensus labels therefore serve as the primary evidence for these dimensions.

%% file: results_tech.tex
\section{Results}
\label{sec:results}

\subsection{RQ1: Technical Support Across the Ecosystem}
\label{sec:results_technical}

No system consistently provided complete, practical support. Google Search and most conversational systems usually surfaced on-topic information, whereas Reddit threads addressed victims' questions less consistently. Once relevant information was available, general-purpose LLMs were more likely to turn recommendations into concrete steps and explain their benefits. However, complete technical guidance remained uncommon across all three ecosystems. For Google Search and Reddit, accuracy, actionability, and persuasiveness are evaluated only on relevant webpages or threads. Conversational-system percentages use all 50 responses per system. We preserve this denominator distinction throughout rather than treating the conditional scores as end-to-end success rates.

\subsubsection{Relevant Support Is Available, but Unevenly Distributed}

\begin{figure}[t]
\centering
\subfloat[Queries with relevant support]{
   \includegraphics[width=0.47\columnwidth]{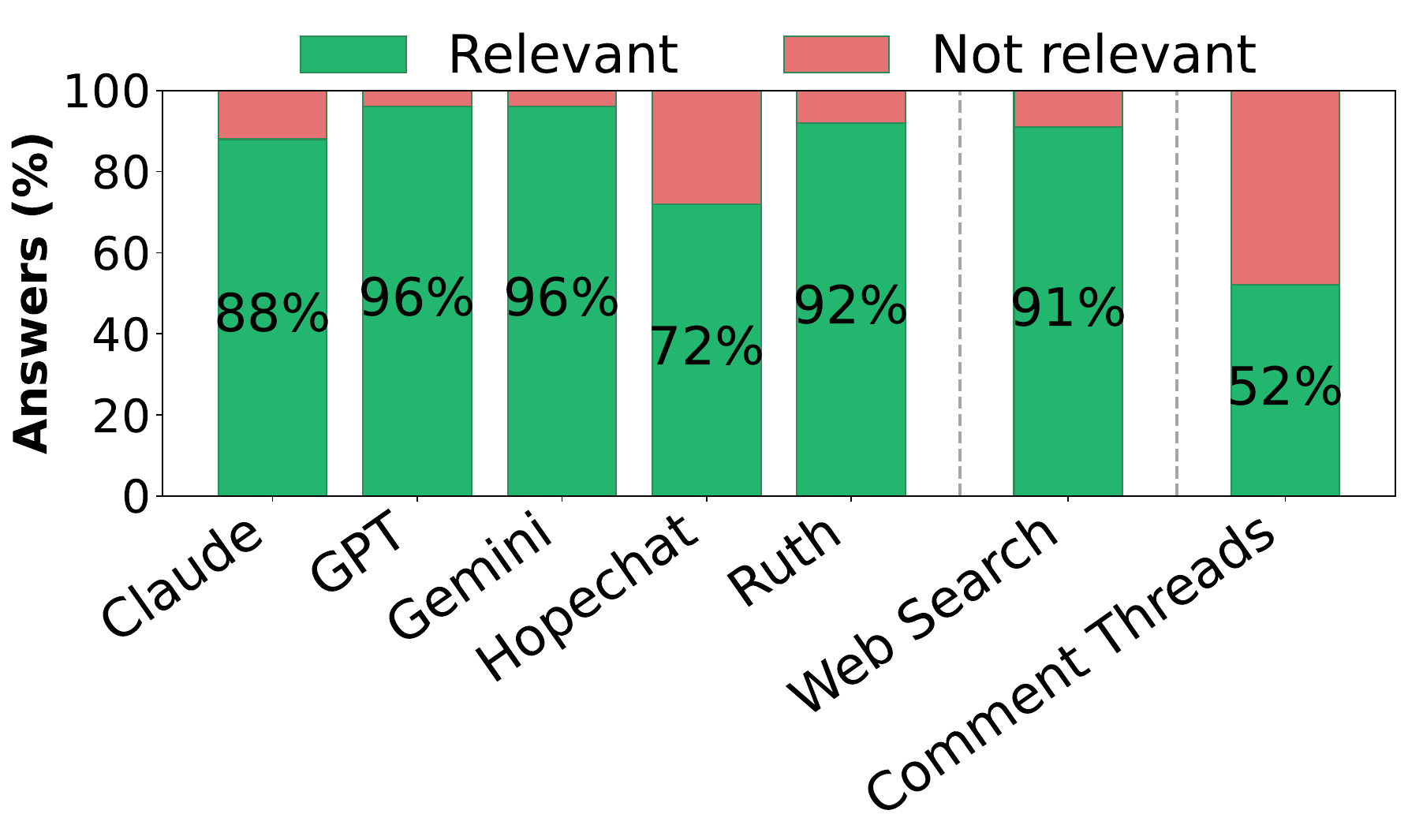}
    \label{cross_platform_relevance}}
\subfloat[Rank of first relevant result]{
    \includegraphics[width=0.45\columnwidth]{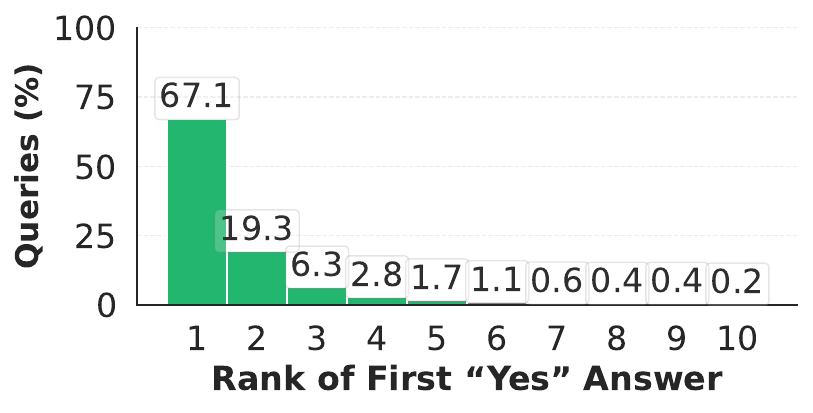}
    \label{fig:web_first_rank}}
\caption{Availability of relevant support across ecosystems. Web Search and Comment Threads \shirin{why do we talk about Reddit here?}\nowshin{updated the chart to include Comment Threads instead of Reddit} values represent query-level coverage, while each conversational system returns one response per query.}
\label{fig:overall_relevance}
\end{figure}

Google Search provided the broadest coverage. Among the 2,797 victim-derived queries, 91\% (2,559) returned at least one relevant webpage in the top ten results (Figure~\ref{fig:overall_relevance}). Among these successful queries, 86.4\% surfaced a relevant webpage within the first two results. Reddit was considerably less consistent: among the 2,476 queries whose source posts received comments, only 52\% had a thread that fully or partially addressed the extracted question. Most conversational systems also remained on topic. Claude, GPT, Gemini, and Ruth produced relevant responses for 88--96\% of queries, whereas HopeChat did so for 72\%. Appendix Table~\ref{tab:llm_ci} reports 95\% confidence intervals for model-level estimates.

\begin{figure*}[t]
\centering
\subfloat[Google webpages]{
    \includegraphics[width=0.24\linewidth]{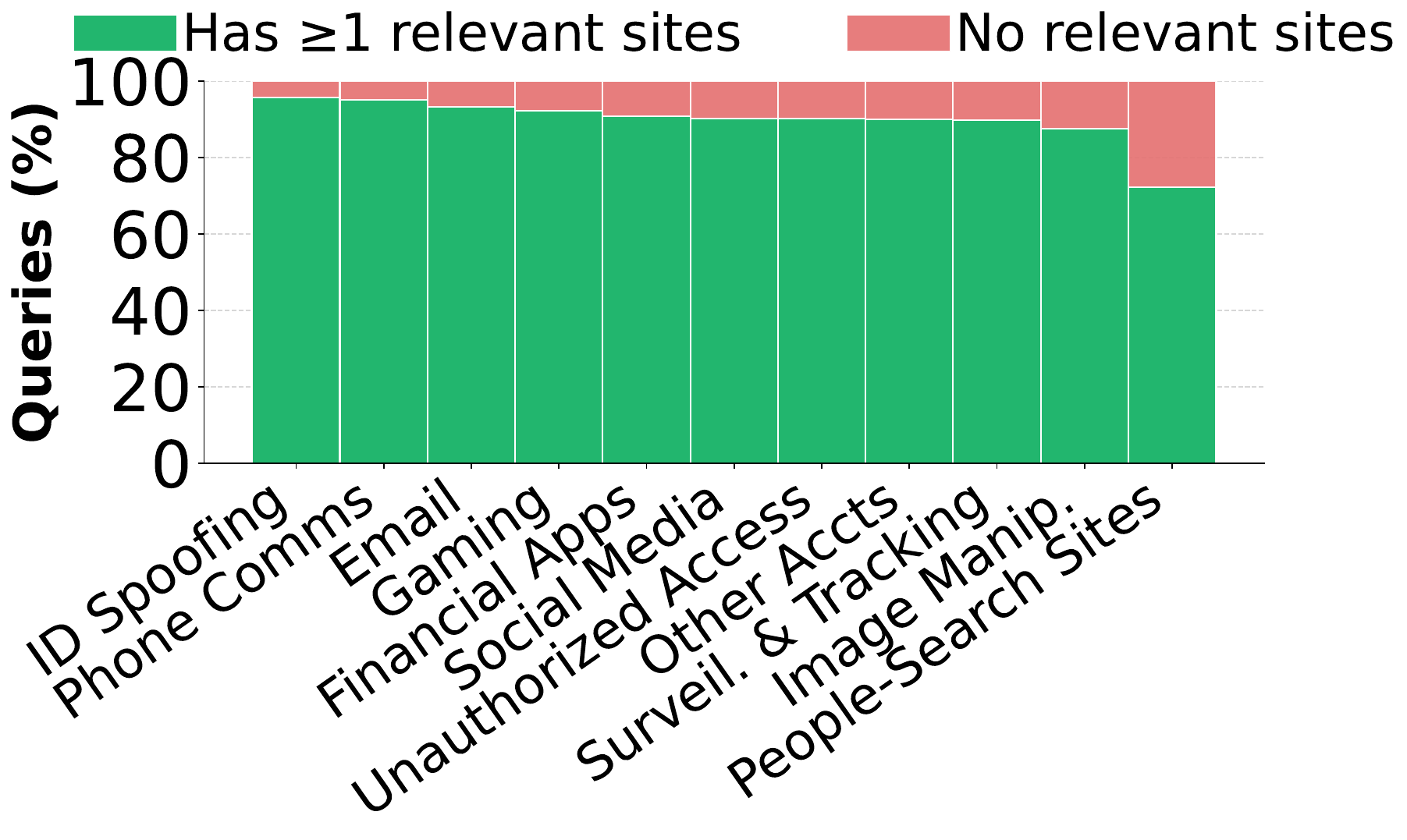}
    \label{fig:web_relevance_by_tech}
}
\subfloat[First-ranked Google webpage]{
    \includegraphics[width=0.21\linewidth]{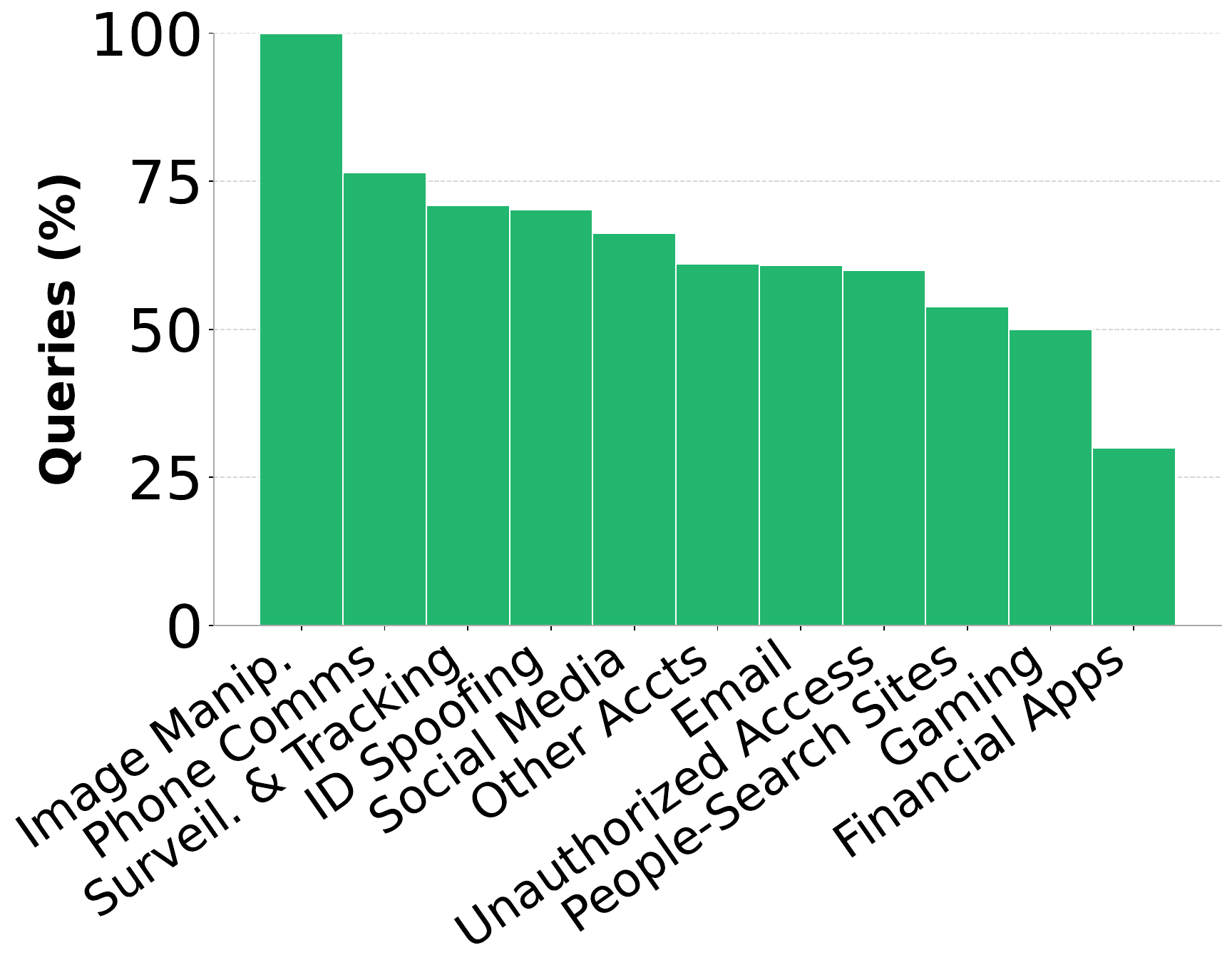}
    \label{fig:web_relevance_first_rank_by_tech}
}
\subfloat[Reddit threads]{
    \includegraphics[width=0.21\linewidth]{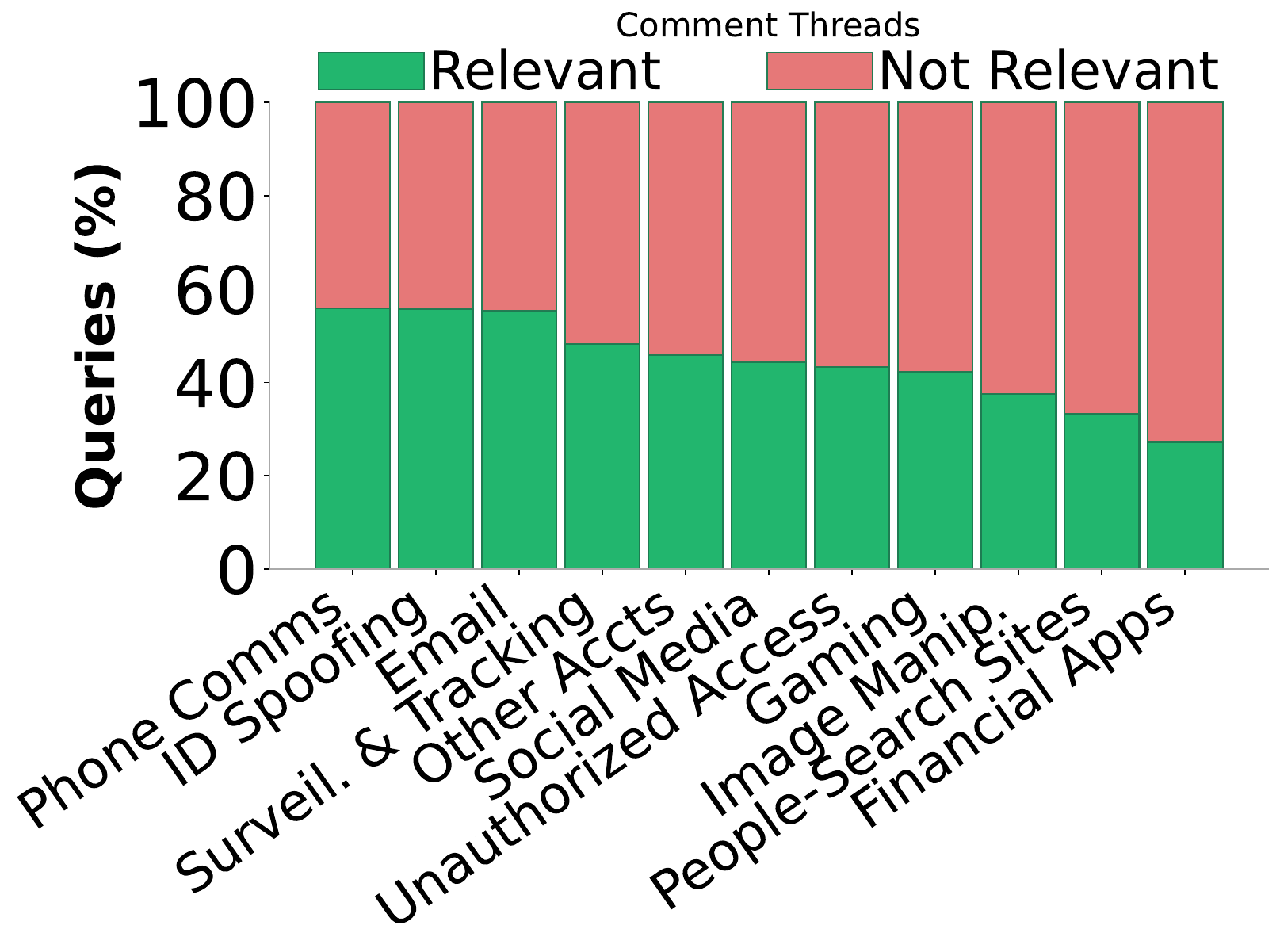}
    \label{fig:reddit_relevance_by_tech}
}
\subfloat[Conversational systems]{
    \includegraphics[width=0.29\linewidth]{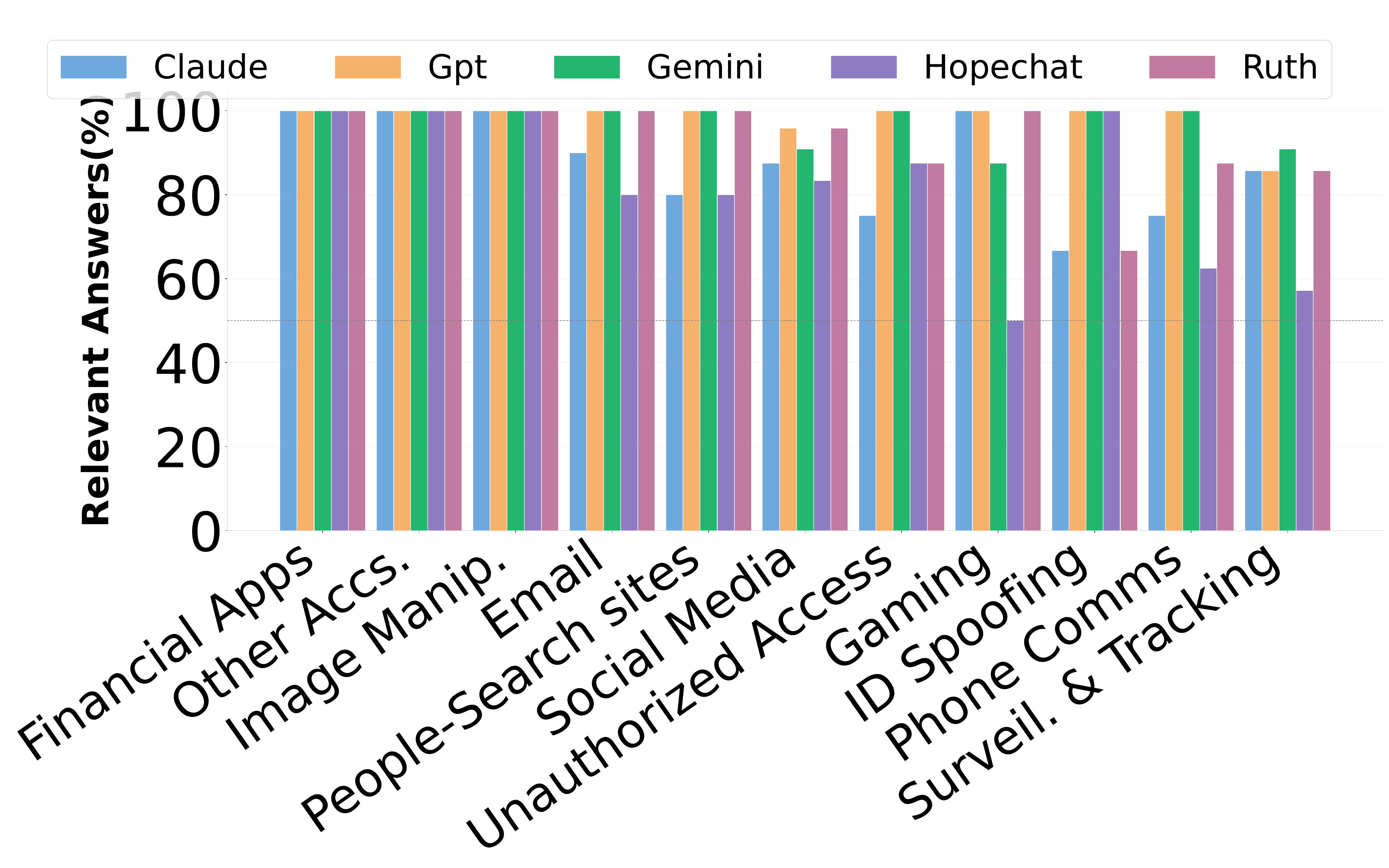}
    \label{fig:llm_relevance_by_tech}
}
\caption{Proportion of queries with relevant support by technology-misuse category.}
\label{fig:relevance_by_tech}
\end{figure*}

Coverage also depended on the technology involved (Figure~\ref{fig:relevance_by_tech}). For Google Search, more than 90\% of queries involving \textit{Identity Obfuscation and Spoofing Technology} returned a relevant webpage, compared with 64\% of queries involving \textit{Data Broker and People-Search Websites}. Although \textit{Social Media and Messaging Platforms} was among the most common categories in the corpus, approximately 85\% of these queries returned relevant results. The position of the first relevant webpage also varied. All answered \textit{Image and Video Manipulation} queries surfaced relevant guidance in the first result, whereas only 30\% of answered \textit{Financial and Payment Platform} queries did so.

The same categories were not uniformly difficult across ecosystems. Reddit relevance ranged from 35\% to 55\% across technology-misuse categories. Conversational systems were generally relevant for \textit{Financial and Payment Platforms}, \textit{Other Online Service Accounts}, and \textit{Image and Video Manipulation}, but were least consistent for \textit{Surveillance and Tracking Technologies}. Thus, high overall relevance did not eliminate topic-specific gaps, particularly for covert monitoring and tracking.

\subsubsection{Relevant Responses Rarely Provide Complete Technical Guidance}

Table~\ref{tab:accuracy_label_distribution} reports the risk-sensitive accuracy labels for 236 relevant Google webpages, 30 relevant Reddit threads, and all 250 conversational-system responses. Fully supported guidance was uncommon, accounting for 7.79\% of webpages and 16.8\% of chatbot responses. None of the evaluated Reddit threads was fully supported. 
For only 3 of the 50 evaluated queries, the first relevant Google webpage was Fully Supported. Instead, \textit{Partially Supported} was the most common label for both webpages (48.1\%) and Reddit threads (63.3\%), indicating that relevant content often omitted important elements of the reference guidance. Conversational responses showed the same pattern in aggregate, with 44.8\% labeled \textit{Partially Supported}. The \textit{Damaging Guidance} category is discussed as a contextual-safety failure in Section~\ref{sec:results_contextual_safety}.

\begin{table}[t]
\centering
\caption{Risk-sensitive accuracy labels. Google and Reddit results are conditional on relevance; conversational-system results include every response.}
\label{tab:accuracy_label_distribution}
\resizebox{0.88\columnwidth}{!}{%
\begin{tabular}{lccc}
\toprule
\textbf{Accuracy Label} & \textbf{Webpages} & \textbf{Threads} & \textbf{Systems} \\
\midrule
Fully Supported & 7.79\% & 0\% & 16.8\% \\
Partially Supported & 48.1\% & 63.3\% & 44.8\% \\
Damaging Guidance & 17.3\% & 13.3\% & 19.6\% \\
Contradictory & 1.7\% & 6.7\% & 4.4\% \\
Not Present & 25.1\% & 16.7\% & 14.4\% \\
\bottomrule
\end{tabular}%
}
\end{table}

\begin{figure*}[t]
\centering
\subfloat[Google webpages by category]{
    \includegraphics[width=0.24\linewidth]{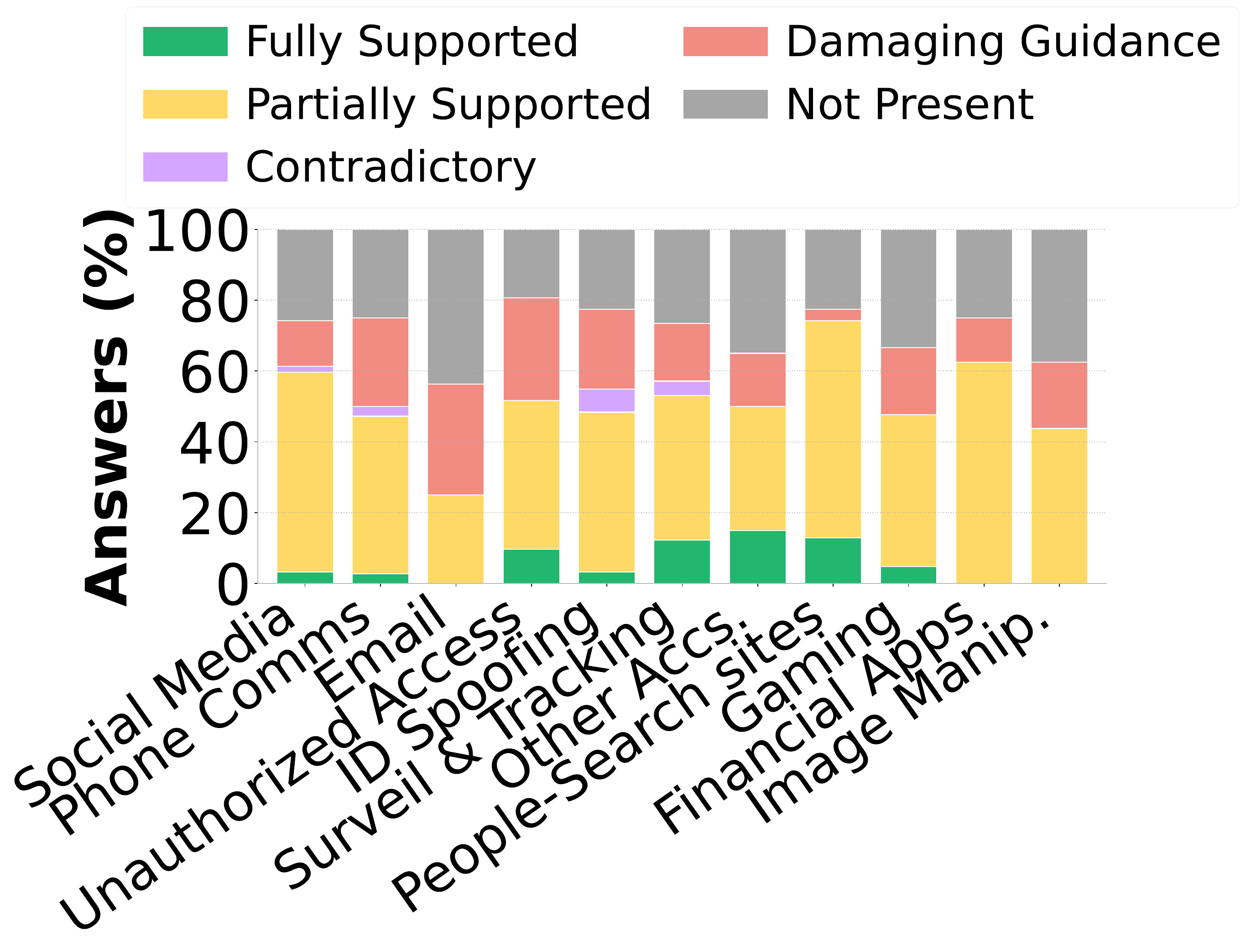}
    \label{fig:google_acc_by_tech}
}
\subfloat[Reddit threads by category]{
    \includegraphics[width=0.24\linewidth]{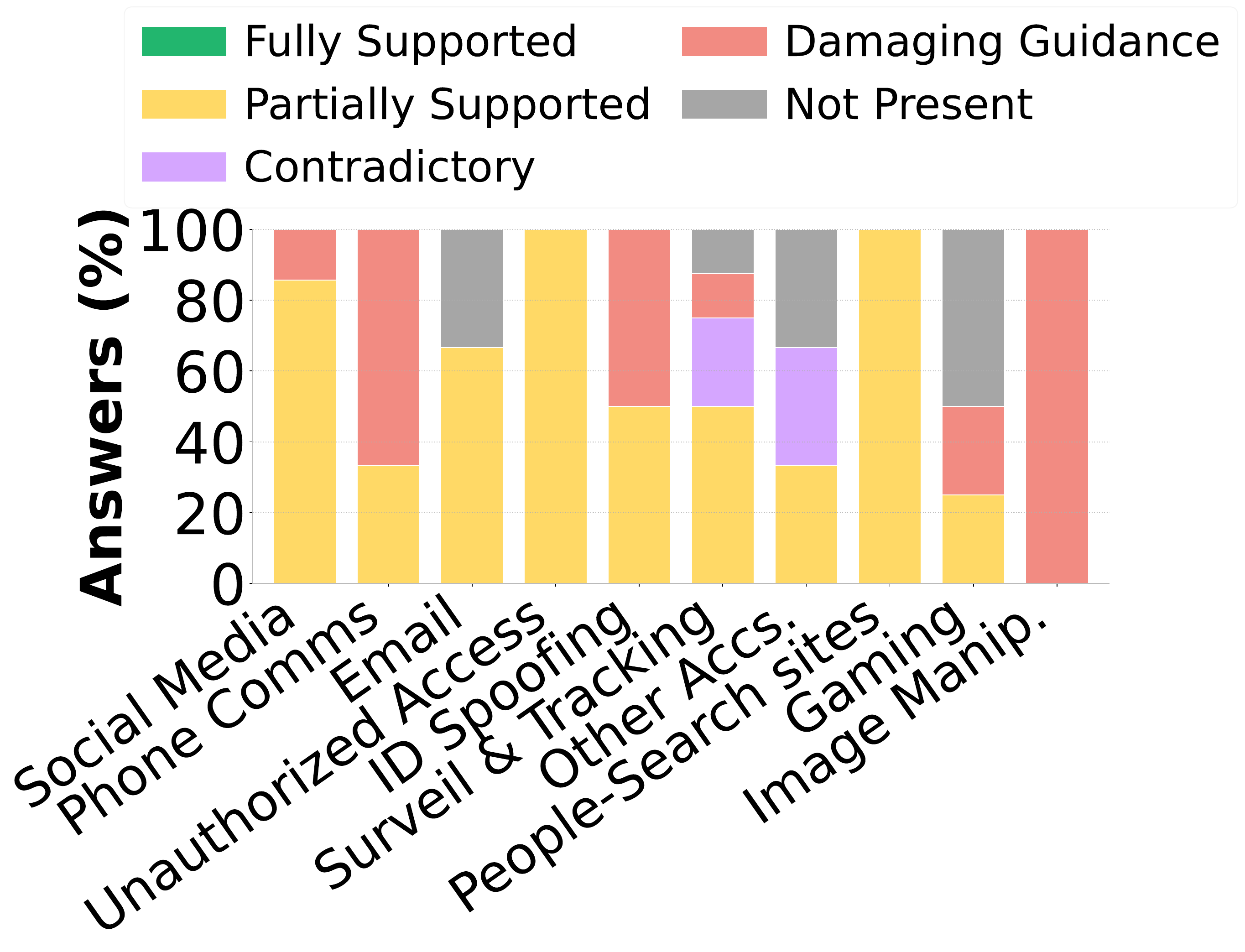}
    \label{fig:reddit_acc_by_tech}
}
\subfloat[LLM responses by category]{
    \includegraphics[width=0.24\linewidth]{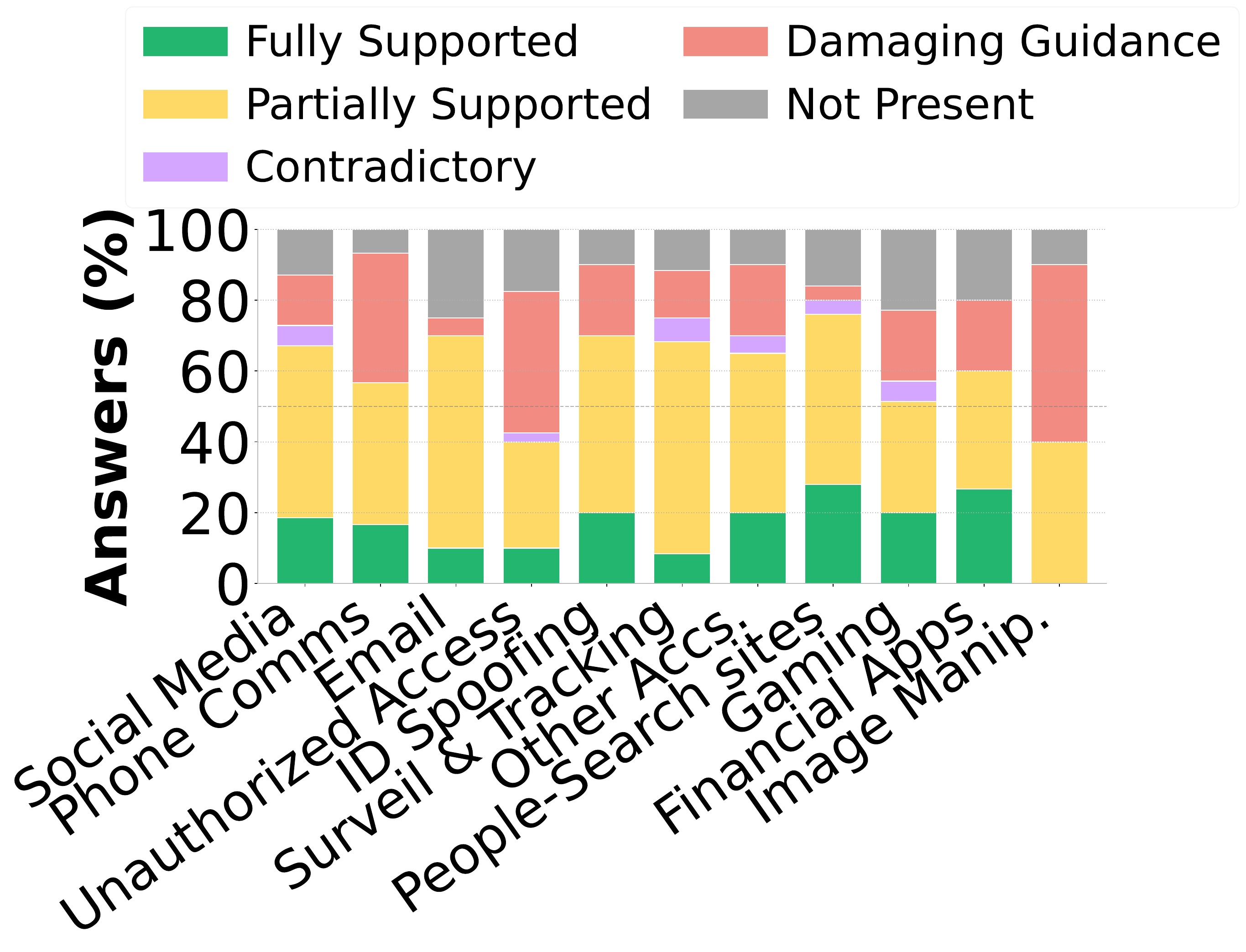}
    \label{fig:llm_acc_by_tech}
}
\subfloat[LLM responses by system]{
    \includegraphics[width=0.22\linewidth]{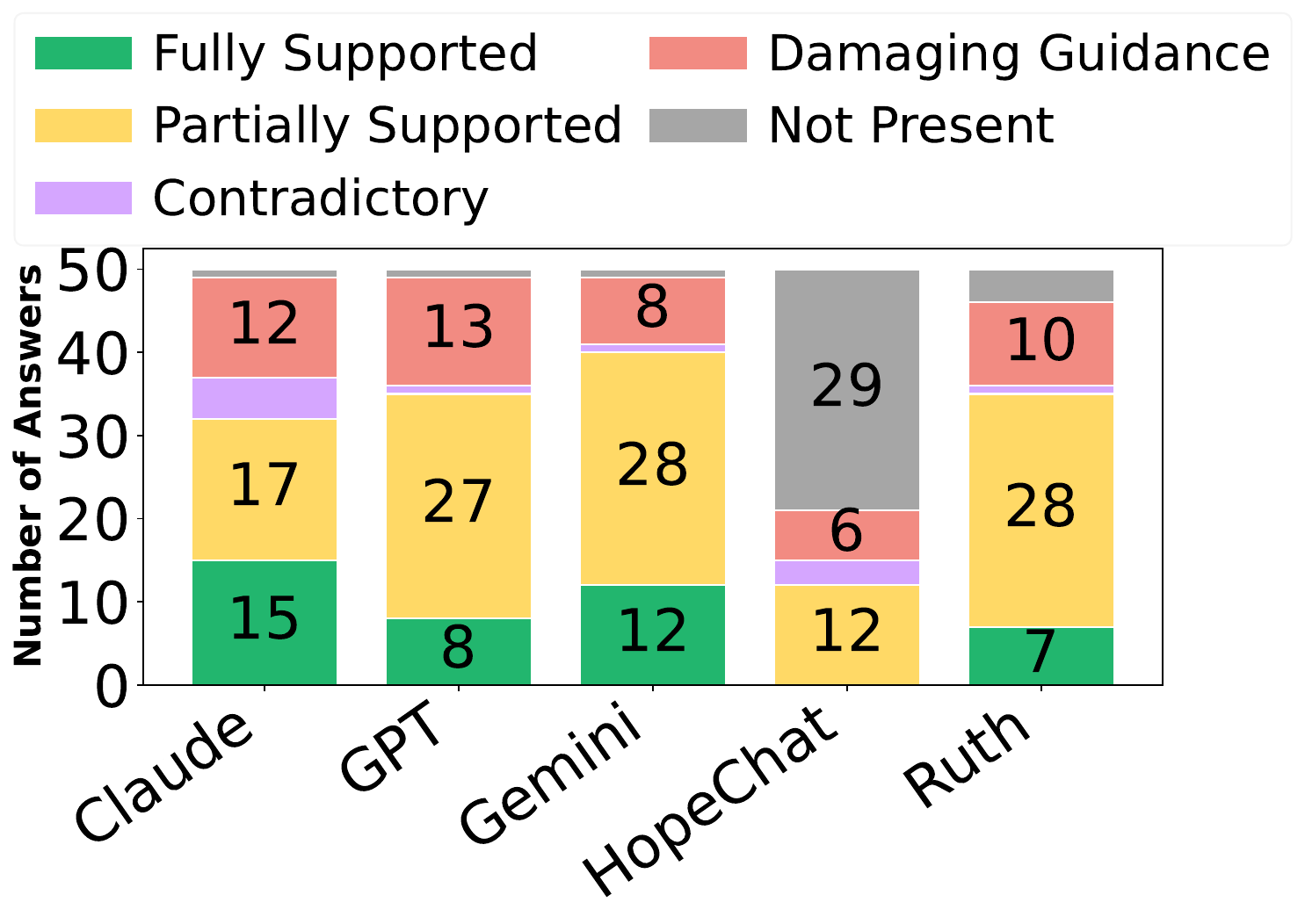}
    \label{fig:model_accuracy_eval}
}
\caption{Risk-sensitive accuracy results by technology-misuse category and conversational system.}
\label{fig:accuracy_by_tech}
\end{figure*}

Performance varied among conversational systems (Figure~\ref{fig:model_accuracy_eval}). Claude, GPT, and Gemini produced fully or partially supported responses in 64--80\% of cases. Ruth reached 70\%, whereas HopeChat reached 24\% and provided no substantive technical guidance in more than half of its responses. General-purpose systems also made occasional factual errors. Claude and GPT stated that identifying an AirTag's owner was impossible, although Apple's documentation explains that scanning an unknown AirTag can reveal the last four digits of the owner's phone number~\cite{apple-airtag}. Ruth incorrectly suggested that a victim could unblock and immediately re-block a Facebook contact, overlooking Facebook's 48-hour waiting period.

Technical completeness also varied by technology (Figure~\ref{fig:accuracy_by_tech}). No evaluated Google webpage for \textit{Email}, \textit{Financial and Payment Platforms}, or \textit{Image and Video Manipulation} were fully supported. Reddit threads were never fully supported in any category. General-purpose LLMs performed relatively well for \textit{Email}, \textit{Identity Obfuscation and Spoofing Technology}, and \textit{People-Search Sites}, but remained uneven elsewhere. \textit{Image and Video Manipulation} was a recurring weakness across the ecosystem. Per-model breakdowns appear in Appendix~\ref{sec:tech_acc_appendix}.

\subsubsection{Conversational Systems Provide More Steps and Rationale}

\begin{figure*}[t]
    \centering
    \subfloat[Google webpages]{
        \includegraphics[width=0.22\linewidth]{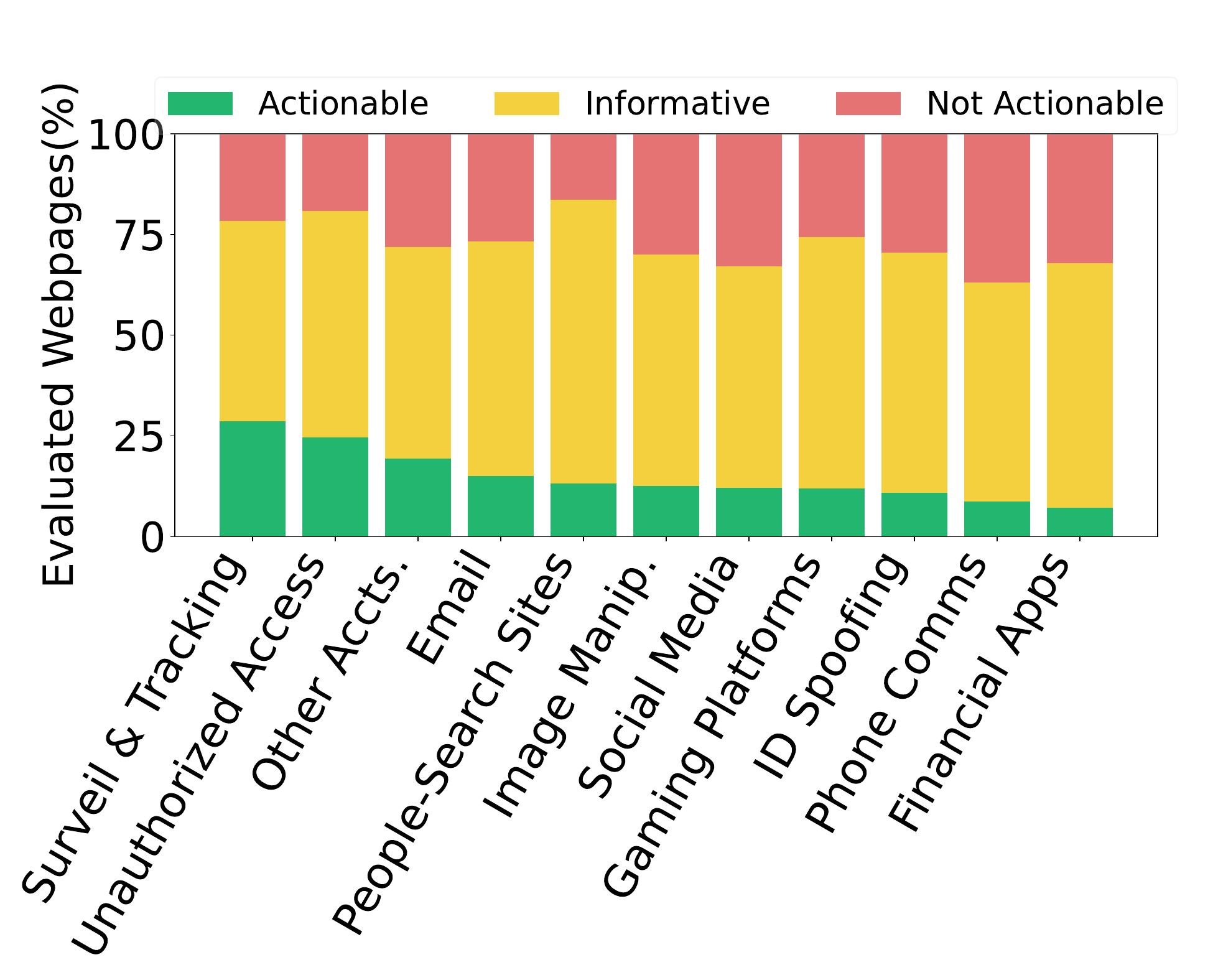}
        \label{fig:google_search_actionability_by_tech_victim}
    }
    \subfloat[Reddit threads]{
        \includegraphics[width=0.22\linewidth]{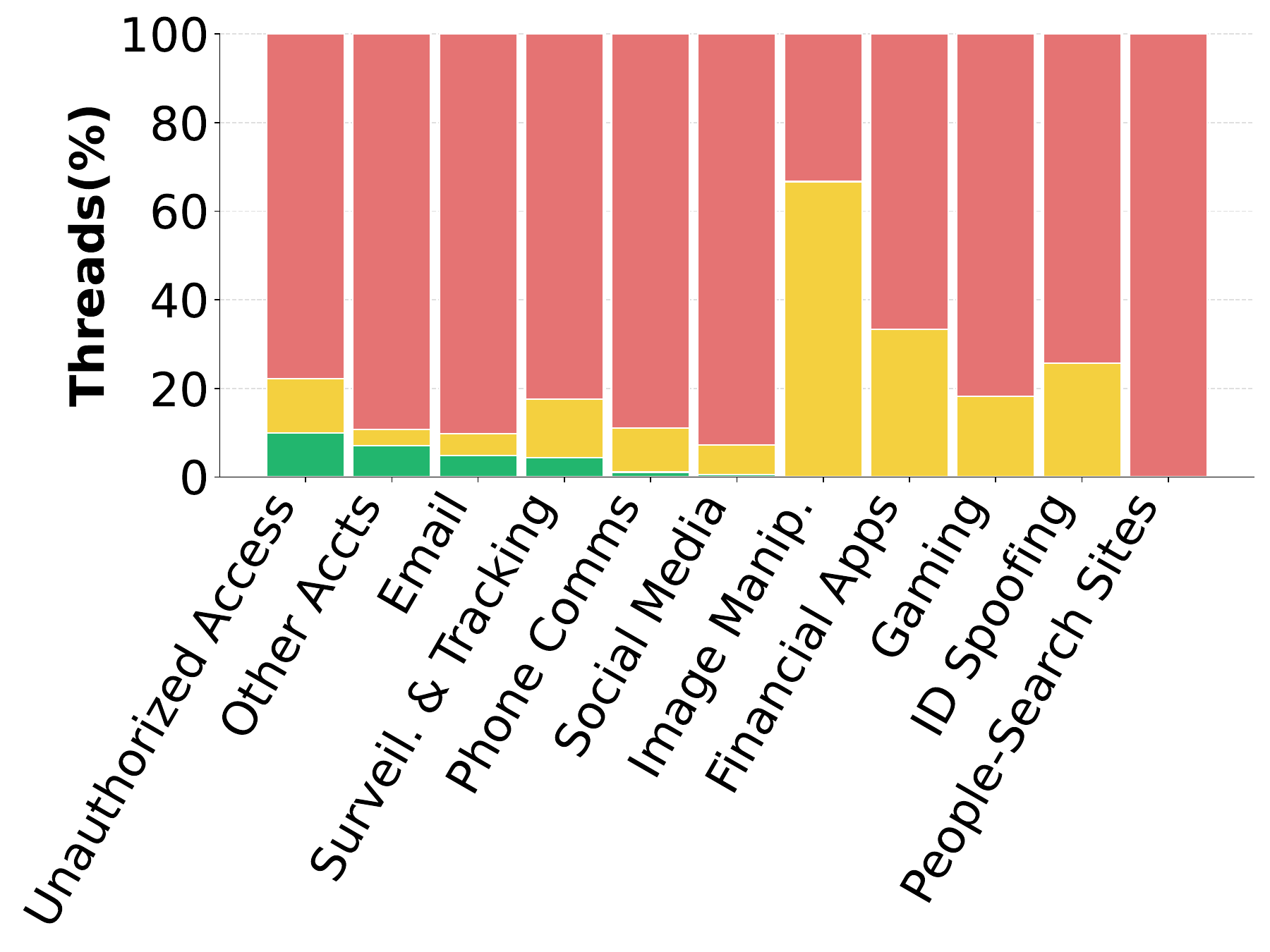}
        \label{fig:reddit_comments_actionability_by_tech_victim}
    }
    \subfloat[Conversational systems]{
        \includegraphics[width=0.26\linewidth]{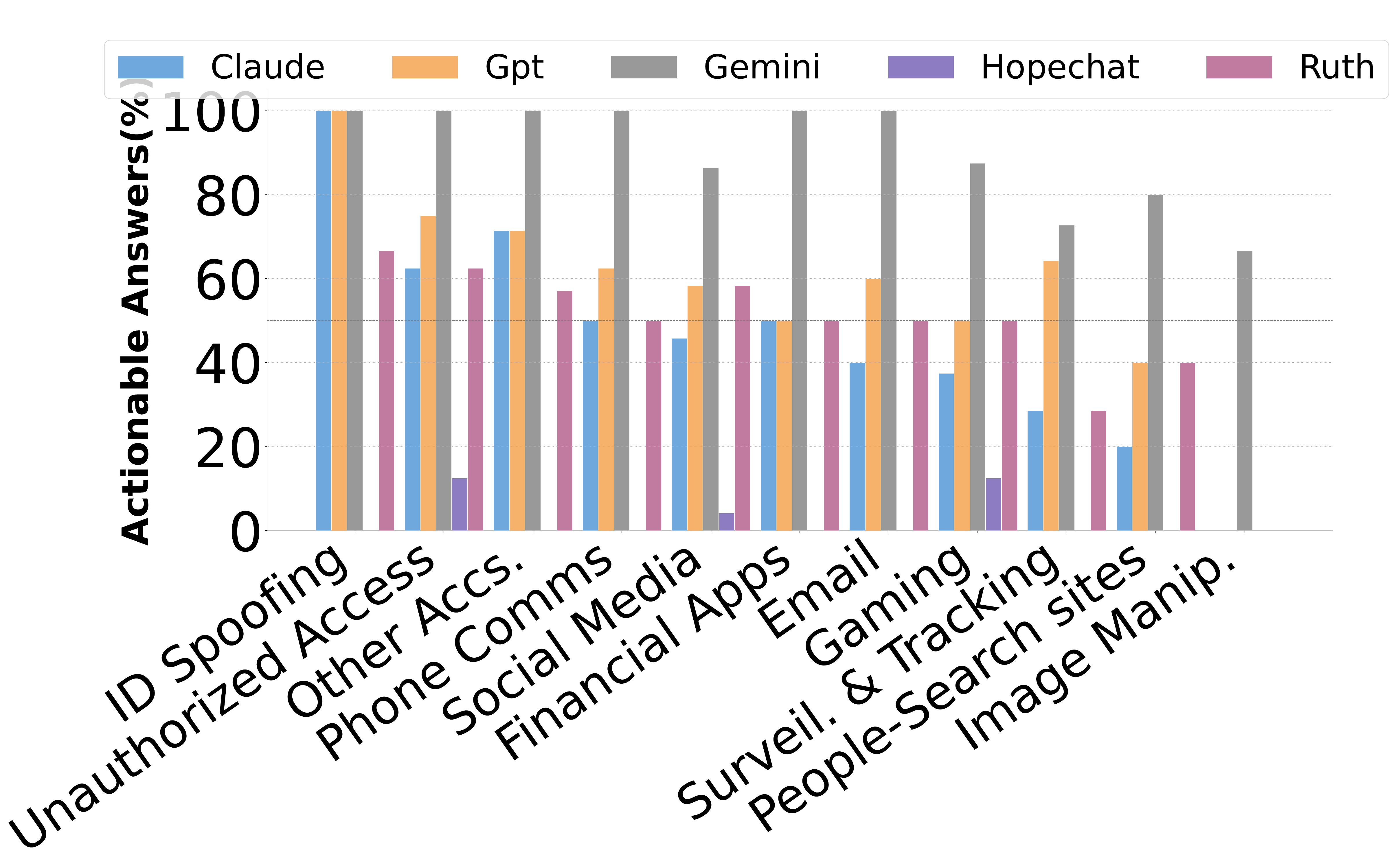}
        \label{fig:actionable_model_tech}
    }
    \subfloat[Cross-ecosystem comparison]{
        \includegraphics[width=0.24\linewidth]{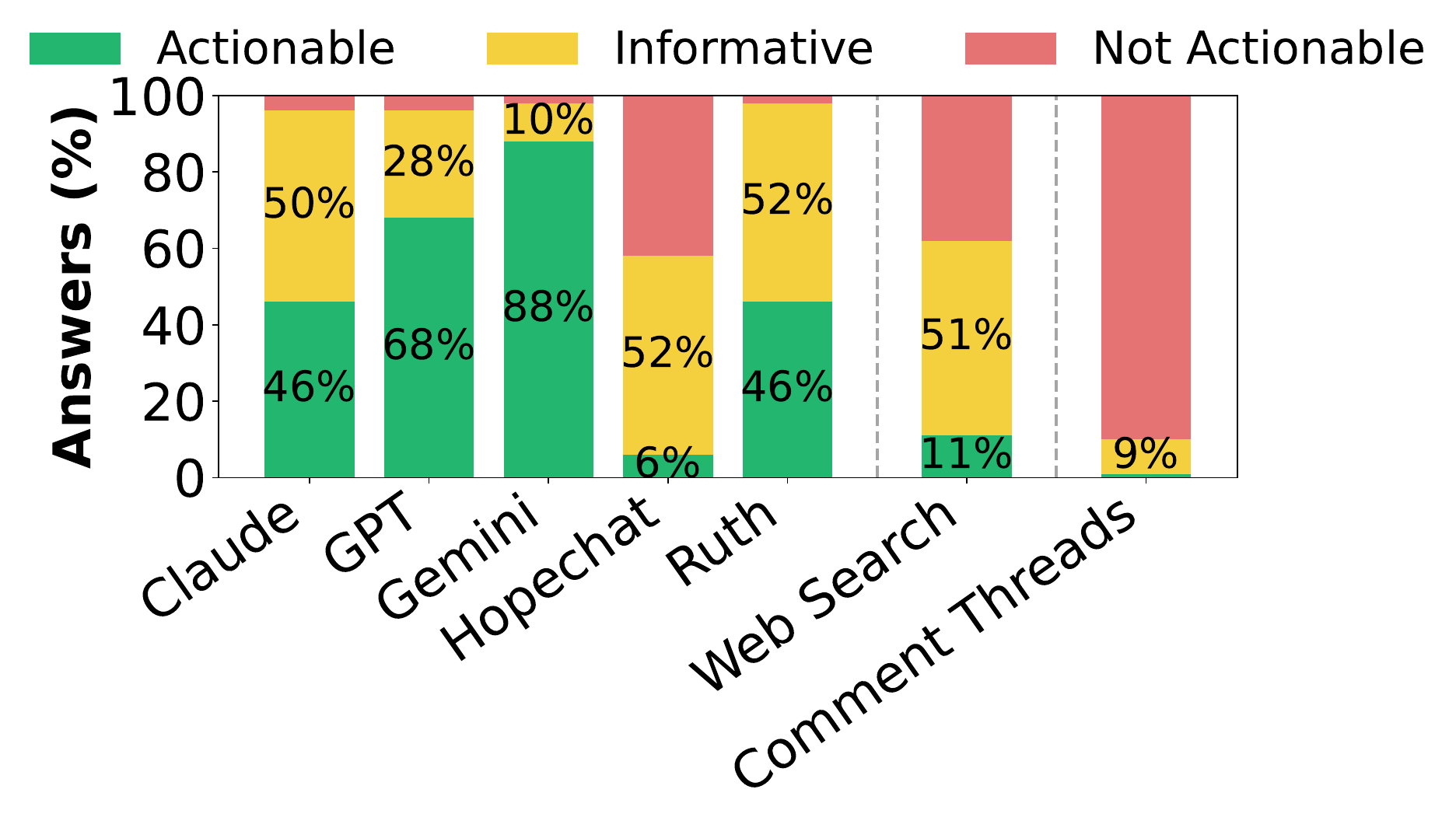}
        \label{fig:cross_platform_actionability}
    }
    \caption{Actionability across ecosystems and technology-misuse categories.}
    \label{fig:actionability_by_tech}
\end{figure*}

The largest cross-ecosystem differences concerned how guidance was presented. Among relevant Google webpages, 11\% were \textit{Actionable}, 51\% were \textit{Informative}, and 38\% were \textit{Not Actionable} (Figure~\ref{fig:actionability_by_tech}). 
At the query level, only 223 queries (7.97\%) fulfilled the condition where the first \textit{Relevant} webpage returned was also \textit{Actionable}.
 Thus, search often supplied background information without instructions that a help seeker could directly follow. Relevant Reddit threads were still less practical: 1\% were \textit{Actionable}, 9\% were \textit{Informative}, and 90\% were \textit{Not Actionable}, and only 13 (0.52\%) of all evaluated queries recieved both relevant and actionable guidance. In contrast, Claude, GPT, and Gemini produced actionable responses for 46--88\% of queries. Ruth reached 46\%, whereas HopeChat reached only 6\% and often relied on general recommendations, such as changing passwords or enabling two-factor authentication, without platform-specific instructions.

Actionability gaps were largest for several high-consequence categories. Google webpages were actionable for 7.14--28.27\% of relevant responses across categories, with the lowest proportion for \textit{Financial and Payment Platforms}. Reddit provided actionable threads only for four categories and provided neither actionable nor informative guidance for \textit{People-Search Websites}. Among conversational systems, \textit{Financial and Payment Platforms}, \textit{People-Search Websites}, and \textit{Surveillance and Tracking Technologies} remained comparatively difficult. Only Gemini produced actionable responses for the \textit{Image and Video Manipulation} queries.

\begin{figure*}[t]
\centering
\subfloat[Google webpages]{
    \includegraphics[width=0.22\linewidth]{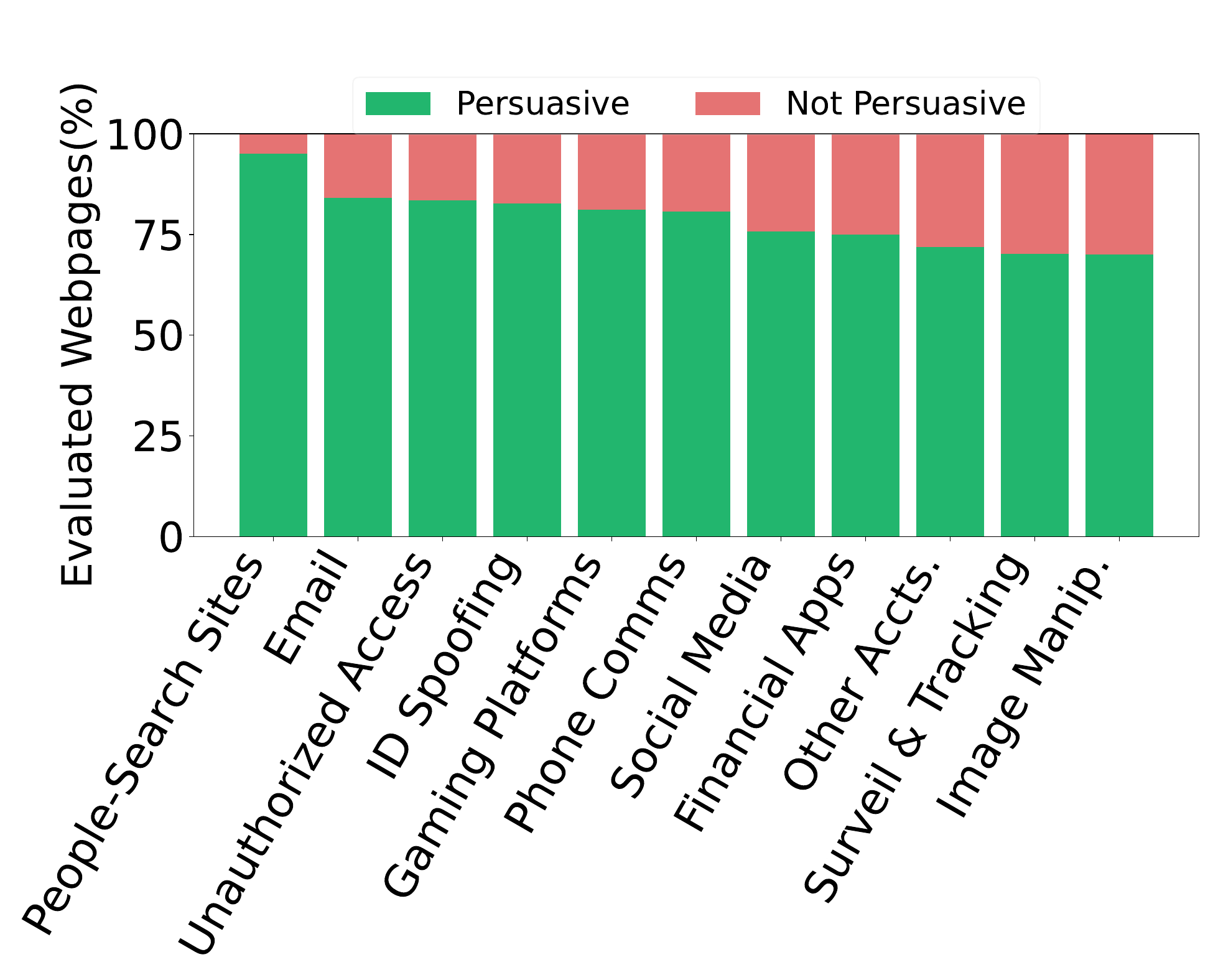}
    \label{google_search_persuasiveness_by_tech_victim}
}
\subfloat[Reddit threads]{
    \includegraphics[width=0.22\linewidth]{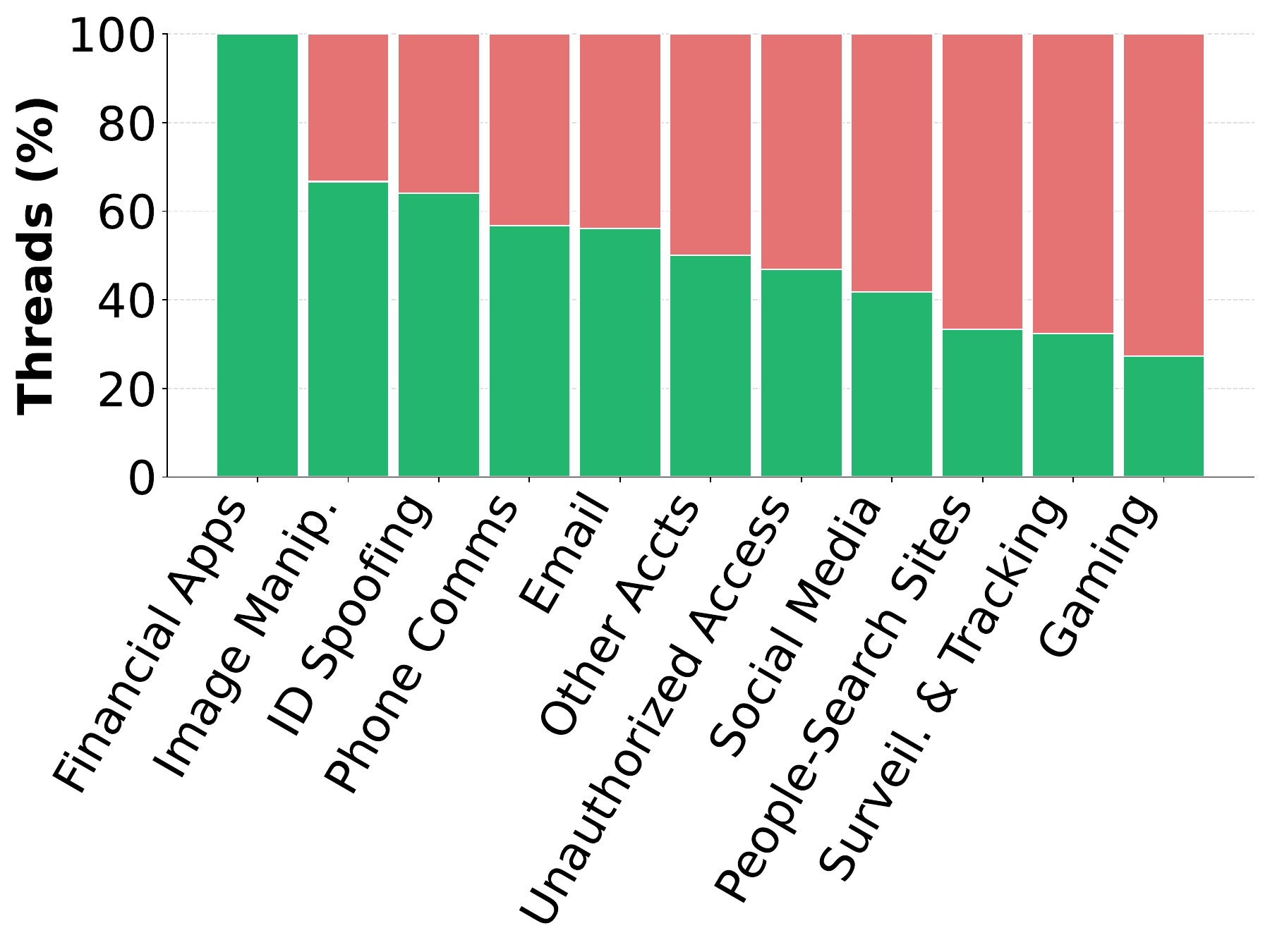}
    \label{reddit_comments_persuasiveness_by_tech_victim}
}
\subfloat[Conversational systems]{
    \includegraphics[width=0.26\linewidth]{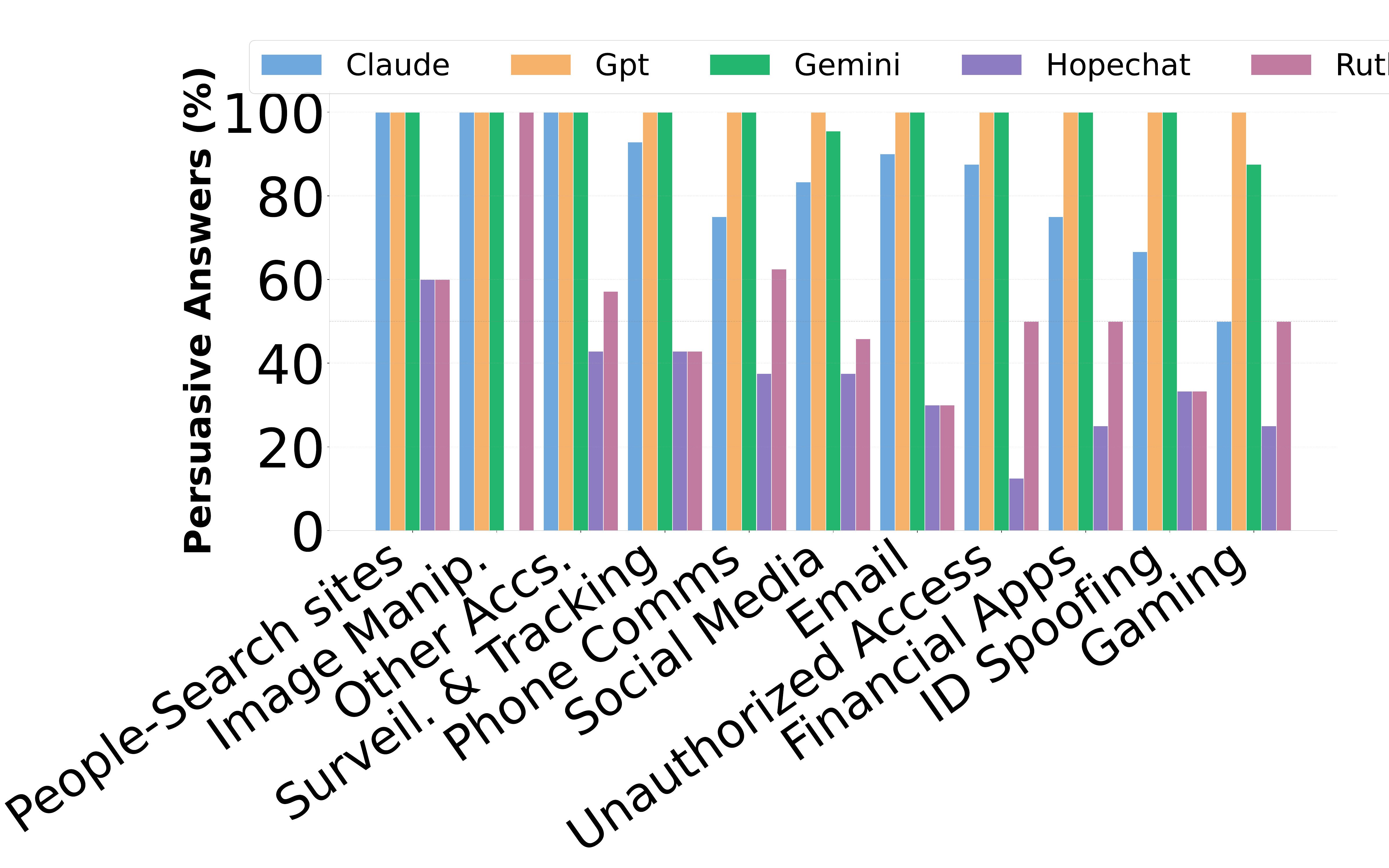}
    \label{fig:persuasive_model_tech}
}
\subfloat[Cross-ecosystem comparison]{
    \includegraphics[width=0.24\linewidth]{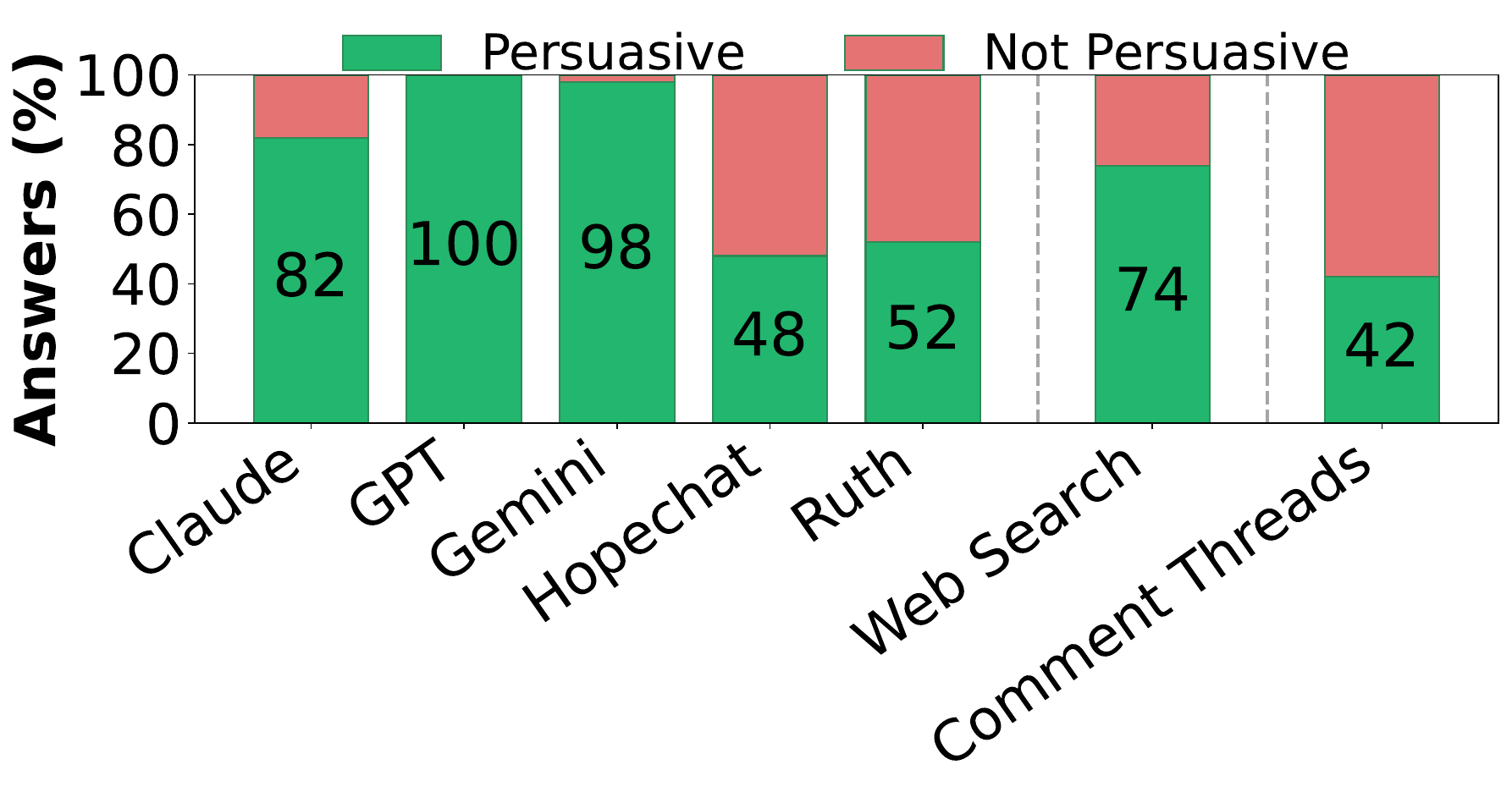}
    \label{fig:cross_platform_persuasiveness}
}
\caption{Persuasiveness across ecosystems and technology-misuse categories.}
\label{fig:persuasiveness_by_tech}
\end{figure*}

The persuasiveness results followed a similar pattern (Figure~\ref{fig:persuasiveness_by_tech}). Among relevant content, 74\% of Google webpages explained why a recommended action could help, compared with 42\% of Reddit threads. Moreover, 66.46\% of all evaluated queries received \textit{Persuasive} guidance in the first relevant Google webpage returned, while 22\% received Reddit comment threads that were both \textit{Relevant} and \textit{Persuasive}. Claude, GPT, and Gemini were persuasive for 82--100\% of queries, whereas HopeChat and Ruth reached 48\% and 52\%, respectively. Google webpages were persuasive for 75--90\% of relevant responses across technology categories. General-purpose LLMs were also consistently persuasive across categories, while the two survivor-support chatbots frequently listed actions without connecting them to the victim's safety or technical goal.

\subsubsection{Readability Does Not Distinguish Safe Support}

Responses were generally readable at or below a high-school level. Google webpages were typically written at grades 8--12, Reddit comments at grades 5--8, and conversational responses at grades 6--12. These results, reported in Appendix Section~\ref{sec:understandability}, suggest that linguistic complexity was not the principal barrier across the three ecosystems. Readability, however, does not indicate whether guidance is complete, actionable, or safe to follow.

%% file: results_social.tex
\subsection{RQ2: Contextual Safety Failures}
\label{sec:results_contextual_safety}

The technical-quality results do not capture whether acting on guidance could increase danger, whether retrieved sources expose users to questionable destinations, or whether a response provides trauma-informed support. We therefore first examine \textit{Damaging Guidance}, the common risk-sensitive label applied across ecosystems, and then report the distinct exposures measured for Google Search, Reddit, and conversational systems. The ecosystem-specific measures are complementary and are not interchangeable safety scores.

\subsubsection{Technically Plausible Actions Can Still Be Damaging}

All three ecosystems contained recommendations that could increase escalation risk or destroy useful evidence. Among the relevant responses included in the accuracy audit, 17.3\% of Google webpages and 13.3\% of Reddit threads were labeled \textit{Damaging Guidance}. 
Moreover, 11 (22\%) queries returned a relevant webpage containing \textit{Damaging Guidance}.  The rate was 19.6\% across all 250 conversational-system responses (Table~\ref{tab:accuracy_label_distribution}). These percentages have different denominators because the Google and Reddit audits are conditional on relevance, while every conversational response was evaluated.

The recommendations often resembled ordinary security advice but omitted the victim's interpersonal safety context. Webpages and Reddit threads recommended changing phone numbers, deleting accounts, removing spyware, or resetting devices without first discussing evidence preservation, surveillance, or escalation. One webpage advised deleting a harasser's voicemails rather than preserving them for possible reporting. Claude, GPT, and Ruth recommended unblocking an abusive Facebook contact to recover messages without considering the risks of re-establishing contact. Damaging guidance was not confined to one technology category. It exceeded 20\% of evaluated webpages for \textit{Phone Communications}, \textit{Email}, \textit{Identity Obfuscation and Spoofing Technology}, and \textit{Unauthorized Access and Device Compromise}. It was also prominent in conversational responses involving \textit{Image and Video Manipulation}, \textit{Phone Communications}, and \textit{Unauthorized Access and Device Compromise}. These results expose the central gap between plausible technical remediation and contextually safe assistance.

\subsubsection{Search Results Can Lead to Flagged Secondary Links}

At least one webpage returned for 1,832 victim queries (65.5\%) contained a secondary URL flagged by two or more VirusTotal engines. Some queries were associated with as many as 400 flagged secondary URLs. This is a source-integrity risk, not evidence that Google directly returned a malicious webpage or that a user followed any secondary link. Phishing was the most common threat category associated with the flagged URLs, accounting for 64.4\% of the category assignments shown in Figure~\ref{fig:contextual_exposure}. Other assignments included \textit{malware}, \textit{suspicious}, \textit{spam}, and \textit{not recommended}. Query-level exposure remained high across technology-misuse categories, ranging from 63.1\% to 69.4\%.

\subsubsection{Peer Threads Sometimes Contain Concentrated Harmful Discourse}

\begin{figure}[t]
\centering
\subfloat[Flagged secondary-URL threat \\ categories]{
    \includegraphics[width=0.47\linewidth]{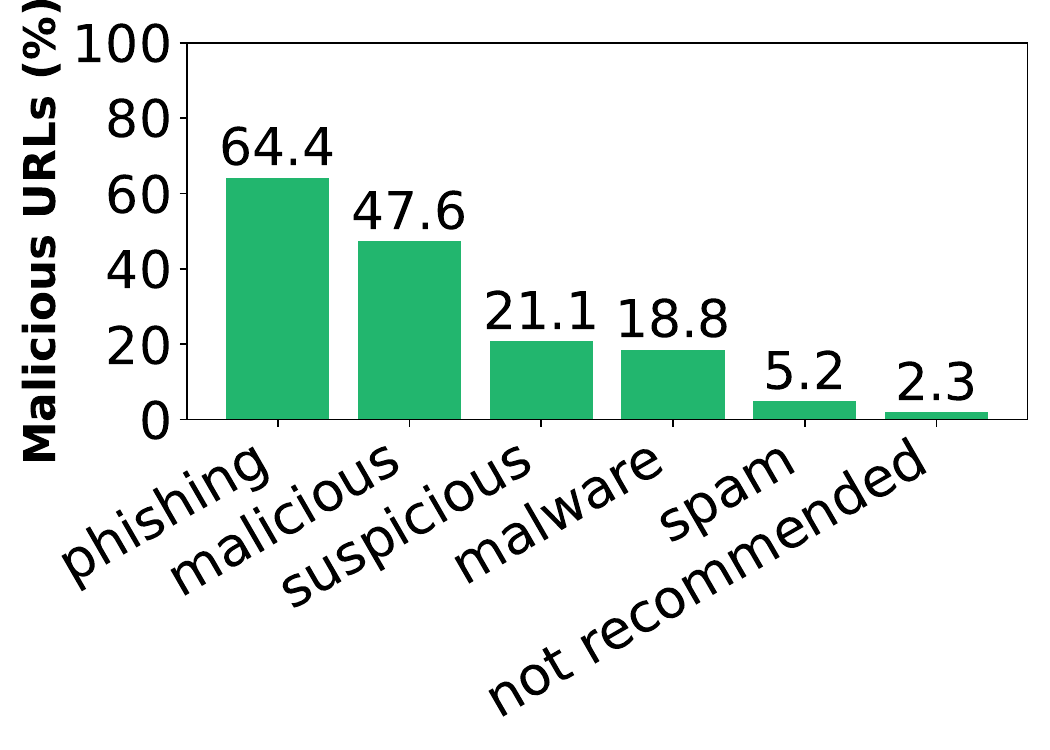}
    \label{fig:malicious_url_victim_categorization}
}
\subfloat[Harmful-language attributes in Reddit threads]{
    \includegraphics[width=0.43\linewidth]{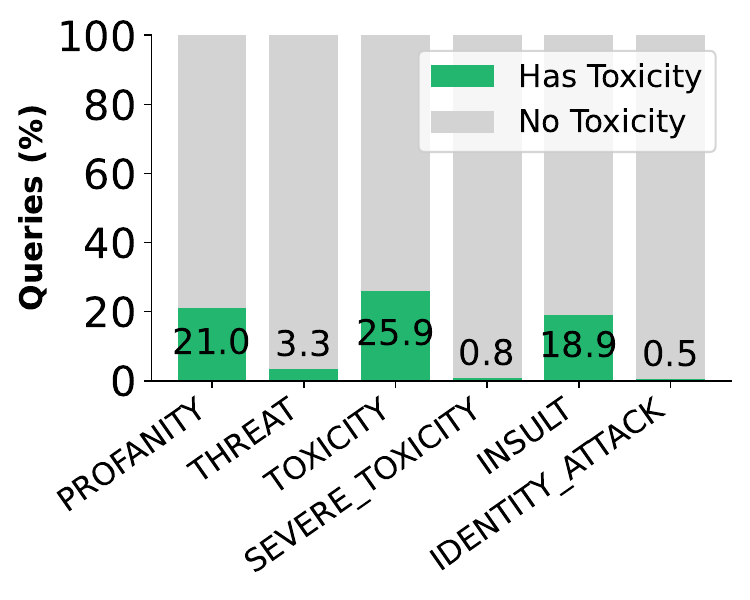}
    \label{fig:ques_toxic_victim}
}
\caption{Ecosystem-specific exposures in Google-retrieved webpages and Reddit threads.}
\label{fig:contextual_exposure}
\end{figure}

More than 20\% of queries with Reddit responses had at least one comment that exceeded the Perspective API threshold for a harmful-language attribute (Figure~\ref{fig:contextual_exposure}). \textit{Toxicity}, \textit{Insult}, and \textit{Profanity} were the most common signals. Most threads were largely supportive, but harmful comments were concentrated in some discussions. One query's source thread contained 31 flagged comments (Appendix Figure~\ref{fig:CDF_toxicity_per_query_victim}). Because a thread can contain both supportive and harmful comments, this analysis measures potential exposure rather than classifying an entire community response as harmful.

\subsubsection{Conversational Systems Often Lack Trauma-Informed Support}

\begin{figure*}[t]
    \centering
    \subfloat[Empathy and Humanization]{
        \includegraphics[width=0.19\linewidth]{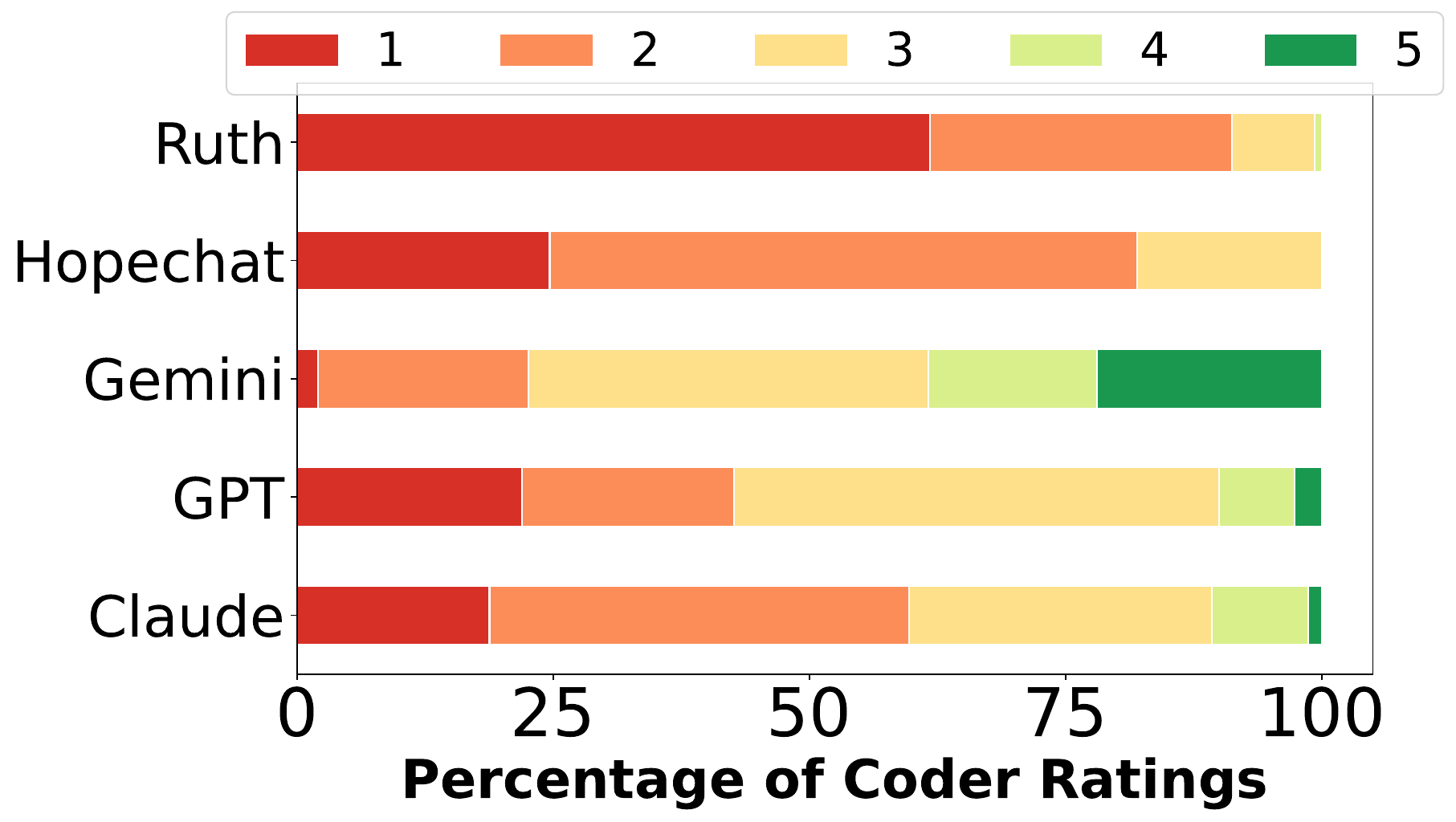}
        \label{fig:likertscale1}
    }
    \subfloat[Voice and Choice]{
        \includegraphics[width=0.19\linewidth]{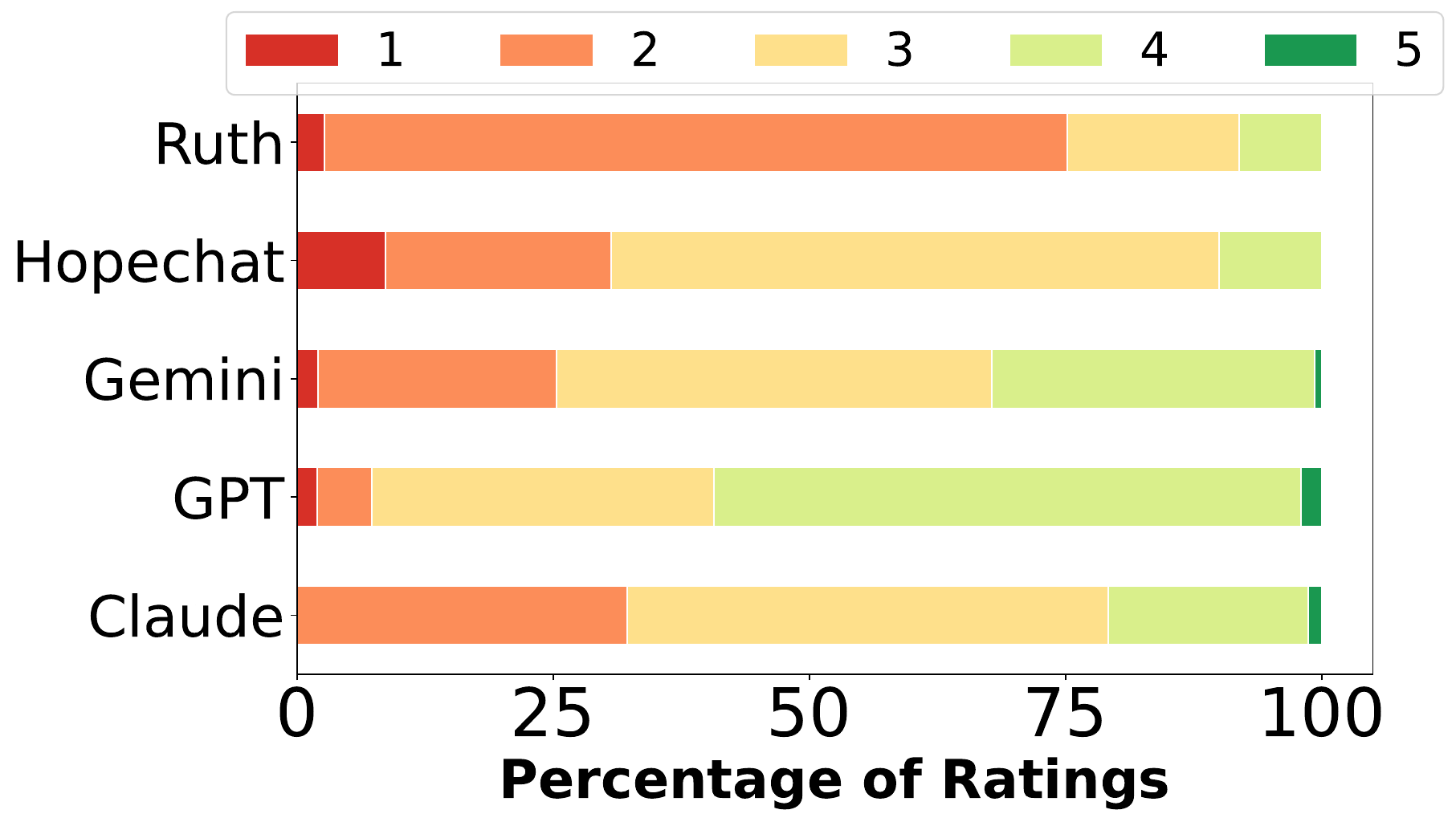}
        \label{fig:likertscale2}
    }
    \subfloat[Bias]{
        \includegraphics[width=0.18\linewidth]{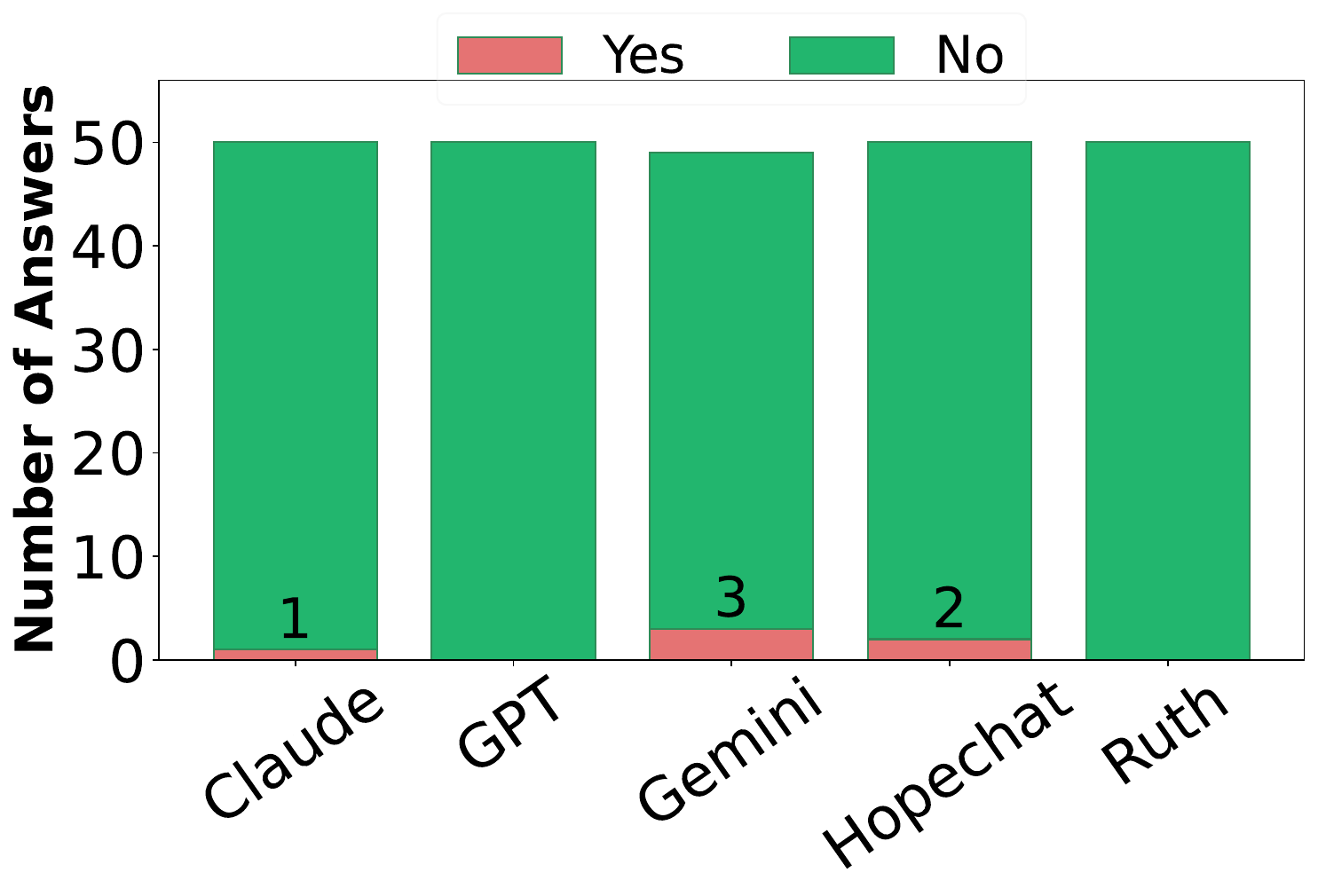}
        \label{fig:bias}
    }
    \subfloat[Risk-Informed Guidance]{
        \includegraphics[width=0.18\linewidth]{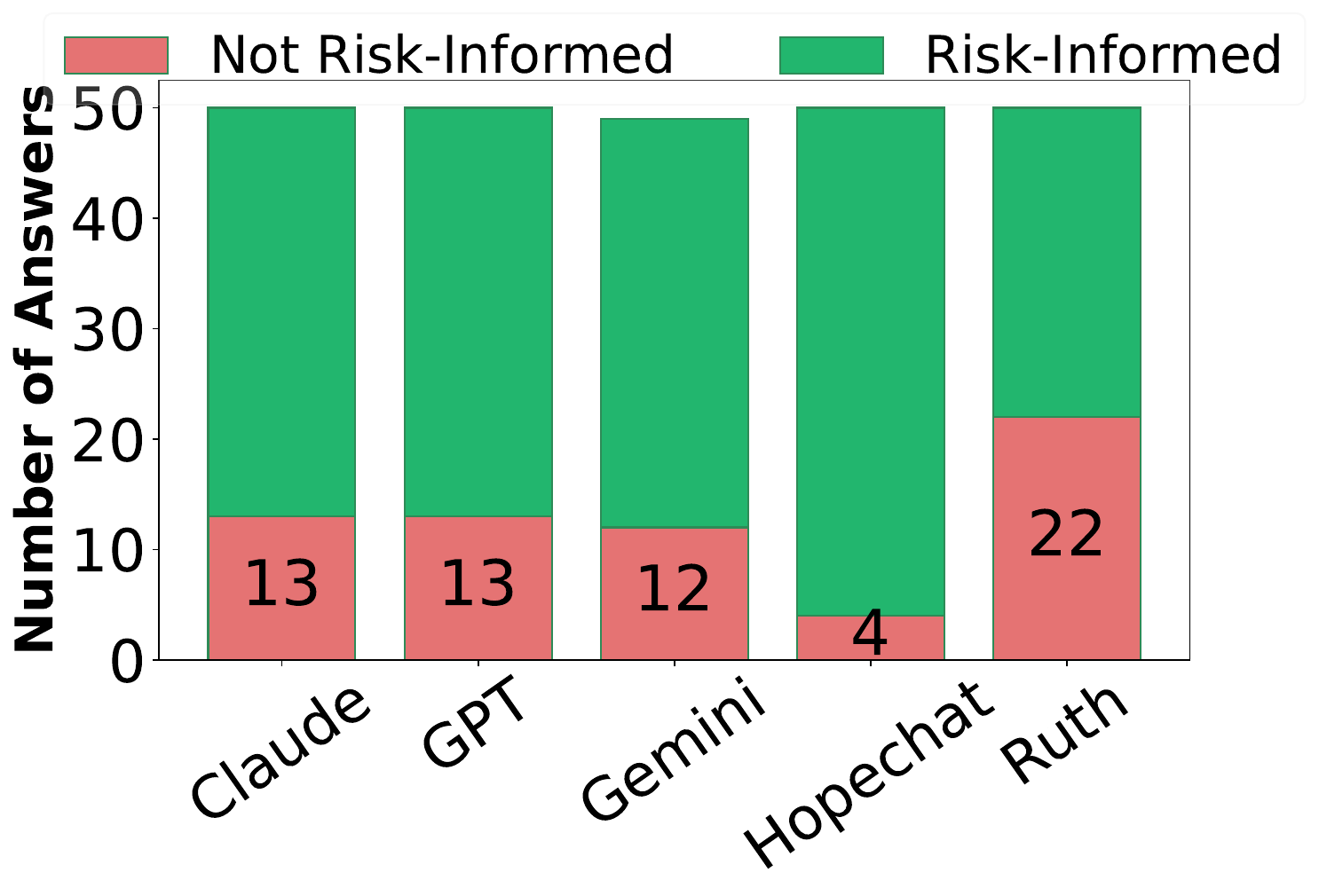}
        \label{fig:safety}
    }
    \subfloat[Support Information]{
        \includegraphics[width=0.18\linewidth]{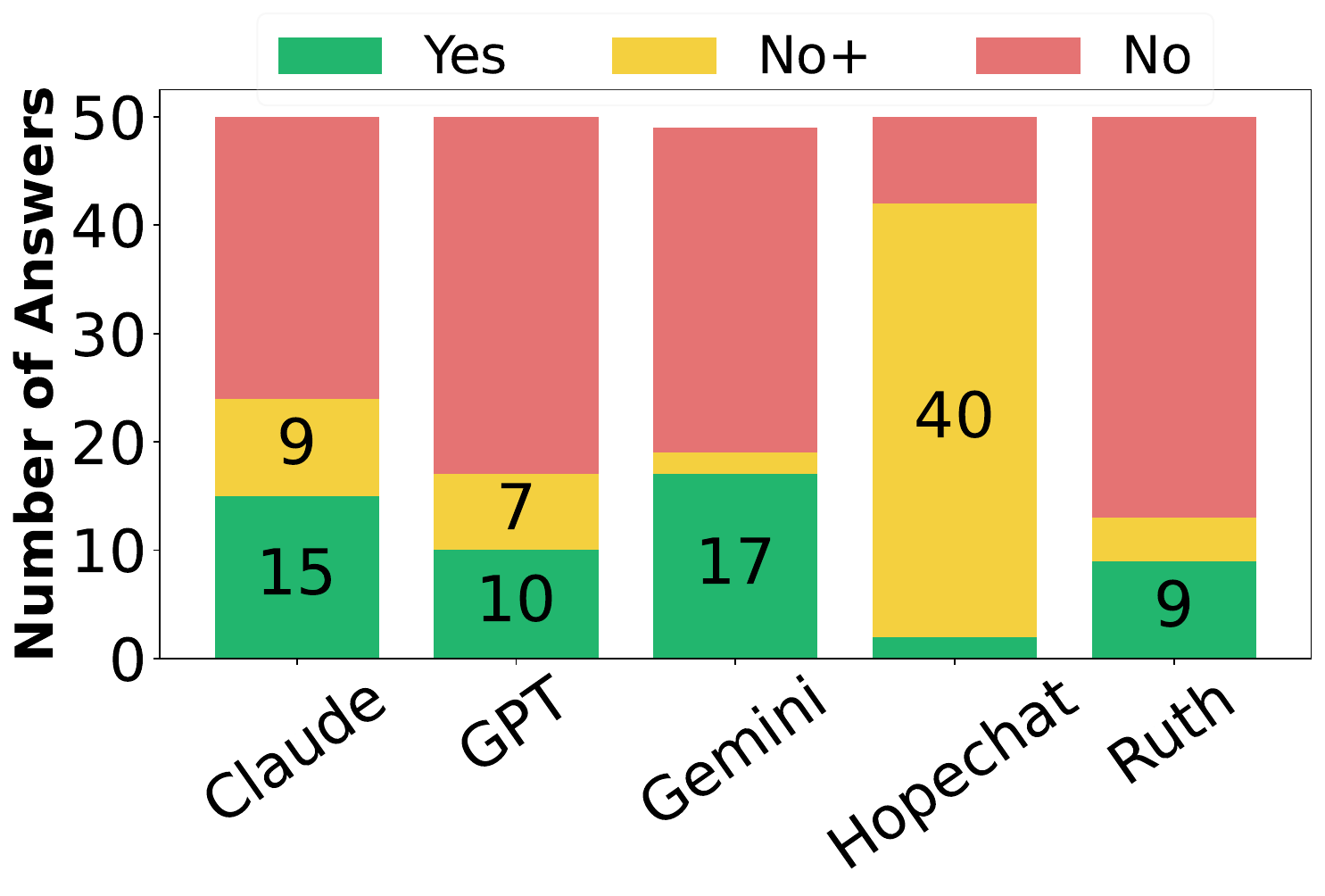}
        \label{fig:support}
    }
    \caption{Trauma-informed support measures for conversational-system responses.}
    \label{fig:socialmetricseval}
\end{figure*}

The conversational systems varied in empathy and support for user autonomy (Figure~\ref{fig:socialmetricseval}). Ruth and HopeChat received the lowest \textit{Empathy and Humanization} ratings. Nearly 90\% of Ruth responses and approximately 80\% of HopeChat responses were rated 1 or 2. Claude responses were concentrated at ratings 2 and 3. Gemini received the highest ratings, but only 38\% of its responses were rated 4 or 5. For \textit{Voice and Choice}, GPT performed best, with 59\% of responses rated 4 or 5. Gemini received ratings of 3--5 for 73\% of responses, while 74\% of Ruth responses were rated 1 or 2.

Explicit bias was uncommon, but several systems introduced unsupported assumptions or victim-blaming language. Claude responded to a question about hidden cameras or trackers by suggesting that feeling watched may reflect a mental-health condition before addressing the safety concern. Gemini assumed the help seeker was a minor and advised the parents to speak, although the query did not identify the victim as a minor. HopeChat suggested that an ex-partner might find a victim's new Reddit account \textit{``especially if you haven't taken steps to protect your privacy,''} shifting responsibility toward the victim. It also declined stalking and harassment questions that did not explicitly mention domestic violence.

All systems sometimes produced responses lacking risk-informed guidance. Common failures included recommending confrontation, disclosure, or technical remediation without discussing escalation or contingency planning. Ruth had the highest proportion of not risk-informed responses, at 44\%. HopeChat appeared to make fewer unsafe recommendations partly because it avoided specific technical guidance and offered general advice. Gemini performed strongly on several technical metrics, empathy, and autonomy, but weaker risk-informed and support-referral results show that these strengths do not necessarily co-occur.

\subsubsection{Support Information Rarely Connected Victims to Direct Services}
Specialized support referrals were rare. Most responses received \textit{No+} (asking whether the user needed resources) or \textit{No} (providing no support resources and not asking about support needs) labels for \textit{Support Information}. Social Work coders noted that, even when support resources were provided, they were often educational or informational materials published by organizations rather than directive referrals that connected victims to crisis-oriented services, advocacy organizations, shelters, hotlines, or legal resources.

\subsection{RQ3: Cross-Ecosystem Differences}
\label{sec:results_cross_ecosystem}

Three cross-ecosystem differences emerge. First, information availability and response delivery vary substantially. Google Search most consistently provided access to relevant information, while general-purpose LLMs most consistently converted advice into concrete steps and explanations. Reddit offered the least consistent query-specific support, and even relevant threads were usually incomplete and non-actionable.

Second, conventional quality measures obscure different forms of contextual risk. Search results can be relevant and persuasive while linking to destinations flagged by multiple security engines. Reddit can provide peer validation while exposing a victim to harmful comments. Conversational systems can produce fluent, actionable, and persuasive guidance while omitting escalation risk, evidence preservation, or specialized support. The common \textit{Damaging Guidance} label shows that unsafe recommendations are not confined to one ecosystem.

Third, domain specialization did not guarantee better support. HopeChat underperformed the general-purpose LLMs on relevance, technical completeness, actionability, persuasiveness, empathy, and autonomy. Ruth was competitive with general-purpose systems on some technical measures, but performed poorly on empathy, autonomy, and risk-informed guidance. Because the ecosystem-specific analyses measure different exposures, these findings do not support a single overall safety ranking. They instead show that users face distinct failure modes as they move among search, peer communities, and conversational systems.

Taken together, the results answer RQ3 by revealing a tradeoff between access, presentation, and contextual safety. The systems that make guidance easiest to find or follow are not necessarily those that best account for the adversarial interpersonal conditions that define TFA.

%% file: discussions.tex
\section{Discussion}
\label{sec:discussion}

\subsection{Technical Quality Is Not Contextual Safety}

Conventional information-quality measures are necessary but insufficient in adversarial interpersonal settings. Relevance and technical accuracy do not establish that following advice is safe. Routine security actions, such as deleting an account, resetting a device, removing spyware, or blocking someone, can alert an abuser, destroy evidence, disrupt an investigation, or provoke escalation. In TFA, the foreseeable consequences of acting on a recommendation are therefore part of its technical quality. 
No ecosystem is uniformly superior. Google Search provides broad access to relevant material but leaves users to assess incomplete pages and linked destinations. Reddit offers experiential support, but its threads are fragmented and sometimes toxic. General-purpose LLMs provide clear guidance, yet their fluency can make unsafe recommendations appear authoritative. Survivor-support chatbots also do not reliably resolve this problem. Each ecosystem combines a different benefit with a different failure mode.
More broadly, security guidance must therefore account not only for whether an action addresses a technical problem, but also for how it changes the victim’s exposure, evidence, and agency.

\subsection{Design Implications}

\textbf{Prefer staged and reversible guidance.} Systems should distinguish low-risk steps from actions that may be irreversible or observable to an abuser. Before recommending account deletion, device reset, tracker removal, blocking, or confrontation, they should explain possible consequences and prompt consideration of evidence preservation, monitored-device risk, and individualized safety planning. With incomplete context, systems should favor safer options and escalation over confident remediation.

\textbf{Preserve user autonomy.} Trauma-informed responses should validate concerns, offer choices, explain their tradeoffs, and avoid unsupported assumptions. Empathy is insufficient when advice removes agency or lacks contingency planning.

\textbf{Treat source and referral integrity as part of support quality.} Search systems should consider destination reputation and provenance, while conversational systems should distinguish general information from specialized support. Referrals should be current, appropriate, and available to the user.

\textbf{Evaluate the complete help-seeking path.} Evaluations should report availability, technical completeness, contextual risk, uncertainty, and escalation separately because high conditional quality can obscure low coverage and actionability can obscure damaging advice. Criteria not reducible to automated classification should receive direct assessment by technical and victim-advocacy experts.

\subsection{Limitations}
Our queries derive from \emph{r/Stalking} and may not represent other survivor populations or support settings; category distributions therefore characterize only our corpus. Question extraction may also omit safety-relevant context. The evaluation has additional constraints: risk-sensitive accuracy covers 50 queries because interdisciplinary reference answers require substantial manual effort; Google and Reddit measures are conditional on relevance, whereas conversational-system measures cover all responses; and automated evaluators remain fallible. In particular, Perspective may misclassify abuse-related language, while VirusTotal flags are vendor signals, not ground truth or evidence of user visits. Finally, because online content and systems change, our results reflect the collection period and measure available content—not victims’ trust, actions, or resulting harms.

%% file: background.tex
\section{Related Work}
\label{sec-related-work}
\textbf{TFA and survivor help-seeking.}
Prior research documents how abusers repurpose  technologies for surveillance, coercion, and harassment~\cite{definingTFA,spywareIPV,webofabuse,stalkerparadise}. This work examines misuse of social and communication platforms~\cite{clinicalTFA,social-media,abuseradvicetiktok,TFAdefinition,singhal2024content}, image-based abuse and deepfakes~\cite{deepfake2,han2025characterizing,deepfake4,sexualTFA}, and dual-use technologies such as IoT devices, location trackers, and financial applications~\cite{iotabuse,iotabuse2,dualuseapps,financialTFA,zhang2025abusability}. Interviews, surveys, and case studies have established how survivors experience these harms. Our focus is complementary: we study the information needs expressed while survivors seek help online and the support they encounter in response.

\textbf{Formal and online support.}
Technology-focused clinics, including CETA, TECC, and Mavs ETA, combine security expertise with survivor advocacy~\cite{clinicalTFA,TECC,tseng2021digital,mavsetaclinic,madisontechclinic}. Their reach is constrained by staffing, geography, and referral-based access~\cite{clinic}. Survivors also face limited technical expertise among service providers, uncertainty about available resources, and concerns about retaliation, stigma, and disbelief~\cite{lackofexpertise,lacoftechadvocates,barriershelp-seeking,disclosureofabuse}. Consequently, many turn to search engines and online communities to interpret technology-related harms and identify possible actions~\cite{policehelp2,policehelp,webofabuse,helpseekingyouth}. We evaluate these online systems as parts of the existing support landscape, not as substitutes for individualized advocacy.

\textbf{Web search and peer-support communities.}
Almansoori et al.~\cite{webofabuse} evaluated Google Search resources for intimate partner surveillance using resource-level measures of relevance, understandability, actionability, and quality. We instead begin with questions extracted from survivor narratives across multiple TFA categories, assess whether retrieved pages answer each question, and examine the integrity of secondary destinations linked from those pages. Whiting et al.~\cite{whiting2023online} and Hui et al.~\cite{hui2023examining} studied the social and informational support offered to abuse survivors in online communities. Our Reddit analysis retains this social-support perspective while also assessing whether complete discussion threads provide query-specific technical guidance and whether they expose help-seekers to harmful discourse.

\textbf{Conversational systems for survivor support.}
Researchers have developed specialized conversational systems such as Ruth, Aimee, and HopeChat~\cite{ruth,aimee,hopechat}. Sanz et al.~\cite{sanz2025empathy} evaluated gender-based violence chatbots, emphasizing empathy, bias, privacy, and data responsibility. Most closely related, Prakash et al.~\cite{prakash2026assessing} evaluated general-purpose and domain-specific LLMs for TFA questions using expert assessment of accuracy, completeness, safety, and actionability, together with a survivor study of perceived actionability. Our objective differs: we situate conversational systems alongside web search and peer discussion, submit the same victim-derived questions to each ecosystem, apply common technical measures where comparison is appropriate, and separately evaluate ecosystem-specific risks. This design reveals how availability, technical quality, trauma-informed support, and structural safety risks vary across the online support ecosystem.

%% file: conclusion.tex
\section{Conclusion}

We presented a victim-centered measurement of digital support for technology-facilitated abuse across Google Search, Reddit, three general-purpose LLMs, and two survivor-support chatbots. Using 2,797 questions derived from \emph{r/Stalking} narratives from 2014--2024, we evaluated technical support alongside damaging guidance and ecosystem-specific risks.

Our results reveal fragmented support and a persistent gap between technical usefulness and contextual safety. Search results can lead to flagged destinations, Reddit threads can contain harmful discourse, and conversational systems can omit escalation risks, evidence-preservation concerns, and specialized support. Neither community knowledge nor domain specialization consistently resolves these failures. Safe TFA assistance must therefore account for the consequences of technical actions within adversarial interpersonal contexts, favoring staged guidance, uncertainty communication, evidence preservation, and appropriate escalation.





%% file: ethics.tex
\section{Acknowledgements}
This material is based upon work supported by the National Science Foundation under grants No. 2309318. Any opinions, findings, conclusions, or recommendations expressed in this material are those  of the author(s) and do not necessarily reflect the views of the National Science
Foundation. We also thank Michael Danilson and Karanjit Unique (M.Sc., University of Texas at Arlington) for their assistance with ground-truth annotations for help-seeking question extraction and expert assessment of the technical accuracy of guidance across the evaluated platforms, respectively. We used GPT-5 to proofread the manuscript and improve clarity. In addition, we used several open-source language models, including Llama, GPT-OSS, and Gemma, for specific evaluation tasks. All model-assisted evaluations were manually validated, and the methodological details are reported in the corresponding task-specific sections of the paper.

\section{Ethical Considerations}
\label{sec:ethics}

This study analyzes public Reddit posts containing personal narratives of technology-facilitated abuse. Although the data are publicly available, we treat all content as sensitive because of re-identification and retraumatization risks. Usernames were not retained during data collection and the data were stored on secure servers with restricted access. The study involved no interaction, intervention, recruitment, or direct contact with any users. Given that this research was a secondary analysis of publicly available data, the study was deemed exempt from human subjects’ oversight.

Most LLM-assisted evaluations used locally executed open-source models. Before submitting queries to the evaluated coversational systems, the 50 queries were reviewed to not include any personally identifying information. Examples used in prompts were anonymized or paraphrased. 

Annotation and evaluation protocols were developed in consultation with social work researchers experienced in victim advocacy and followed a trauma-informed approach.

\textbf{Sharing Artifacts.} We release anonymized research artifacts, including code, prompts, codebooks, and sample processed data, at: \url{https://shorturl.at/UVsZU}
Because the study concerns a sensitive population, we provide a representative sample rather than the complete dataset, balancing reproducibility with participant privacy.

%% file: appendix.tex
\appendices

\section{Corpus Construction and Classifier Validation}
\paragraph{\textbf{Technology-Misuse and Poster-Role Classifier Performance}}
Table~\ref{tab:tech_misused} shows the performance for each classifier for the technology-misuse categorization.
\begin{table}[t]
\centering
\caption{Performance of the technology-misuse and poster-role classifiers.}
\label{tab:tech_misused}
\resizebox{\linewidth}{!}{
\begin{tabular}{l c c c c }
\hline
\textbf{Classification label}              & \textbf{Acc.} & \textbf{Prec.} & \textbf{Rec.} & \textbf{F1} \\ \hline
Phone Communications         & 0.93              & 0.9                & 0.87            & 0.89              \\ 
Surveillance
and Tracking Technologies      & 0.94              & 0.85               & 0.83            & 0.84              \\ 
Unauthorized Access and Device Compromise                    & 0.97              & 0.89               & 0.89            & 0.89              \\ 
Identity Obfuscation and Spoofing Tools               & 0.97              & 0.91               & 0.88            & 0.89              \\ 
Email-Based Abuse                        & 0.98              & 0.88               & 0.98            & 0.93              \\ 
Social Media and Messaging Platforms                 & 0.94              & 0.97               & 0.95            & 0.96              \\ 
Other Online Service Accounts        & 0.96              & 0.78               & 0.88            & 0.82              \\ 
Online Gaming Platforms             & 0.98              & 0.86               & 0.91            & 0.89              \\ 
Data Broker and People-Search Websites & 0.9               & 0.85               & 0.82            & 0.81              \\ 
Financial and Payment Platforms       & 0.99              & 0.96               & 0.96            & 0.96              \\ 
Image and Video Manipulation        & 0.99              & 0.85               & 1               & 0.92              \\
Other        & 0.85              & 0.87               & 0.85               & 0.86             \\ \hline
Poster role & 0.96              & 0.97               & 0.96               & 0.96             \\ \hline
\end{tabular}
}
\end{table}

\paragraph{\textbf{Classifiers for TFA-relatedness and Poster-Role and Evaluation Rubrics}}
\label{prompt2}
\label{llm_metrics} The full technology-misuse classification prompts are provided with the research artifact. Tables~\ref{tab:prompt_TFA} and~\ref{tab:prompt_poster} present the prompts used for the two inclusion decisions (TFA-relatedness \& Poster-role) that directly determine the evaluation corpus.

\begin{table}[h]
\centering\scriptsize
\caption{TFA-relevance classifier prompt.}
\label{tab:prompt_TFA}
\resizebox{\linewidth}{!}{
\begin{tabular}{|p{0.93\linewidth}|}
\hline
Classifier: does the post involve Technology-Facilitated Abuse?

\textit{TFA}: any use of technology, digital platforms, or electronic devices in abuse, harassment, or intimidation.

\textbf{yes}: technology, digital platforms, or devices are explicitly mentioned or clearly implied in the abuse.

\textbf{no}: no technology abuse is mentioned, or the post concerns in-person abuse or stalking only.

\textit{Output only the label. Do not explain.}\\
\hline
\end{tabular}
}
\end{table}

\begin{table}[h]
\centering\scriptsize
\caption{Poster-role classifier prompt.}
\label{tab:prompt_poster}
\resizebox{\linewidth}{!}{
\begin{tabular}{|p{0.93\linewidth}|}
\hline
Classify a stalking or harassment forum post as written by an abuser or a victim.

\textbf{abuser}: the poster seeks advice on how to stalk, harass, intimidate, or manipulate someone, or confesses to such behavior.

\textbf{victim}: the poster seeks help coping with or protecting themselves from stalking, harassment, or abuse.

\textbf{na}: neither category clearly applies.

\textit{Output only the label. Do not explain.}\\
\hline
\end{tabular}
}
\end{table}

\section{Question Extraction and Quality}
\label{sec:q_quality}
\label{prompt1}
\paragraph{\textbf{Prompt for Question Extraction}}
Table~\ref{prompt_implied_question} presents the prompt used to extract explicit and implicit help-seeking questions from the Reddit narratives.

\begin{table}[h]
\centering\scriptsize
\caption{Prompt for help-seeking question extraction.}
\label{prompt_implied_question}
\resizebox{\linewidth}{!}{
\begin{tabular}{|p{0.93\linewidth}|}
\hline
You are assisting in my research on Technology-Facilitated Abuse (TFA) by analyzing questions victims ask on Reddit.
Your task is to extract the core question(s) the victim is asking in the provided post.

\vspace{0.5em}
\textbf{INSTRUCTIONS:}
\begin{itemize}
  \item Look for explicit questions. If no explicit question exists, determine whether there is an implied question based on the context.
  \item If multiple questions exist, separate them using a semicolon (;). Each question should remain contextual and include key details from the post.
  \item If there is no question (explicit or implied), output: \texttt{No questions implied by poster.}
  \item Output \emph{only} the extracted question(s), without any additional text.
  \item Keep the extracted question simple while preserving important contextual details.
  \item If the post mentions specific technologies, online platforms, websites, or devices being abused (e.g., Google, PS4, social media, spyware, hacking tools), incorporate them into the question for clarity.
\end{itemize}

\vspace{0.5em}
\textbf{EXAMPLES:}
[Examples omitted for brevity]

\vspace{0.5em}
\textbf{YOUR TASK:}
Given the following post, extract the core question(s) according to the instructions above.
Output only the extracted question(s) without any additional text.
\\
\hline
\end{tabular}
}
\end{table}

\textbf{Human Audit Criteria.}
\label{sec:question_audit}
Table~\ref{tab:question_audit_criteria} summarizes the five binary criteria used to audit each source post and its complete Llama3.3 question set for validating the Question extraction quality.

\begin{table}[h]
\centering
\scriptsize
\caption{Criteria for the human audit of extracted questions.}
\label{tab:question_audit_criteria}
\begin{tabular}{p{0.24\linewidth}p{0.68\linewidth}}
\toprule
\textbf{Criterion} & \textbf{Satisfied when} \\
\midrule
Faithfulness & Every extracted question is supported by the source post and introduces no unsupported details. \\
Help-seeking & Every question represents an explicit or implicit information need expressed by the poster. \\
Self-containedness & Every question contains enough context to be understood without consulting the source post. \\
Retention decision & The question set is correctly retained as victim-oriented and TFA-related. \\
Information-need coverage & The complete question set captures all major help-seeking needs expressed in the source post. \\
\bottomrule
\end{tabular}
\end{table}

\paragraph{\textbf{Question Uniqueness Analysis}}
\label{sec:query_uniqueness}
To assess whether the extracted victim TFA queries represented distinct information needs or semantically similar variations of recurring questions, we performed a semantic clustering analysis. We generated embeddings for each query using the \textit{all-mpnet-base-v2}~\cite{mpnet, song2020mpnet} sentence-transformer model and applied agglomerative clustering~\cite{kurita1991efficient} separately on all questions globally using cosine distance and complete linkage. We used a cosine-similarity threshold of $0.80$ for identifying strongly semantically related queries, consistent with prior semantic-clustering studies~\cite{rehbein2025americans, stefanovitch2023graph, sajeva2024clustering} and allowed the algorithm to determine the number of clusters automatically. Under complete linkage, queries are grouped together only when the least-similar pair within a cluster satisfies the similarity threshold, providing a conservative criterion for identifying semantically similar questions. 

Across the 2,797 queries, 2,075 were singleton clusters, i.e. they were not grouped with any other query. The remaining 722 questions (25.8\%) were distributed across 307 multi-question clusters. Multi-question clusters contained between two and eight questions, with a median size of two, producing 2,382 distinct semantic clusters overall when both singleton and multi-question clusters were counted. 

We also applied the same clustering analysis within each technology misuse category as well. For example, the \emph{Email} category contained 74 queries. After clustering, 66 were singleton clusters (unique queries), while the remaining 8 queries formed four clusters of two semantically similar questions each, producing 70 distinct semantic clusters. Cluster 1 contained:
\textit{`How did my ex-boyfriend gain access to my email account without me ever sharing the login credentials with him?'} and
\textit{` How did my ex-boyfriend gain access to my emails, phone, and passwords?'}
Similarly, Cluster 2 contained: \textit{`What steps can I take to lock down my personal information and protect my privacy from my stalker, considering my work location and email are publicly available?'}
and \textit{`How can I balance keeping my job with limiting the information my stalker can access through my work email?'}
Although questions within the same cluster reflect a similar underlying information need, they differ in how the query is formulated and in the contextual details provided by the victim. Since prior work has shown that different formulations of the same underlying information need can produce different search and retrieval outcomes~\cite{alaofi2022queries,penha2022evaluating}, these contextual and linguistic differences may also influence the relevance, safety, and appropriateness of the responses returned by online support systems. We therefore retained all 2,797 victim TFA queries for subsequent evaluation.

\section{Technical-quality, contextual-safety and Trauma-informed support metrics}
Tables~\ref{tab:techmetrics} and ~\ref{tab:socialmetrics} summarize the evaluation criteria used to assess the quality and safety of responses across the three support ecosystems. Table~\ref{tab:techmetrics} defines the technical-quality and contextual-safety dimensions, including \textit{Accuracy}, \textit{Relevance}, \textit{Actionability}, \textit{Persuasiveness}, and \textit{Understandability}, together with the label definitions used to distinguish different levels or types of response quality. Table~\ref{tab:socialmetrics} presents the trauma-informed support dimensions used for conversational systems, including \textit{Empathy \& Humanization}, \textit{Voice \& Choice}, \textit{Risk-Informed Guidance}, \textit{Bias}, and \textit{Support Information}.

\begin{table}[h]
\centering\footnotesize
\caption{Technical-quality and contextual-safety metrics and labels.}
\label{tab:techmetrics}
\setlength{\tabcolsep}{3pt}
\renewcommand{\arraystretch}{1.05}
\resizebox{\linewidth}{!}{
\begin{tabular}{p{0.19\linewidth} p{0.21\linewidth} p{0.50\linewidth}}
\toprule
\textbf{Metric} & \textbf{Label} & \textbf{Description} \\
\midrule
\multirow{4}{=}{Technical Accuracy}
  & Fully supported      & Covers all critical elements of the reference answer. \\
  & Partially supported  & Provides correct guidance but omits critical elements. \\
  
  & Damaging guidance    & Recommends an action that may increase danger, destroy evidence, or heighten exposure without appropriate safeguards. \\
  & Contradictory        & Contains substantive technical claims that conflict with the reference answer. \\
  & Not present          & Provides no substantive technical guidance that addresses the query. \\
\midrule
\multirow{2}{=}{Relevance}
  & Relevant     & Directly addresses the victim's question. \\
  & Not relevant & Diverts or provides unrelated advice. \\
\midrule
\multirow{3}{=}{Actionability}
  & Actionable     & Offers detailed, executable instructions. \\
  & Informative    & Provides useful advice without executable steps. \\
  & Not actionable & Contains no concrete technical guidance. \\
\midrule
\multirow{2}{=}{Persuasiveness}
  & Persuasive     & Explains why a recommendation is useful in the stated situation. \\
  & Not persuasive & Lists recommendations without explaining their purpose or benefit. \\
\midrule
Understandable & Grade 0--18 &
  0--3: kindergarten; 3--6: elementary; 6--9: middle; 9--12: high school;
  12--15: college; 15--18: post-grad. \\
\bottomrule
\end{tabular}
}
\end{table}

\begin{table}[h]
\centering\footnotesize
\caption{Trauma-informed support metrics and labels.}
\label{tab:socialmetrics}
\setlength{\tabcolsep}{2pt}
\renewcommand{\arraystretch}{1.05}
\resizebox{\linewidth}{!}{
\begin{tabular}{p{2.2cm} l p{4.0cm}}
\toprule
\textbf{Metric} & \textbf{Labels} & \textbf{Description} \\
\midrule
Empathy \& Humanization & Likert 1--5 &
  Warmth, compassion, and acknowledgment of the victim's experience. \\
\midrule
Voice \& Choice & Likert 1--5 &
  Presents safe options and preserves the victim's autonomy. \\
\midrule
\multirow{2}{*}{\makecell[l]{Risk-Informed\\Guidance}}
  & Yes & Considers escalation, evidence preservation, and other safety consequences. \\
  & No  & Recommends potentially risky actions without appropriate warnings or safety planning. \\
\midrule
\multirow{2}{*}{Bias}
  & Yes & Contains stereotypes, victim-blaming, or unsupported assumptions. \\
  & No  & Contains no identified bias or unsupported assumption. \\
\midrule
\multirow{3}{*}{\makecell[l]{Support\\Information}}
  & Yes & Provides a specific external support resource. \\
  & No+ & Offers to help identify support or asks for information needed to provide a referral. \\
  & No  & Provides no external support information or referral assistance. \\
\bottomrule
\end{tabular}
}
\end{table}

\section{Automated Evaluator Validation}

Table~\ref{tab:per_class_performance_tech_eval} reports per-class performance for the automated technical evaluators. The webpage validation set included forum-style pages such as Reddit \& Quora posts with comment threads returned by Google Search, but the automated evaluators was not directly validated on \emph{r/Stalking} threads.

\shirin{Please fill/fix: Add the direct Reddit-thread validation results when available and clarify the averaging scheme used for the aggregate precision, recall, and F1 scores in the main paper.}\nowshin{added the details about macro average}

\begin{table}[t]
\centering
\caption{Per-class performance of automated evaluation pipelines.}
\label{tab:per_class_performance_tech_eval}
\resizebox{\columnwidth}{!}{%
\begin{tabular}{p{0.24\linewidth}p{0.16\linewidth}p{0.20\linewidth}cccc}
\toprule
\textbf{Response set} & \textbf{Metric} & \textbf{Class} &
\textbf{Prec.} & \textbf{Rec.} & \textbf{F1} & \textbf{Support} \\
\midrule

\multirow{7}{=}{Conversational-system responses}
& \multirow{2}{*}{Relevance}
& Relevant       & 0.93 & 0.96 & 0.94 & 43 \\
& & Not Relevant & 0.67 & 0.57 & 0.62 & 7 \\
\cmidrule(lr){2-7}

& \multirow{2}{*}{Persuasiveness}
& Persuasive       & 0.88 & 0.97 & 0.92 & 37 \\
& & Not Persuasive & 0.89 & 0.62 & 0.73 & 13 \\
\cmidrule(lr){2-7}

& \multirow{3}{*}{Actionability}
& Actionable       & 0.91 & 0.83 & 0.87 & 24 \\
& & Informative    & 0.72 & 0.84 & 0.78 & 19 \\
& & Not Actionable & 0.83 & 0.71 & 0.77 & 7 \\

\midrule

\multirow{7}{=}{Webpages and forum-style pages}
& \multirow{2}{*}{Relevance}
& Relevant       & 0.757 & 0.903 & 0.824 & 31 \\
& & Not Relevant & 0.943 & 0.848 & 0.893 & 59 \\
\cmidrule(lr){2-7}

& \multirow{2}{*}{Persuasiveness}
& Persuasive       & 1.000 & 0.867 & 0.929 & 83 \\
& & Not Persuasive & 0.607 & 1.000 & 0.756 & 17 \\
\cmidrule(lr){2-7}

& \multirow{3}{*}{Actionability}
& Actionable       & 0.828 & 0.774 & 0.800 & 31 \\
& & Informative    & 0.800 & 0.762 & 0.780 & 42 \\
& & Not Actionable & 0.806 & 0.926 & 0.862 & 27 \\

\bottomrule
\end{tabular}%
}
\end{table}

\section{Understandability/Readability}
\label{sec:understandability}
\textbf{Understandability of Online Support Systems. }We operationalize Understandability using the Flesch--Kincaid Grade Level, which estimates the U.S. school grade needed to understand a text based on sentence length, word complexity, and syllable count~\cite{flesch-kincaid}.

\textbf{Evaluation Methodology}
We computed scores using the \textit{readability} Python module~\cite{readability-python}. Source-specific preprocessing removed Unicode artifacts and links from webpages, restored punctuation in YouTube transcripts using XLM-RoBERTa~\cite{1800badcode2023punctuators}, and removed URLs, hashtags, emojis, and punctuation-only text from Reddit comments. Table~\ref{tab:techmetrics} summarizes score interpretation; responses at or below grades 9--12 were treated as understandable for general audiences.
\begin{figure}[h]
\centering

\subfloat[Google Websites]{
    \includegraphics[width=0.47\linewidth]{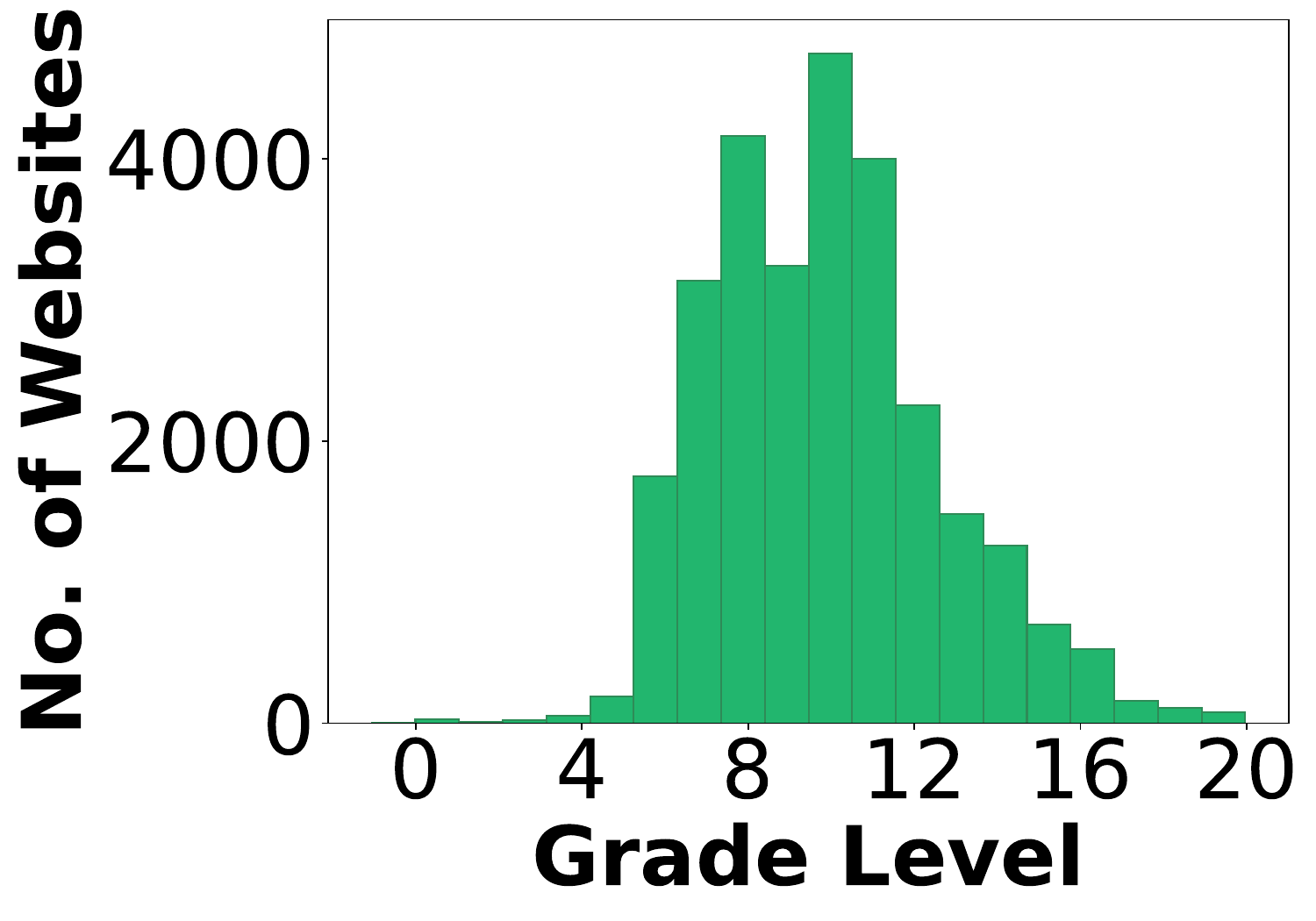}
    \label{google_search_understandability_victim}
}
\subfloat[Reddit Comments]{
    \includegraphics[width=0.47\linewidth]{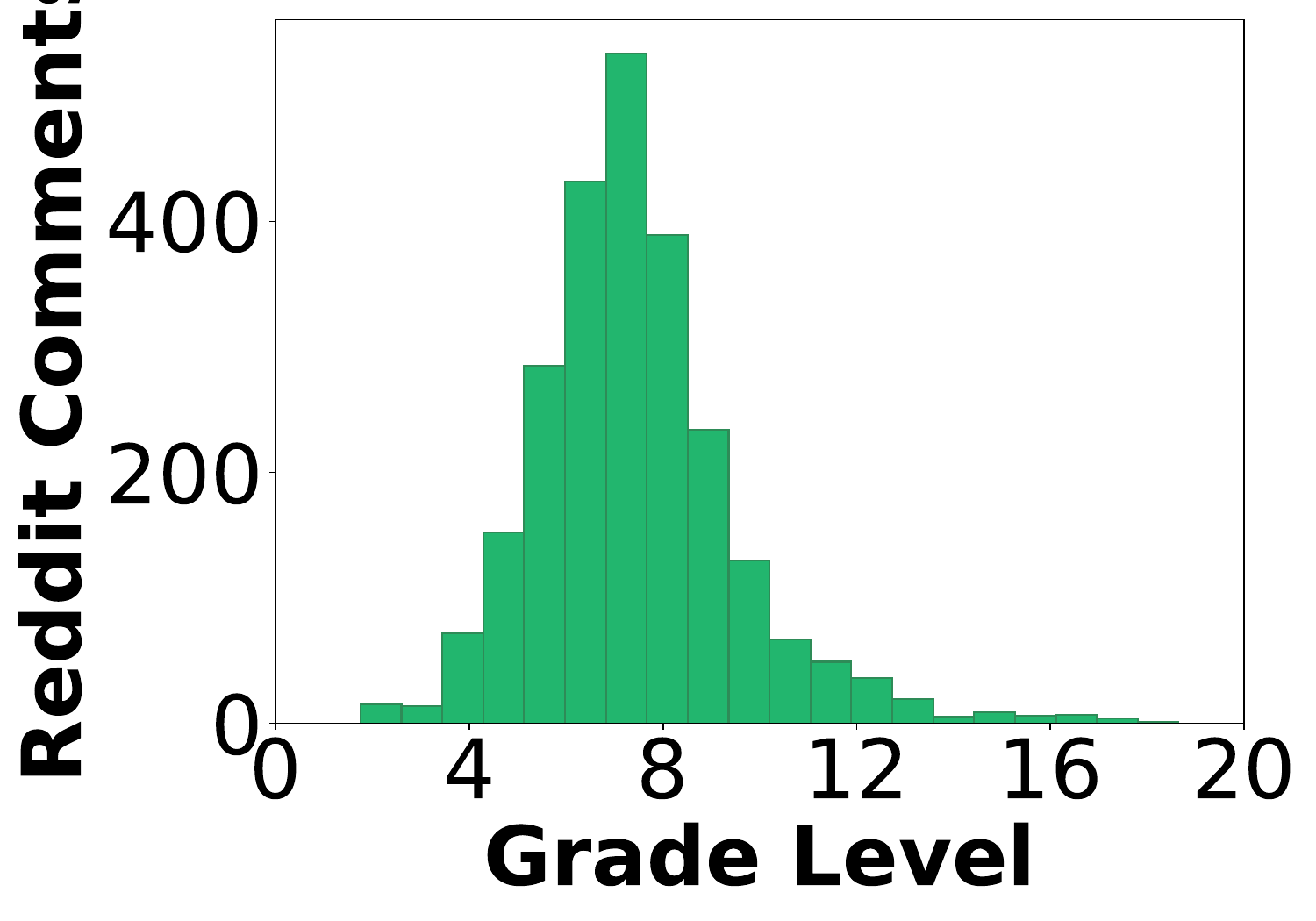}
    \label{reddit_comments_understandability_victim}
}
\hfill
\subfloat[LLM/Chatbots]{
    \includegraphics[width=0.47\linewidth]{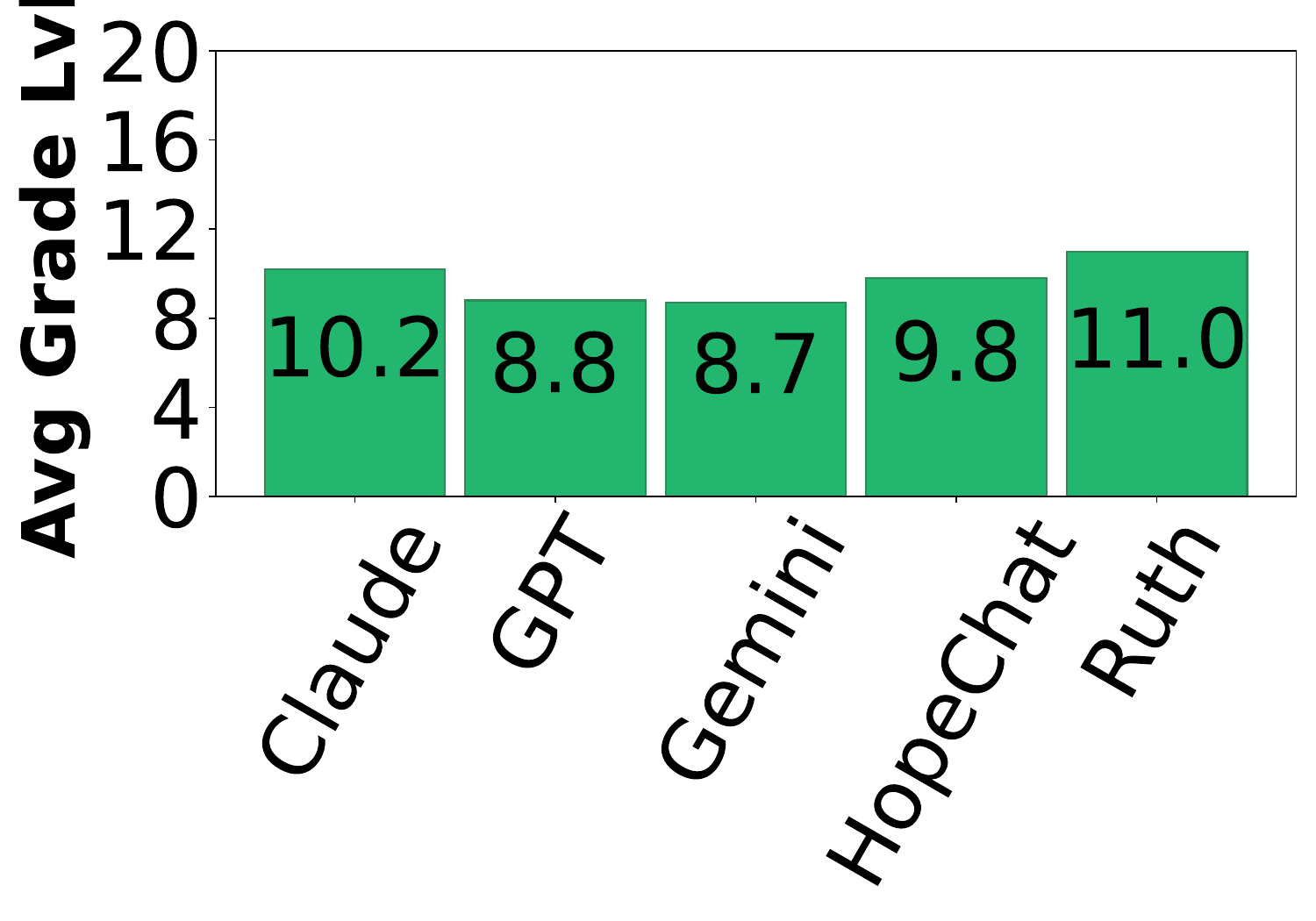}
    \label{understandability_by_model}
}
\caption{Grade level scores for Understandability evaluation}

\label{fig:understandability}
\end{figure}

\textbf{Results}
Figure~\ref{fig:understandability} shows that Google Search webpages were typically written at grades 8--12, with some requiring college-level reading. Reddit comments were generally easier to read (grades 5--8), while LLM responses fell mostly between grades 6 and 12, indicating broadly accessible guidance to general population.

\section{Uncertainty for Conversational-System Estimates}

Table~\ref{tab:llm_ci} reports 95\% Wilson confidence intervals for outcomes measured on the 50 responses from each conversational system. The intervals emphasize the uncertainty associated with this sample size and discourage overinterpretation of small differences between systems.


\begin{table*}[h]
\centering
\caption{Model-level estimates with 95\% Wilson confidence intervals.}
\label{tab:llm_ci}
\resizebox{0.92\textwidth}{!}{%
\begin{tabular}{lccccc}
\toprule
\textbf{Metric} &
\textbf{Claude} &
\textbf{GPT} &
\textbf{Gemini} &
\textbf{HopeChat} &
\textbf{Ruth} \\
\midrule

Relevant
& 88 [76.2--94.4]
& 96 [86.5--98.9]
& 96 [86.5--98.9]
& 72 [58.3--82.5]
& 92 [81.2--96.8] \\

Fully or Partially Supported
& 64 [50.1--75.9]
& 70 [56.2--80.9]
& 80 [67.0--88.8]
& 24 [14.3--37.4]
& 70 [56.2--80.9] \\

Damaging Guidance
& 24 [14.3--37.4]
& 26 [15.9--39.6]
& 16 [8.3--28.5]
& 12 [5.6--23.8]
& 20 [11.2--33.0] \\

Actionable
& 46 [33.0--59.6]
& 68 [54.2--79.2]
& 88 [76.2--94.4]
& 6 [2.1--16.2]
& 46 [33.0--59.6] \\

Persuasive
& 82 [69.2--90.2]
& 100 [92.9--100.0]
& 98 [89.5--99.6]
& 48 [34.8--61.5]
& 52 [38.5--65.2] \\

\bottomrule
\end{tabular}%
}
\end{table*}

\section{Supplemental Exposure Distributions}

\begin{figure}[h!]
\centering
\hfill
\subfloat[Flagged secondary URLs]{
    \includegraphics[width=0.37\linewidth]{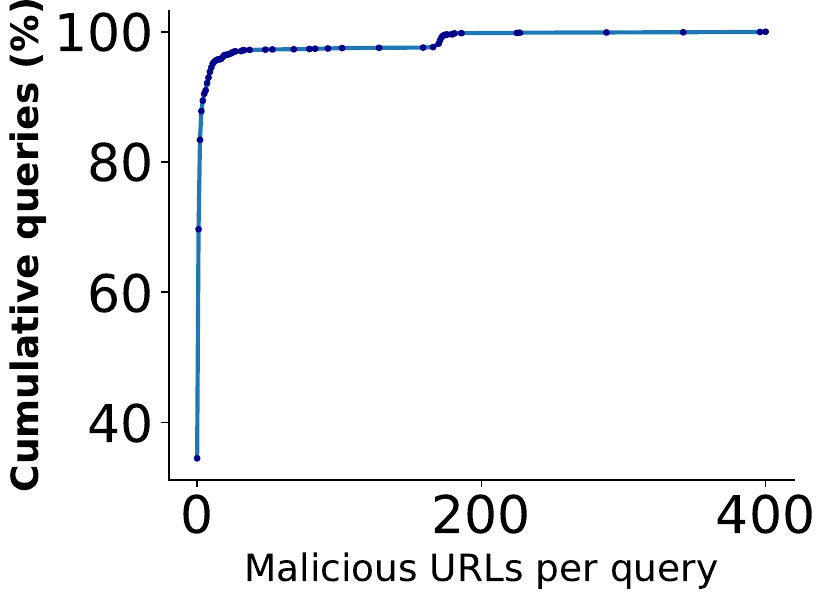}
    \label{fig:percent_queries_malicious_victim}
}%
\hfill
\subfloat[Comments flagged for toxicity]{
    \includegraphics[width=0.43\linewidth]{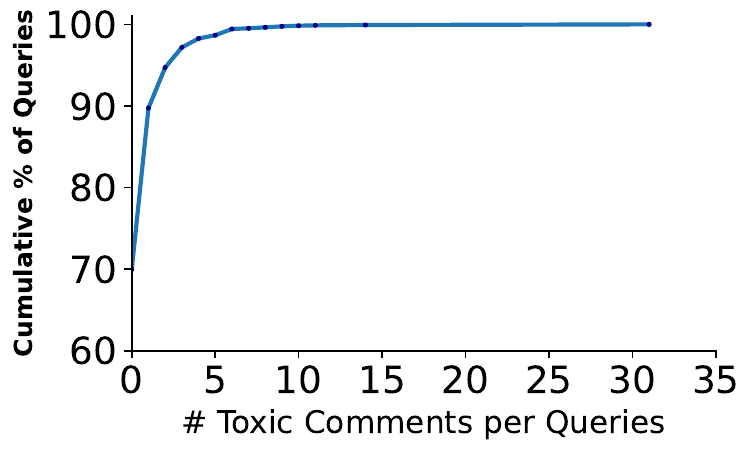}
    \label{fig:CDF_toxicity_per_query_victim}
}
\caption{Query-level distributions of flagged secondary URLs and comments flagged for toxicity.}
\label{fig:cdf}
\end{figure}

Figure~\ref{fig:cdf} presents the query-level  cumulative distributions of the two platform-specific safety risks examined for Google Search and Reddit. For Google Search, the figure shows how many victim queries were exposed to secondary URLs that were flagged as malicious by VirusTotal engines across the retrieved webpages. For Reddit, it shows the distribution of comments flagged for toxic or harmful discourse within the comment threads associated with each victim query.

\section{Technical Accuracy of LLMs by Tech-Misuse Categories}
\label{sec:tech_acc_appendix}

\begin{figure}[h]
    \centering

    \subfloat[Claude]{
        \includegraphics[width=0.47\linewidth]
        {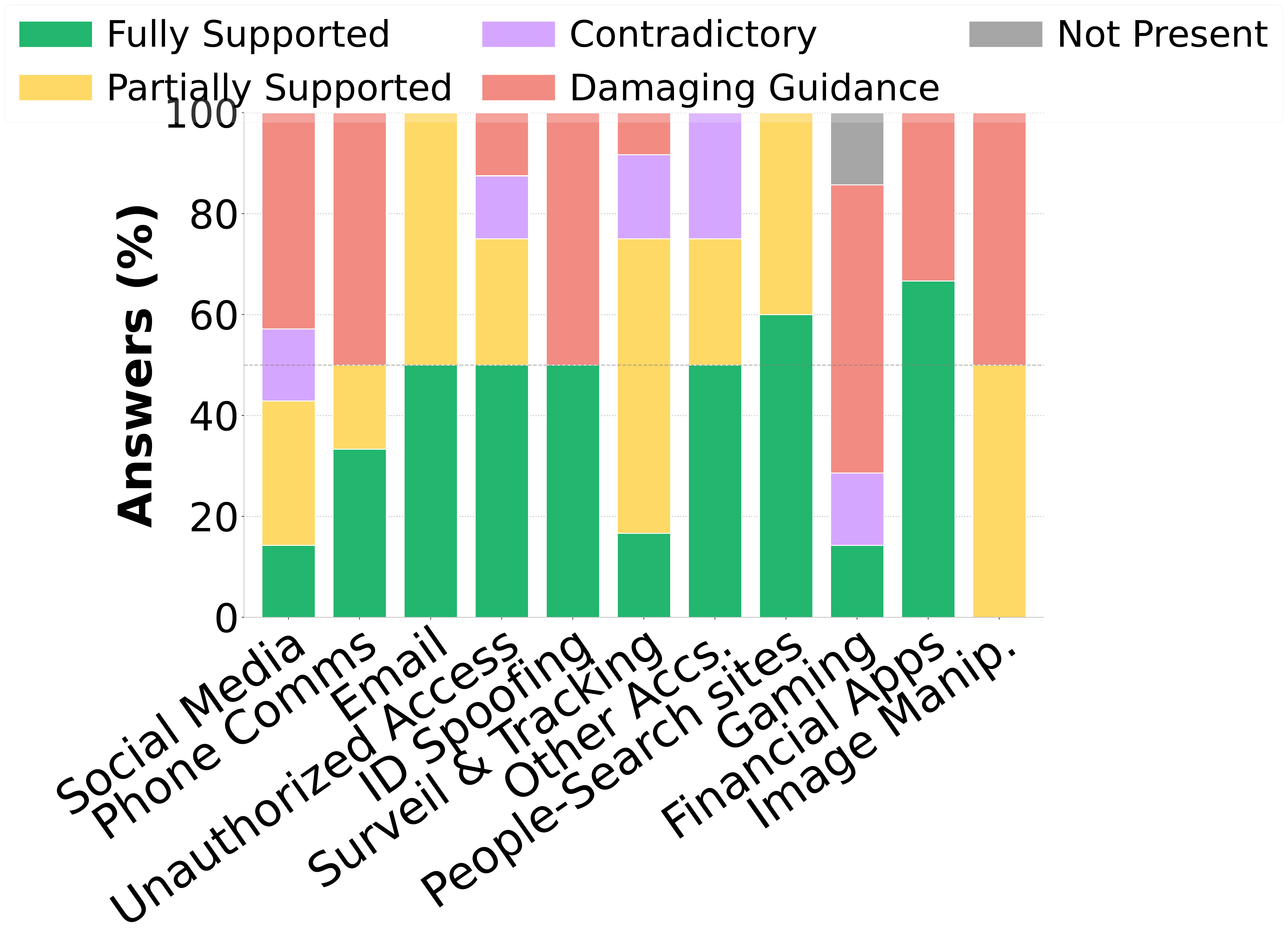}
        \label{fig:accuracy_tech_claude}
    }
    \hfill
    \subfloat[GPT]{
        \includegraphics[width=0.47\linewidth]
        {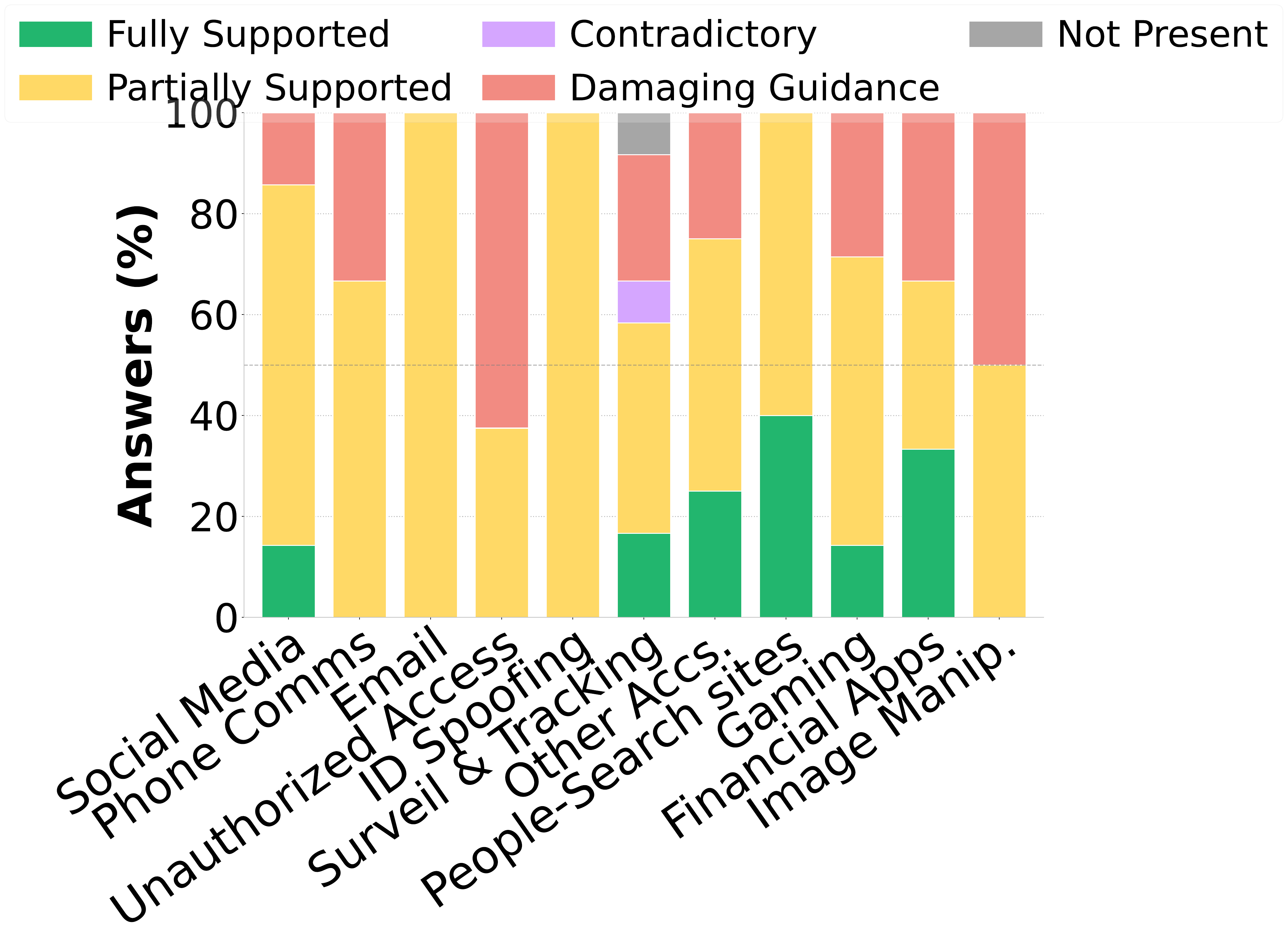}
        \label{fig:accuracy_tech_gpt}
    }
    \hfill
    \subfloat[Gemini]{
        \includegraphics[width=0.47\linewidth]
        {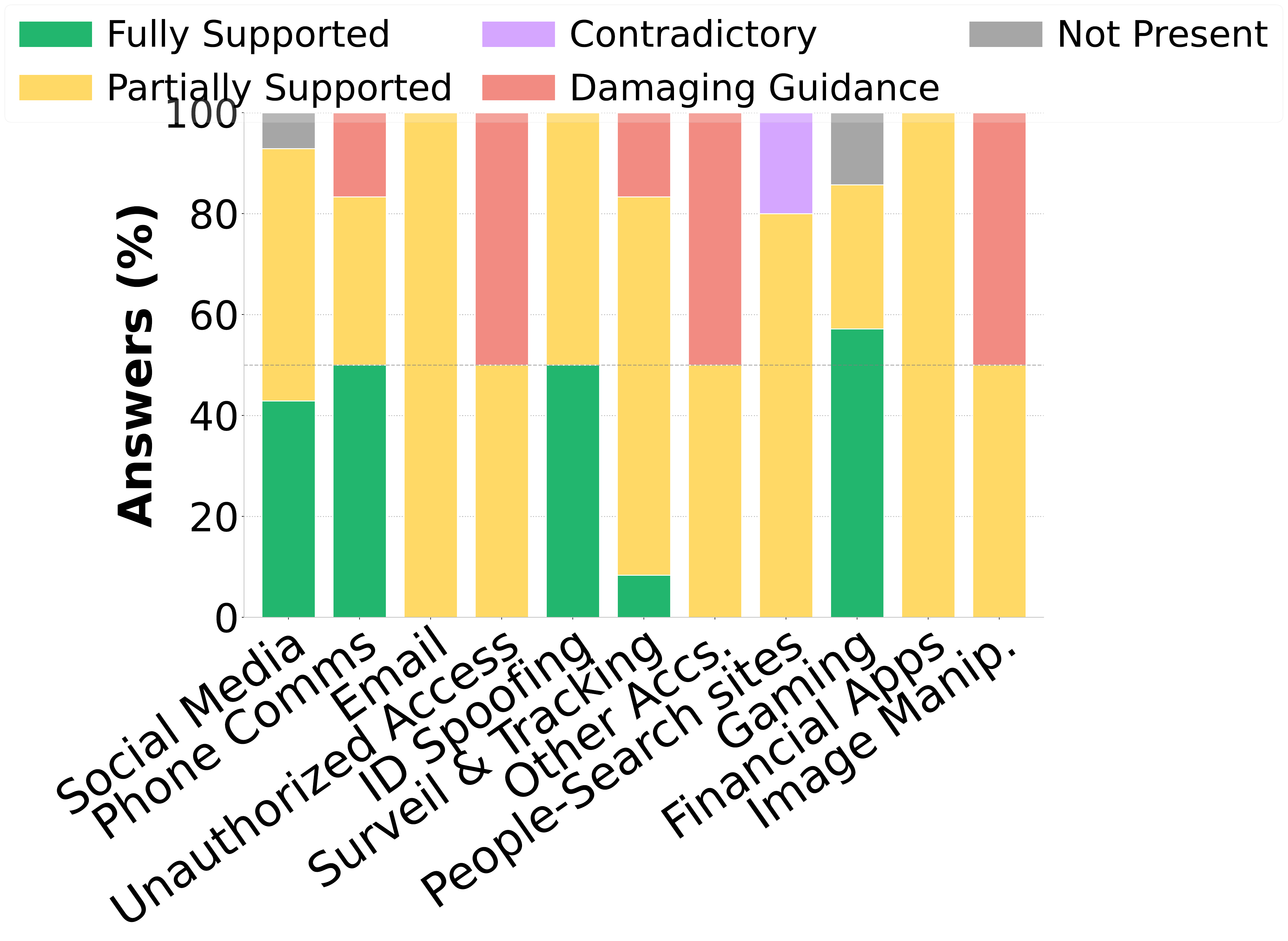}
        \label{fig:accuracy_tech_gemini}
    }

    \hfill
    \subfloat[HopeChat]{
            \includegraphics[width=0.45\linewidth]
            {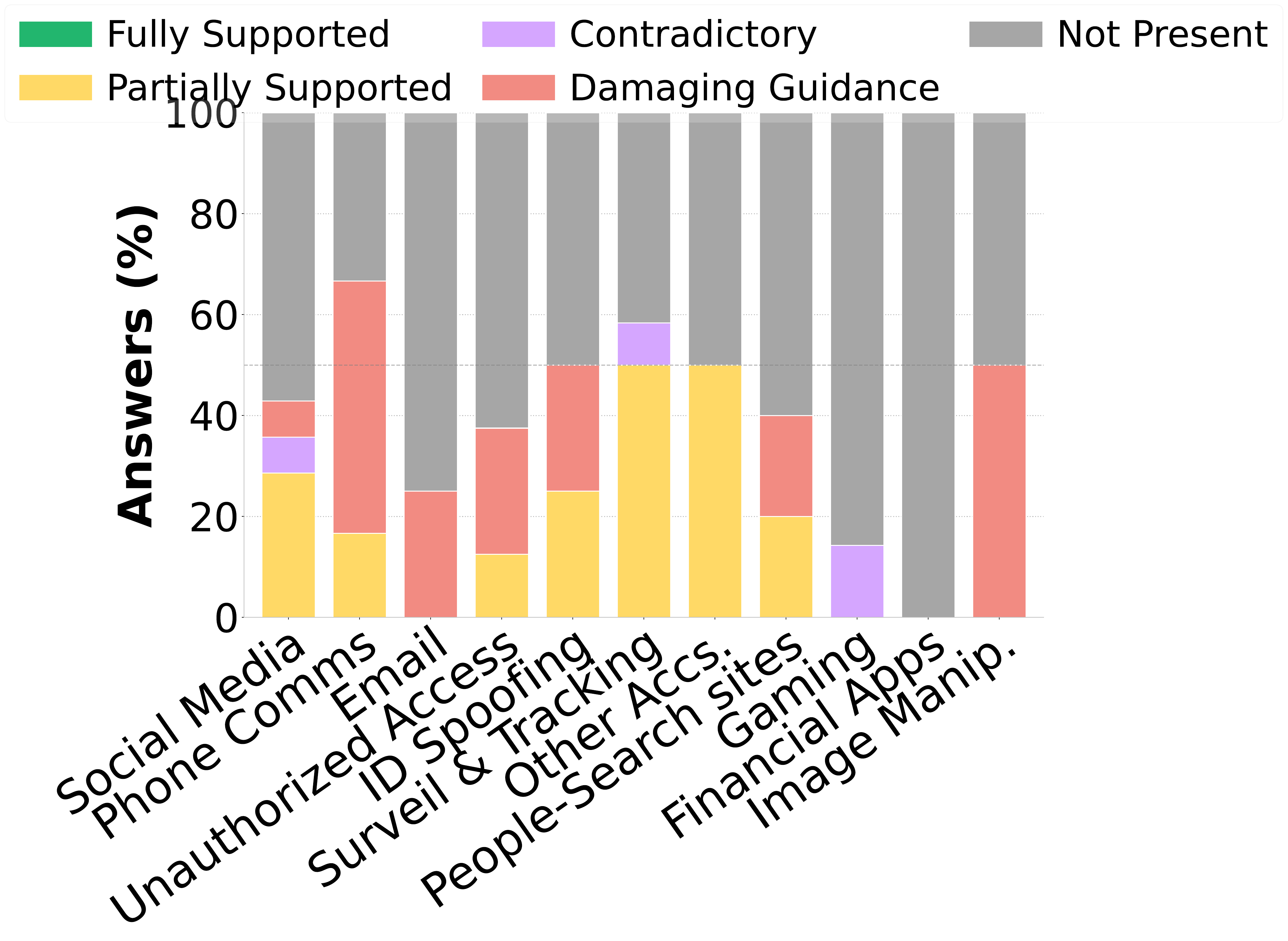}
            \label{fig:accuracy_tech_hopechat}
        }
        \hfill
        \subfloat[Ruth]{
            \includegraphics[width=0.45\linewidth]
            {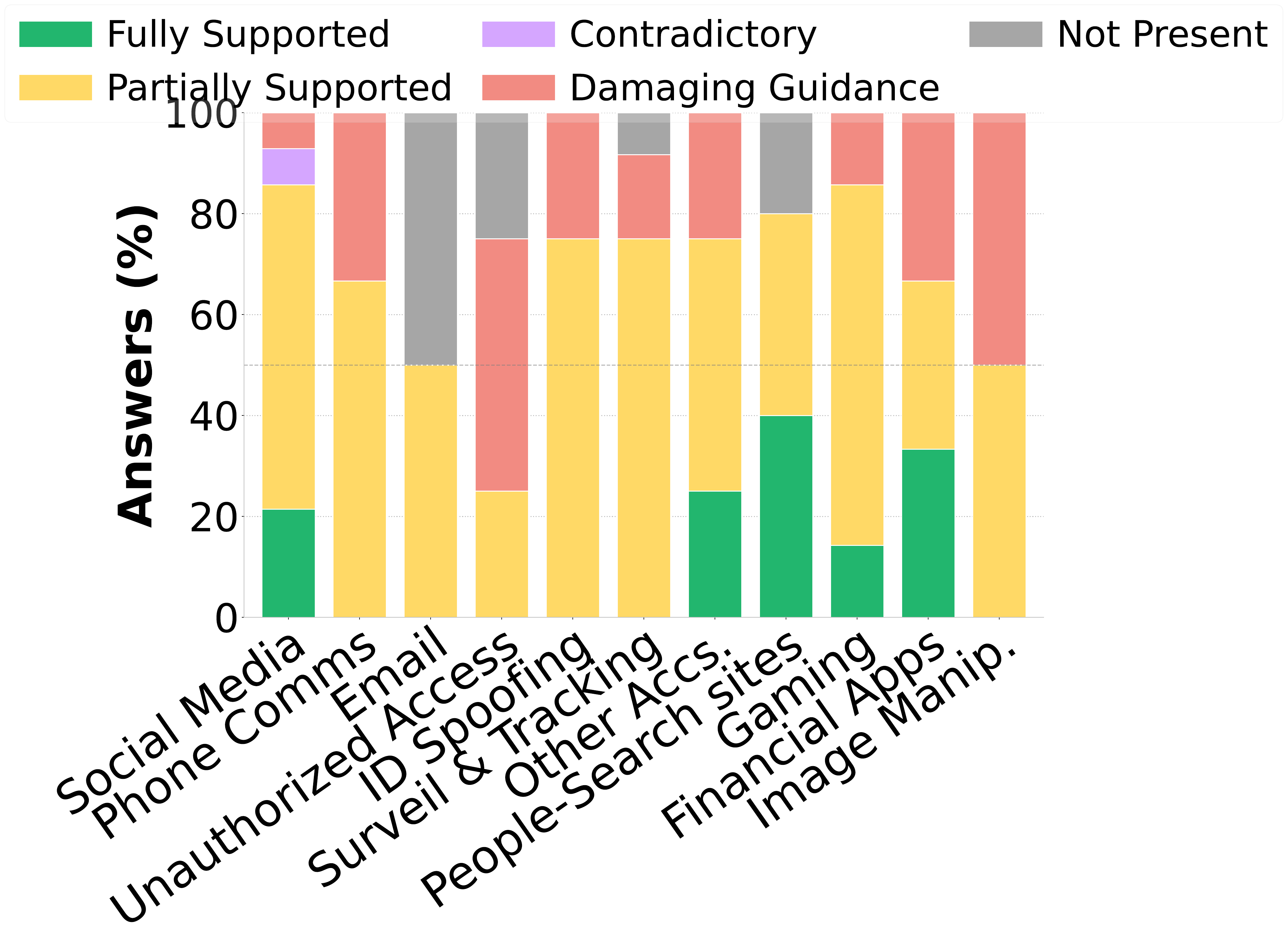}
            \label{fig:accuracy_tech_ruth}
        }

    \caption{Accuracy of LLMs across tech-misuse categories.}
    \label{fig:llm_accuracy_by_technology}
\end{figure}

Figure~\ref{fig:llm_accuracy_by_technology} presents the accuracy of each conversational system across the different technology-misuse categories represented in the victim queries. Across the three general-purpose LLMs, \textit{Social-Media} and \textit{Gaming Platforms} related queries showed the strongest and most consistent accuracy, whereas \textit{Image \& Video Manipulation} showed the weakest performance and a higher prevalence of damaging guidance. Among the domain-specific chatbots, performance was comparatively strongest for categories such as \textit{Surveillance and Tracking Technologies} and \textit{Other Online Service Accounts}, while \textit{Unauthorized Access and Device Compromise} was again among the weakest categories, particularly for HopeChat.
